\documentclass[10pt]{article}
\usepackage{multicol}
\usepackage{pdflscape}
\usepackage{gensymb}
\newcommand{\etal}{\emph{et al~}}

\newcommand{\Ukm}{\ensuremath{\mathrm{\;km}}}
\newcommand{\UkmZ}{\ensuremath{\mathrm{km}}}
\newcommand{\Ug}{\ensuremath{\mathrm{\;g}}}

\newcommand{\UdayZ}{\ensuremath{\mathrm{day}}}

\newcommand{\UyrZ}{\ensuremath{\mathrm{yr}}}
\newcommand{\Ukg}{\ensuremath{\mathrm{\;kg}}}
\newcommand{\UkgZ}{\ensuremath{\mathrm{kg}}}

\newcommand{\UminZ}{\ensuremath{\mathrm{min}}}
\newcommand{\Uft}{\ensuremath{\mathrm{\;ft}}}

\newcommand{\Um}{\ensuremath{\mathrm{\;m}}}

\newcommand{\UFL}{\ensuremath{\mathrm{\;FL}}}
\newcommand{\UFLZ}{\ensuremath{\mathrm{FL}}}
\newcommand{\Umm}{\ensuremath{\mathrm{\;mm}}}

\newcommand{\Ucm}{\ensuremath{\mathrm{\;cm}}}
\newcommand{\UcmZ}{\ensuremath{\mathrm{cm}}}

\newcommand{\UumZ}{\ensuremath{\mathrm{\mu m}}}

\newcommand{\UsZ}{\ensuremath{\mathrm{s}}}
\newcommand{\Uh}{\ensuremath{\mathrm{\;h}}}
\newcommand{\UhZ}{\ensuremath{\mathrm{h}}}
\newcommand{\USv}{\ensuremath{\mathrm{\;Sv}}}
\newcommand{\USvZ}{\ensuremath{\mathrm{Sv}}}
\newcommand{\UmSv}{\ensuremath{\mathrm{\;mSv}}}

\newcommand{\UuSv}{\ensuremath{\mathrm{\;\mu Sv}}}
\newcommand{\UuSvZ}{\ensuremath{\mathrm{\mu Sv}}}

\newcommand{\UpSvZ}{\ensuremath{\mathrm{pSv}}}
\newcommand{\UJZ}{\ensuremath{\mathrm{J}}}

\newcommand{\UkeV}{\ensuremath{\mathrm{\;keV}}}

\newcommand{\UGeV}{\ensuremath{\mathrm{\;GeV}}}
\newcommand{\UGeVZ}{\ensuremath{\mathrm{GeV}}}

\newcommand{\UGyZ}{\ensuremath{\mathrm{Gy}}}

\newcommand{\PGg}{\ensuremath{\mathrm{\gamma}}\ }
\newcommand{\PGpp}{\ensuremath{\mathrm{\pi^{+}}}\ }
\newcommand{\PGpm}{\ensuremath{\mathrm{\pi^{-}}}\ }
\newcommand{\PGmp}{\ensuremath{\mathrm{\mu^{+}}}\ }
\newcommand{\PGmm}{\ensuremath{\mathrm{\mu^{-}}}\ }
\newcommand{\PGppm}{\ensuremath{\mathrm{\pi^{\pm}}}\ }
\newcommand{\Pem}{\ensuremath{\mathrm{e^{-}}}\ }
\newcommand{\Pep}{\ensuremath{\mathrm{e^{+}}}\ }
\newcommand{\Pepm}{\ensuremath{\mathrm{e^{\pm}}}\ }

\newcommand{\PGmpm}{\ensuremath{\mathrm{\mu^{\pm}}}\ }
\newcommand{\Pp}{\ensuremath{\mathrm{p}}\ }
\newcommand{\Pn}{\ensuremath{\mathrm{n}}\ }

\usepackage{hyperref}
\hypersetup{
    bookmarks=true,         
    unicode=false,          
    pdftoolbar=true,        
    pdfmenubar=true,        
    pdffitwindow=false,     
    pdfstartview={FitH},    
    pdftitle={Radiation Dose Charts for Long Geodetic and Polar Flights with CARI-7A.},    
    pdfauthor={F Qui\~nonez, L A N\'u\~nez, E A Casallas, V S Basto-Gutierrez, J L Gonz\'alez-Arango, P A Ospina-Henao, A J Hern\'andez-G\'oez, C Y P\'erez-Arias. },     
    pdfsubject={Radiation Protection, Medical Physics},   
    pdfcreator={F. Qui\~nonez},   
    pdfproducer={F. Qui\~nonez}, 
    pdfkeywords={EARD, EARDR, CARI-7A}, 
    pdfnewwindow=true,      
    colorlinks=true,       
    linkcolor=black,          
    citecolor=black,        
    filecolor=magenta,      
    urlcolor=blue           
}
\usepackage[all]{hypcap}

\usepackage{slashed}
\usepackage{setspace}
\usepackage{multirow}
\usepackage{graphicx}
\usepackage{subfig}

\usepackage{pifont}
\usepackage{titletoc}
\usepackage{capt-of}
\usepackage{calc}
\usepackage{morefloats}
\usepackage[top=3cm, bottom=3cm, left=3cm, right=3cm]{geometry}

\usepackage[dvipsnames]{xcolor}

\begin{document}
\title{\textbf{Radiation Dose Charts for Long Geodetic and Polar Flights with CARI-7A.}}
\author{F Qui\~nonez${}^{1}$, L~A N\'{u}\~{n}ez${}^{2}$, E~A Casallas${}^{1}$, 
\\V~S Basto-Gonzalez${}^{3}$, J L Gonz\'alez-Arango${}^{3}$, P~A Ospina-Henao$^{1,4}$,\\A~J Hern\'andez-G\'oez${}^{2}$ and C~Y P\'erez-Arias${}^{2}$
 \\\\${}^{1}$ \emph{Grupo de Investigaci\'on en Electr\'onica y Tecnolog\'ias para la Defensa - TESDA,}
\\\emph{Fuerza A\'erea Colombiana, CO 250030, Madrid, Colombia.} \\\\
${}^{2}$ \emph{Grupo de Investigaci\'on en Relatividad y Gravitaci\'on.} 
\\\emph{Escuela de F\'isica, Universidad Industrial de Santander,} 
\\\emph{A.A. 680002, Bucaramanga, Colombia.}\\\\
${}^{3}$ \emph{Grupo de Investigaci\'on INTEGRAR. Departamento de F\'isica,}
\\\emph{Universidad de Pamplona, Pamplona, Colombia.} \\\\
${}^{4}$ \emph{Grupo de Investigaci\'on en Ciencias B\'asicas y Aplicadas.}
\\\emph{Departamento de Ciencias B\'asicas. Universidad Santo Tom\'as.} 
\\\emph{CO 680001, Bucaramanga, Colombia.}
}
\date{December, 2019}

\maketitle

\begin{abstract}
We have calculated by using CARI-7A the effective and ambient radiation doses absorbed by a reference human phantom inside aircrafts with cruise speed, and averaged ascent/descent rates as presented by real airplanes such as: Airbus 380, Boeing 777, 787, Hercules C-130 and Twin Otter DHC-6, in a sample of flight aerial routes in a range of flight levels. In the sample, fifteen routes are between the longest non-stop flight around the world, where four flights pass over the Arctic and we have considered one air route involving the Antarctica with two additional special cases. The Earth curvature based on model WGS84 was taken into account for every point in each route by using the INVERSE3D software. Then we constructed radiation dose charts according to the ICRP-103 conditions by using the radiometric results coming from the CARI-7A simulations. These charts present the effective absorbed radiation dose as a function of flight level and flight time in a year of normal operations of the crew as recommended by the FAA. We have also characterized geographically and kinematically the cosmic effective radiation absorption rates at Earth's atmosphere in terms of the structure and order of polynomials in the altitude. From the flight with the leading rate of irradiation we found a thumb rule to estimate the effective absorbed radiation dose and the ambient equivalent dose that gives a superior threshold for any flight of the sample and perhaps on Earth, valid in altitude interval [20000, 50000] ft.
\end{abstract}
%
\section{Introduction}
The study of the interaction of radiation with materials presents both, the good and bad aspects of it. 
The good use of radiation involves its application in purification of water, plants, and prevention of foodborne pathogens \cite{Salunkhe1961}, 
detection of illegal weapons and drugs in ports and airports, particle therapy in cancer treatments, genetics, 
the improvement of materials used in aerospace technology and shielding, etc. see \cite{IAEA2014}, \cite{DuranteNature2017}. 
The research in regard to the bad aspects of radiation is focused  mostly on the cellular 
damage and the consequent diseases in humans
\cite{ICRP60}, \cite{ICRP91},\cite{ICRP103}, \cite{Durante2011}, \cite{MaaloufDuranteForay2011}.
In the past, radiation was the designation given to the 
photons (gamma radiation), electrons and positrons (beta radiation) and
He nuclei (alpha radiation) emitted by the radioactive elements, but in the present time we have to consider 
all the zoo of particles that have energy above or below a well known established threshold for each one of them. 
We also have to include the cosmic rays (CR) in the designation of radiation, which are mostly light atomic nuclei. 
Cosmic ray spectrum is composed of free protons 79\%, alpha particles 15\%,
and the rest are atomic nuclei from $Z=3,4\ldots$, see \cite{PDG2016,PDG2018}.  
The cosmic rays affects principally to the astronauts, and when CR have not decayed freely or by means of collisions 
with either the astronaut molecules or the atmosphere molecules or something else, they are called \emph{primary particles} or simply \emph{primaries}.
After their first decay the particle daughters are called \emph{secondary particles} or \emph{secondaries}.
For aviation purposes, we will deal with secondaries travelling in the altitude range between $20000\Uft$ and $50000\Uft$. 
Secondaries that hit a human phantom can pass entirely through it without do any decay or without any interaction with the particle constituents of the 
human phantom. On the other hand, we could have cases when the incoming secondary particle gives energy to the human phantom by means of 
a very important kinematic variable called \emph{stopping power}.

\section{Methods}
\label{ConceptsMethods}
\subsection{Radiation dose}
From the particle physics literature \cite{PDG2016}
we found that stopping power can be decomposed as the sum of three types of interactions
$ 
S = S_{\Pem} + S_{N} + S_{rad},
$
where the subindices $\Pem$, $N$, and $rad$ mean that the incoming particle can lose energy
given it to: 
the atomic electrons, 
the nucleons, 
and the medium with consequent atomic or molecular radiative de-excitation of the medium, respectively.
Stopping power $S=d\epsilon/dl$ has units of energy per length. If we divide it by the medium density $\rho$, we
obtain the \emph{mass stopping power}, which has units of cross section times energy per mass, i.e. units of area times radiation dose.
When an amount of radiation of type $R$ with infinitesimal energy $d\epsilon$ is absorbed by a body 
of infinitesimal mass $dm$, we can define the
absorbed radiation dose due to $R$ as 
$
D_{R} = d\epsilon/dm.
$
In The International System of Units, the radiation dose is either measured in 
\emph{Grays} $\UGyZ=\UJZ/\UkgZ$
when the body is not biological; or in \emph{Sieverts} 
$\USvZ=\UJZ/\UkgZ$, when the body is biological.
Similarly, if the body is a tissue with mass $m_{T}$ and absorbs
an averaged energy $\epsilon_{T}$, the \emph{tissue absorbed radiation dose}  
due to $R$ is
\begin{equation}
D_{T,R} = \frac{\bar{\epsilon}_{T}}{m_{T}}.
\label{eq:RD}
\end{equation}
\begin{table}[h!]
\centering
\begin{tabular}{cc}
\hline
Particle & $w_{R}$ \\
\hline
\PGg and X-rays & 1 \\
\Pepm and \PGmpm & 1 \\
\Pp and \PGppm & 2 \\
\Pn & 2-20 \\
$\alpha$ and HI & 20 \\
\hline
\end{tabular}
\caption{Statistical radiation weights, taken from \cite{ICRP103}. We could see in this table that the values are greater or equal than unity.
Certainly radiation weights for radiation composed of neutrinos give us a value near zero, this is less than unity.
For muons at high altitude occurs a situation like presented by neutrinos, 
$w_{R} \approx 0$, because the small size of the 
spherical reference human phantom, and the fact that a muon of $1 \UGeV$ has a mean free path of $5 \Um$ in water,
so the value $w_{R}=1$ can be considered as a conservative value, also taking into account that muons are 
produced at $6 \UGeV$ approximately. 
The standard specifications established that select $w_{R}=20$ for all types and energies of heavy charged particles
seems to be a conservative estimate sufficient for general application in radiological protection.
For analysis in the low Earth orbit (LEO) and beyond into the space, scientists 
have to take a more realistic approach in order to find $w_{R}$, and for that they are using (\ref{eq:HwithQ})
and (\ref{eq:Href}).
}
\label{tab:radiationweights}
\end{table}
Other quantities had been defined in order to calibrate the instruments that measure the radiation and
for monitoring it and analyzing it, such as
the \emph{tissue equivalent absorbed radiation dose}
\begin{equation}
H_{T} = \sum\limits_{R} w_{R} \, D_{T,R},
\label{eq:EqARD}
\end{equation}
With the \emph{statistical radiation weights} $w_{R}$, generally taking any value, but most of the cases this
quantities are greater than one, $w_{R} \geq 1$.
The statistical radiation weights tell us that each kind of particle interacts with the human phantom constituents in a different manner, it indicates us
the proportion of energy that the human phantom absorbs from the kind of particle or radiation $R$, being in correspondence one to one, for the
most of particle types, see Table~\ref{tab:radiationweights}. 
Tissue equivalent absorbed radiation dose (\ref{eq:EqARD}) assumes the human phantom is composed of one kind of a uniform tissue that absorbs all the 
incoming radiation,
which is not the case for the real problem. 
\begin{table}[h!]
  \centering
\begin{tabular}{cc}
\hline
Tissue & $w_{T}$ \\
\hline
Breast, Bone Marrow, Lung, Colon, Stomach  & 0.12 \\
Gonads & 0.08 \\
Bladder, Liver, Esophagus, Tyrhoid & 0.04 \\
Bone Surface, Brain, Salivary Glands, Skin & 0.01 \\
Remainder & 0.12 \\
\hline
\end{tabular}
\caption{Statistical tissue weights, taken from \cite{ICRP103}.
This values are obtained by using a parameter called
the lifetime attributable risk (LAR),
which is an approximation of the risk of exposure-induced death (REID),
it describes the excess deaths (or disease cases) over a follow-up period with population background rates determined 
by the experience of unexposed individuals.}
\label{tab:tissueweights}
\end{table}
The real problem considers the human phantom is composed of different tissues, each one with different radiation absorption represented by the 
\emph{statistical tissue weights} $w_{T}$. 
Therefore is defined the \emph{effective absorbed radiation dose} (EARD) as
\begin{equation}
E = \sum\limits_{T} w_{T}\,H_{T} 
= \sum\limits_{T}\sum\limits_{R}  w_{T}w_{R} \, D_{T,R},
\label{eq:EARD}
\end{equation}
where the normalization of the statistical tissue weights requires
$
\sum_{T} w_{T} = 1.
$
Tables~\ref{tab:radiationweights},~\ref{tab:tissueweights} 
show the values of the statistical weights for different
types of radiation and tissues. 
The use of statistical tissue weights
in (\ref{eq:EARD}) can be understood as the fact that not all 
the deterministic effects of the absorbed radiation dose, coming from GCR or any other source of radiation, 
may effectively produce cellular damage to the tissues. 
When the human phantom has planned to get some time of exposure to radiation, it is convenient to define
the \emph{commitment effective absorbed dose} in an interval of time $\tau$, given by
\begin{equation}
E_{c}(\tau) = \int\limits_{0}^{\tau} dt\; \frac{dE(t)}{dt},
\label{eq:CD}
\end{equation}
where $dE/dt$ is the \emph{effective absorbed radiation dose rate} (EARDR); $dE/dt$ is approximately equals to $dH/dt$ that is given by the monitoring devices usually in
$\UuSvZ/\UhZ$ units. For pilots the FAA established a maximum of $\tau=1000 \Uh$ of flight in a year.
\subsection{Statistical Weights}
Indeed, the algorithm to determine the statistical weights is much more complicated than the specified here.  
To determine the \emph{Lifetime Attributable Risk} (LAR), it takes into account 50 years of observations for adults, 
70 years for children, as well as genetic and environmental factors averaged over both genders and all ages. 
Initially, the \emph{equivalent absorbed dose} was merely the radiation dose times the quality factor, i.e. 
\begin{equation}
H = DQ \, ,
\label{eq:H=DQ}
\end{equation}
with $D$ taken as in the eq.~(\ref{eq:RD}) and the \emph{quality factor} $Q$ as a function of the 
\emph{linear energy transfer} (LET) $L=d\epsilon/dl$ \cite{Durante2014, DurantePaganetti2016}, i.e. 
\begin{equation}
Q(L) = 
\left\{
\begin{array}{c@{\quad}l}
1, & L < 10\UkeV/\UumZ \, , \\
0.32\, L - 2.2, & 10 \leq L \leq 100 \UkeV/\UumZ\, , \textrm{and} \\
300\,L^{-1/2}, & L>100\UkeV/\UumZ \, ; \\
\end{array}
\right.
\label{eq:QualityFactor}
\end{equation}
However, eq.~(\ref{eq:H=DQ}) has a better accuracy, if we implement
\begin{equation}
H_{T} = \frac{1}{m} \int dm \int dL \; L \, Q(L) \, F_{T}(L),
\label{eq:HwithQ}
\end{equation}
where $F_{T}(L)$ is the particle fluence through the tissue $T$. 
The integral (\ref{eq:HwithQ}) must be calculated numerically at least for 33 points and, any experimental device designed to track particles across an organ (or tissue) should implement this number of points \cite{Durante2011}. 
The calibration for  $w_R$, $w_T$, and $Q$ are in such a way that deaths by a  long time risk radiation exposure should not be higher than 3\% in 10 years. 
The \emph{International Commission on Radiological Protection}, ICRP \cite{ICRP60}, \cite{ICRP103}, had defined the radiation assessment quantities 
based on a human phantom reference called the 
\emph{ICRU sphere}, with diameter of $30 \Ucm$, density of $1 \Ug/\UcmZ^{3}$, made by oxygen 76.2\%, carbon 11.1\%, hydrogen 10.1\%, and nitrogen 2.6\% \cite{ICRU33}, \cite{ICRU39}. 
The \emph{ambient  equivalent radiation dose}, $H^{*}(10)$, indicates the absorbed radiation dose of the radiation field aligned radially onto the ICRU sphere with a 
penetration distance of $10\Umm$. 
\\\\We can also can get the EARD averaged over a group of people exposed to a radiation field. At first, we need the total number of people that absorb an effective radiation dose in the range between $E_{a}$ and $E_{b}$ in a time interval $\Delta t$,
\begin{equation}
N (E_{a}, E_{b}, \Delta t) = \int\limits_{E_{a}}^{E_{b}} dE \;\Bigg[\frac{dN}{dE}\Bigg]_{\Delta t}
\approx \sum\limits_{i: E_{a}<E_{i}<E_{b}} N_{i},
\end{equation}
where $[\frac{dN}{dE}]_{\Delta t}$ is the distribution of $E$ with frequency equals to the number of people $N$ irradiated by $E_{i}$ in the time interval $\Delta t$ and in the energy range $E_{a}<E_{i}<E_{b}$. Secondly, we need the collective effective dose
\begin{equation}
S (E_{a}, E_{b}, \Delta t) = \int\limits_{E_{a}}^{E_{b}} dE \;E\,\Bigg[\frac{dN}{dE}\Bigg]_{\Delta t}
\approx \sum\limits_{i: E_{a}<E_{i}<E_{b}} E_{i}N_{i},
\end{equation}
with SI units $[S]=\USv\cdot\mathrm{person}$. Consequently, we can get the averaged EARD
as
\begin{equation}
\bar{E}(E_{a}, E_{b}, \Delta t) = \frac{S(E_{a}, E_{b}, \Delta t)}{N(E_{a}, E_{b}, \Delta t)}.
\end{equation}
Finally, we can get the variables used as threshold reference values by averaging male (M) and female (F) genders:
\begin{equation}
\bar{E}_{ref} = \frac{1}{2}(\bar{E}_{M} + \bar{E}_{F}) 
\label{eq:Eref}
\end{equation}
\begin{equation}
H_{ref} = \frac{1}{2}(H_{M} + H_{F}).
\label{eq:Href}
\end{equation}
Table~\ref{tab:doses} shows the reference values for each situation of radiation exposure for $\bar{E}_{ref}$, for  low Earth orbit (LEO) and beyond into the space, we  have to take a more realistic approach to find $w_{R}$, correspondint to equations (\ref{eq:HwithQ}) and (\ref{eq:Href}) (further details about these concepts can be gathered from \cite{Durante2011,Durante2014, DurantePaganetti2016} and references therein). 
\begin{table}[h!]
	\centering
	\begin{tabular}{cc}
		\hline
		Case & Radiation \\
		\hline
		Effective Dose People on Earth & $1\UmSv/\UyrZ$. \\
		Effective Dose Radiation Workers on Earth & $5\UmSv/\UyrZ$. \\
		Dose Rate in Low Earth Orbit (LEO) is & $1\UmSv/\UdayZ$. \\
		Dose Rate in Mars is & $100-200$\UmSv/\UyrZ. \\
		Dose Rate in Moon is & $350\UmSv/\UyrZ$. \\
		\hline
	\end{tabular}
	\caption{Threshold values for each situation of radiation exposure. These values are taken by using (\ref{eq:Eref}) presented in \cite{ICRP103}.}
	\label{tab:doses}
\end{table}
\subsection{Solar Radiation and Galactic Cosmic Rays.}
\begin{figure}[h!]
  \centering
\includegraphics[width=6cm]{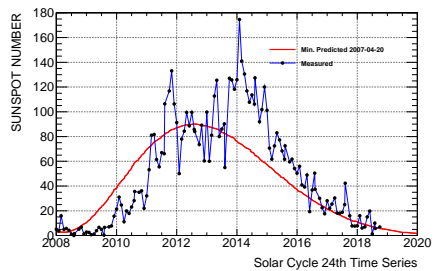}
\caption{Solar cycle 24. In 2018, showed an early minimum of solar radiation. 
Data taken from \cite{SWPC/NOAA} and \cite{SWPC/NOAA2}.}
\label{fig:SolarCycle24}
\end{figure}
The study of the transport of galactic cosmic rays (GCRs) concerns to space meteorology or \emph{Space Weather}.  The spatial extension and temporal duration may classify the physical transport mechanisms involved.  Among the first are the processes related to the variations of the heliospheric magnetic field and large-scale transient phenomena such as coronal mass ejections. Regarding its duration, there are the events that require periods form several decades (dynamics of the solar dynamo) up to a few hours, such as those produced by the solar wind (see for example \cite{SchrijverSiscoe2010} and references therein).
\\\\In this work, we shall consider cosmic ray energies up to $100\UGeV$. 
At energies below some $\UGeVZ$ the flow of CR is dominated by the solar wind, 
while solar modulation can influence cosmic rays even with energies of a 
few hundred $\UGeVZ$ \cite{Asorey2011a, Dasso2012}. 
\\On the other hand, the effect of the terrestrial geomagnetic field 
--depending on the latitude/longitude of the place considered and the direction 
of propagation of the incident particles-- causes a diffuse cut 
for energies of the order of $10$ \UGeV.
\\\\According to the forecast and measurements of the sunspot numbers, presently Sun modulation activity is at the minimum of its cycle \cite{Hathaway2015}. 
Figure~\ref{fig:SolarCycle24} illustrates that there is less activity than forecasted, with two marked minima of the 24th solar cycle ending on March of the 2018 year.  
\\We did all our calculations for  March 1 of 2018, the real solar minimum, 
and from Figure~\ref{fig:SolarCycle24}, we could see the model forecasts the maximum average sunspot number of 90 for May of 2013, 
but instead, the maximum was in April of 2014 and the minimum is for 2020. So it is convenient to
redo this analysis in the future with some new version ($>7$A) of CARI that incorporates the recent behavior of the Sun in the model.    
Because of the geomagnetic uncertainty, it is not convenient at this moment to do the simulations with CARI-7A for year 2020.
%
\\\\GCR reaches us on Earth modulated by the solar activity, and the present solar radiation minima bring us great opportunities to measure cosmic radiation 
less attenuated by the heliocentric potential or the solar radiation itself. For recent measurements of the fluence of CR taken by the International Space Station on a big range in rigidity see  \cite{AMS-proton-2015}, \cite{AMS-He-2016}, \cite{AMS-antiproton-2016}, \cite{AMS-B-C-2016}. 
%
\subsection{CARI-7A.}
CARI is a software developed by the  Civil Aerospace Medical Institute of the Federal Aviation Administration (FAA) of the United States of America.
From release 6 CARI is able to calculate the whole-body radiation dose from the particle fluence \cite{CARI-7A}.
CARI-7A 
(Version 7, released in 2017) stands as a reference,  
used for academic/research-focused purposes. 
CARI-7A calculates the effective absorbed dose of galactic cosmic radiation received by an individual phantom on an aircraft. 
Users can choose several models of Galactic Cosmic Rays and may calculate four defined types of radiation doses 
ICRP-60 \cite{ICRP60}, ICRP-103  \cite{ICRP103},  ambient $H^{*}(10)$  \cite{ICRP103}, whole-body \cite{CARI-7A} and also can print the particle fluence (particles/area). 
It is also possible to choose the date and time for the calculation and even known solar storms. 
\subsection{Initial Data.}
\begin{figure}[h!]
	\centering
	\includegraphics[width=15cm]{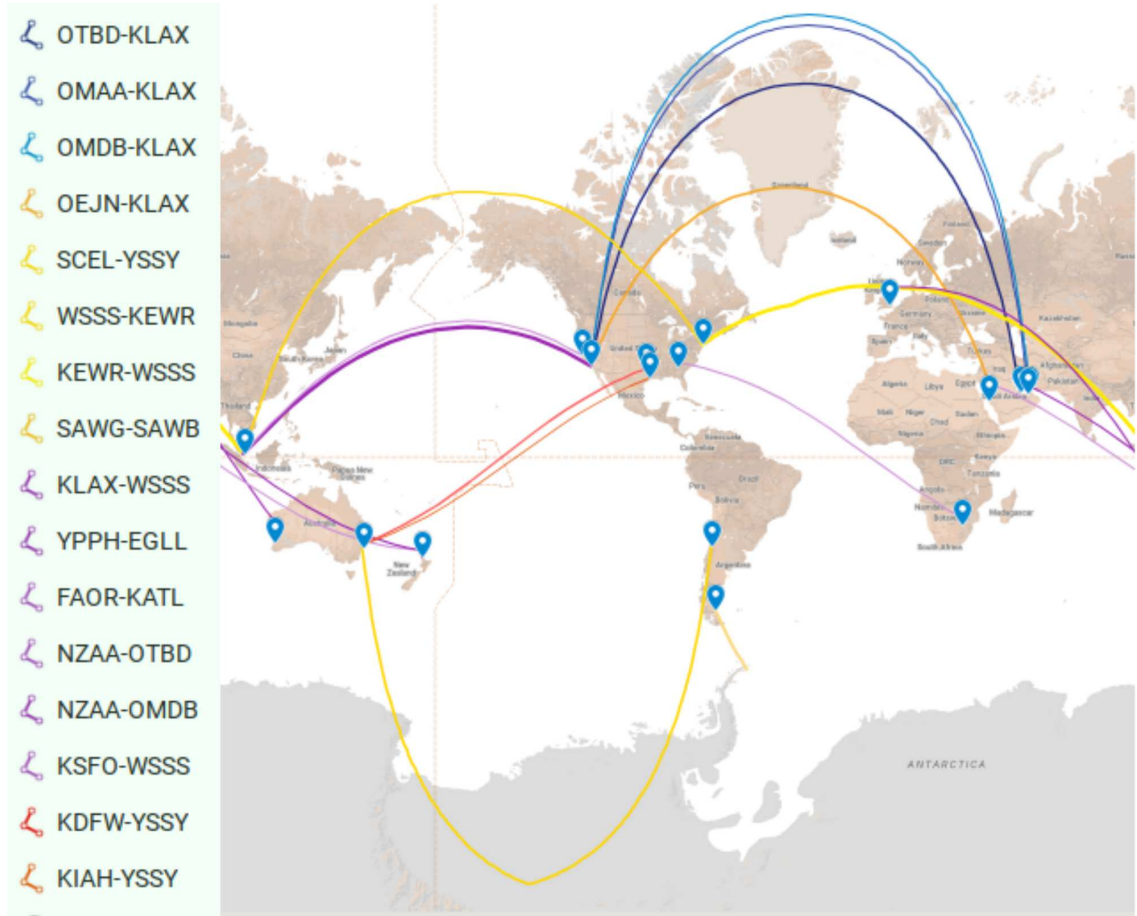}
	\caption{Flight routes studied in this work. 
The figure shows a great circle representation of the routes built by using My Maps from Google. 
Using the International Air Transport Association airport codes (ICAO) we labeled flight as: 
\textbf{OTBD-KLAX} for the route Doha (Qatar) - Los Angeles (USA); 
\textbf{OMAA-KLAX} for Abudabi (UAE) - Los Angeles (USA); 
\textbf{OMDB-KLAX} for Dubai (UAE) - Los Angeles (USA); 
\textbf{OEJN-KLAX} for Jeddah (UAE) -Los Angeles (USA); 
\textbf{SCEL-YSSY} for Santiago (Chile) -Sydney (Australia);
\textbf{KEWR-WSSS} for New York (USA) - Singapore (Singapore); 
\textbf{WSSS-KEWR} for Singapore (Singapore) - New York (USA); 
\textbf{SAWG-SAWB} for Rio Gallegos (Argentina) - Marambio (Antartica); 
\textbf{KLAX-WSSS} for Los Angeles (USA) - Singapore (Singapore); 
\textbf{YPPH-EGLL} for Perth (Australia) - London (UK); 
\textbf{FAOR-KATL} for Johsnnesburg (South Africa) - Atlanta (USA).
\textbf{NZAA-OTBD} for Auckland (New Zealand) - Doha (Qatar);  
\textbf{NZAA-OMDB} for Auckland (New Zealand) - Dubai (UAE); 
\textbf{KSFO-WSSS} for San Francisco (USA) - Singapore (Singapore); 
\textbf{KDFW-YSSY} for Dallas (USA) - Sydney (Australia); 
\textbf{KIAH-YSSY} for Houston (USA) - Sydney (Australia).
Blue colors stand for EARDR $\dot{E}(z)=k_{1}\,z + k_{3}\,z^{3} + k_{5}\,z^{5}$,
yellowish colors stand for $\dot{E}(z)=k_{0}+k_{1}\,z + k_{2}\,z^{2} + k_{3}\,z^{3} + k_{4}\,z^{4}$,
violet colors stand for $\dot{E}(z)=k_{0}+k_{1}\,z + k_{2}\,z^{2} + k_{3}\,z^{3}$,
reddish colors stand for $\dot{E}(z)=k_{0}+ k_{2}\,z^{2} + k_{4}\,z^{4}$, with $z$ the altitude and $k$'s are constants given in Table~\ref{tab:fits}.
}
\label{fig:globo}
\end{figure}

Firstly, we have got  data from the \emph{Flight Plan Database} webpage \cite{flightplandatabase}, checking for the longitude and latitude at every point of the flight route and confront them with the real data registered on the  \emph{Flight Aware} \cite{flightaware} 
and \emph{Flight Radar 24} \cite{flightradar24}  webpages.  After comparison, 
we hold on common points and complete our flight routes synthetically. 
Finally, we use the airports and routes presented in Figure~\ref{fig:globo}.
%
%
%
%
\subsection{Earth Curvature.}
In this work, we had to be very careful when considering the Earth real curvature, since we are dealing with long geodetic flight trajectories,  and also considering that some flight routes cross the actual North Pole. CARI uses the software INVERSE3D, which calculates the geodetic distance between pairs of consecutive points of the route by taking into account the Earth curvature in the basis of the WGS84 model \cite{Bessel1826, Vincenty1975, Vincenty1976, NGS/NOAA}. This way, the user chooses only the two ending points that represent the airports, but this approach does not allow us to choose the flight route points that define each flight route. Therefore, we had to download INVERSE3D and execute it manually, introducing the pair of consecutive points that compose each flight route, and tabulate them with the distance just computed and the time elapsed between consecutive points according to the speed of a specified aircraft.  
\subsection{Aircrafts Velocity Profile.}
We treat the general problem of the aircraft's velocity profile as follows: every aircraft has a regular procedure for takeoff and landing expressed regarding their constant speed of ascent and descent $v_{a}=v_{d}=0.54\cdot v_{c}$. We also assumed that all flights have  constant cruise speed, $v_{c}$, for each level considered. For  Airbus 380 and Boeing 777, we consider $v_{c}=907 \Ukm/\UhZ = 49595 \Uft/\UminZ$;  concerning Boeing 787 $v_{c}=913 \Ukm/\UhZ = 49923 \Uft/\UminZ$; in the case of Hercules C-130,  $v_{c}=540 \Ukm/\UhZ = 29527.838 \Uft/\UminZ$  and finally  $v_{c} = 15200 \Uft/\UminZ$ for Twin Otter DHC-6.  Lastly, we consider the vertical ascent or descent as $v_{z}$ with values  $1830\Uft/\UminZ$ for Hercules C-130, $2300\Uft/\UminZ$ for Airbus 380, Boeing 787/777; and  $1600\Uft/\UminZ$ for Twin Otter DHC-6.  
\subsection{Simulation Settings.}
In this work we have selected the CARI option 4 of GCR, ISO TS15390:2004-MSU-NYMMIK model, which uses a potential heliocentric model of high precision and perform the calculation for the year of 2018 averaged for January. Our scheme of simulation computes the fluence for all particles in the cosmic ray spectrum,  over squared centimeters, as well as ICRP-103, ICRP-60, and ICRU $H^{*}(10)$. The body absorbs doses in units of microSievert.
%
\subsection{Altitude and Flight Level.}
We are going to deal with two definitions for the expression ``flight level'' One for a physical unit and a second one as a synonym for altitude in the WGS84 model.
The unit meaning is equivalent to one hundred feet, i.e. $ 1 \UFL = 100 \Uft,  $ e. g. for an altitude of $43000\Uft$, we could write $z = 430 \UFL$. 
In the aviation argot, control operators ask pilots for their flight level and they just answer a number that corresponds to the magnitude of the altitude in units of one hundred feet, e. g. for an altitude of $43000\Uft$ the pilot should answer ``flight level four three zero'' or ``flight level four hundred thirty'' $ FL=430.$ 

\section{Results}
\begin{figure}[h!]
  \centering
	\begin{tabular}{ccc}
	\subfloat[OTBD-KLAX]{
		\includegraphics[width=3.8cm]{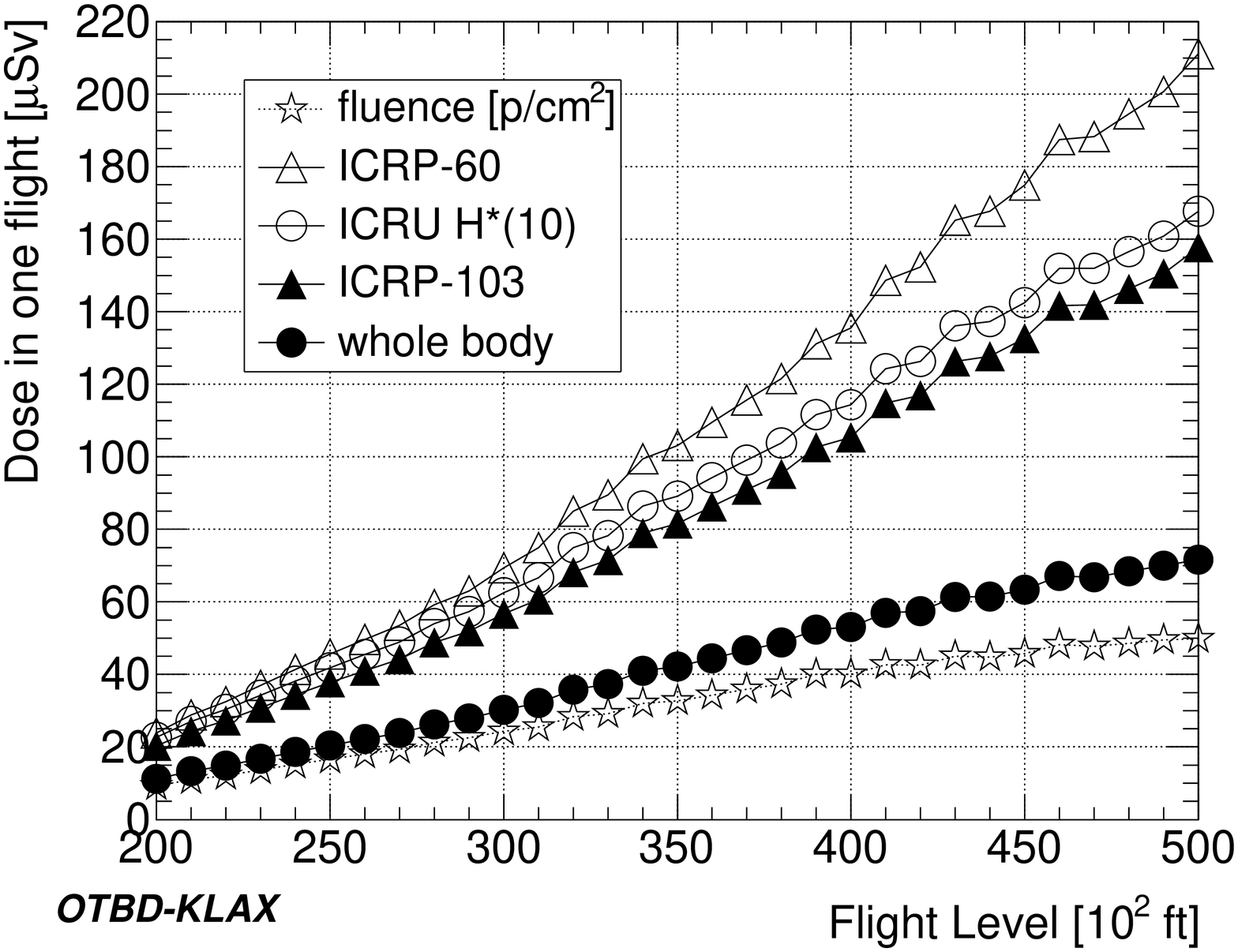}
		\label{fig:OTBD-KLAX_doses}
	} 
	&
 	\subfloat[OMDB-KLAX]{
		\includegraphics[width=3.8cm]{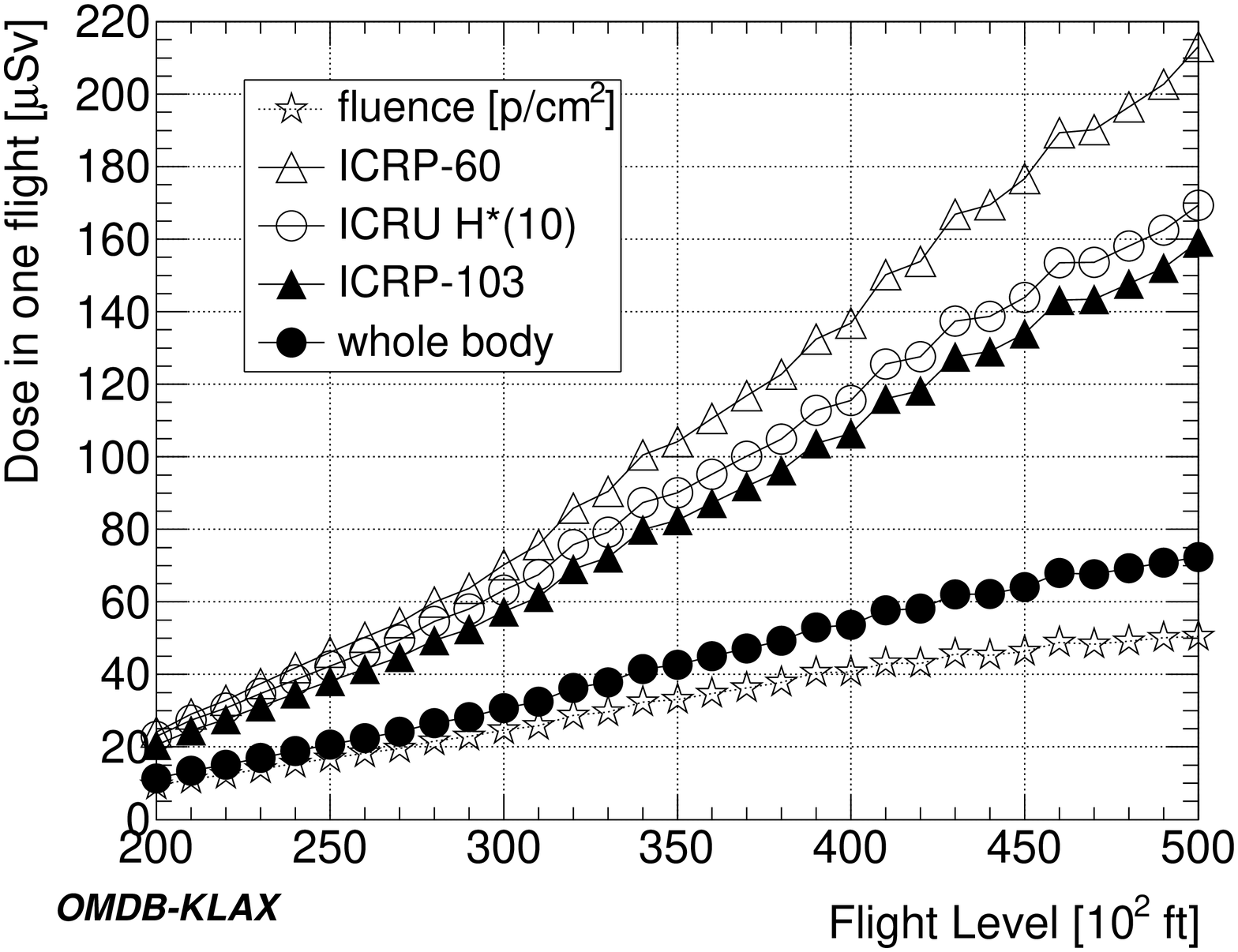}
		\label{fig:OMDB-KLAX_doses}
	} 
	&
	\subfloat[OMAA-KLAX]{
		\includegraphics[width=3.8cm]{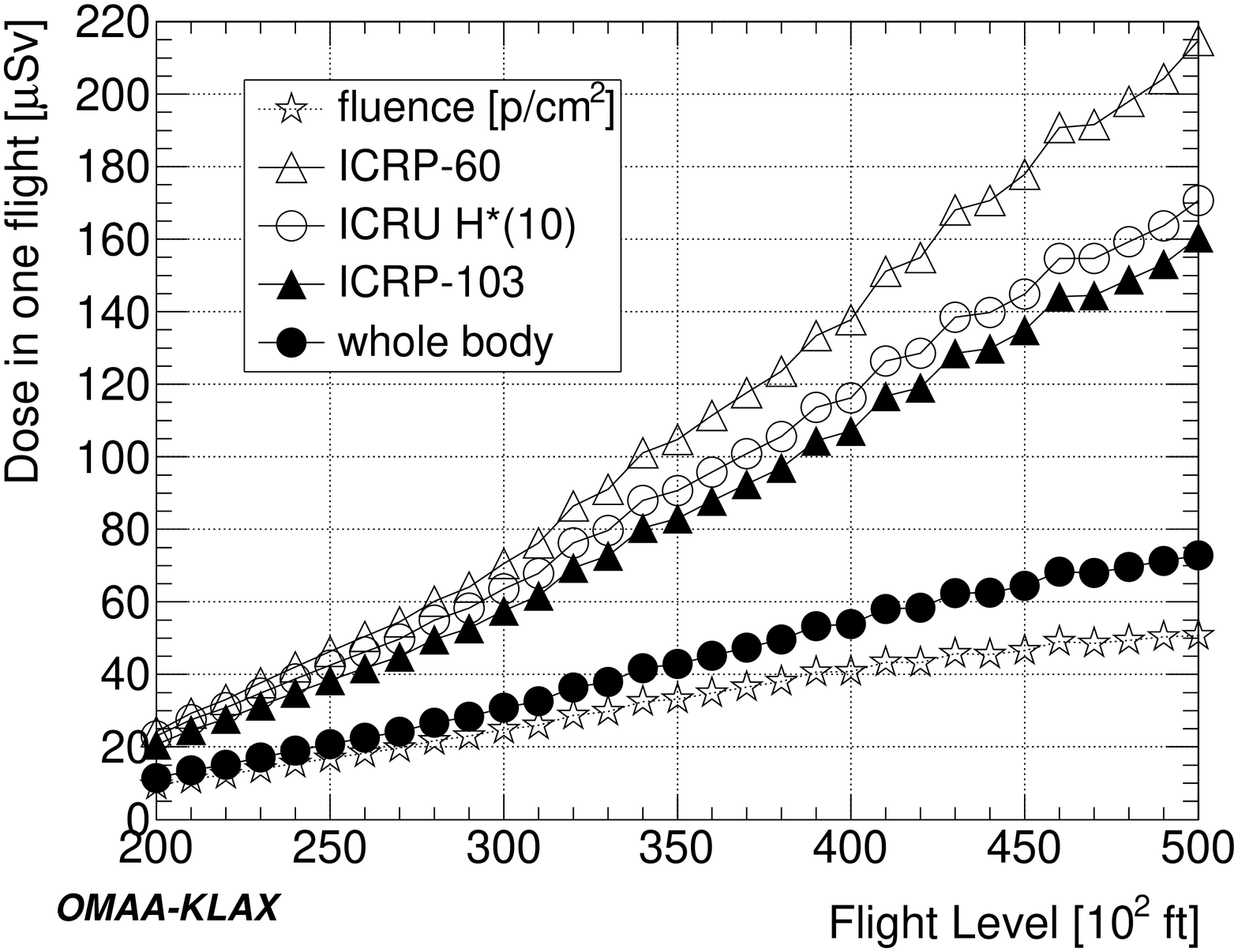}
		\label{fig:OMAA-KLAX_doses}
	} 
	\\
	\subfloat[OEJN-KLAX]{
		\includegraphics[width=3.8cm]{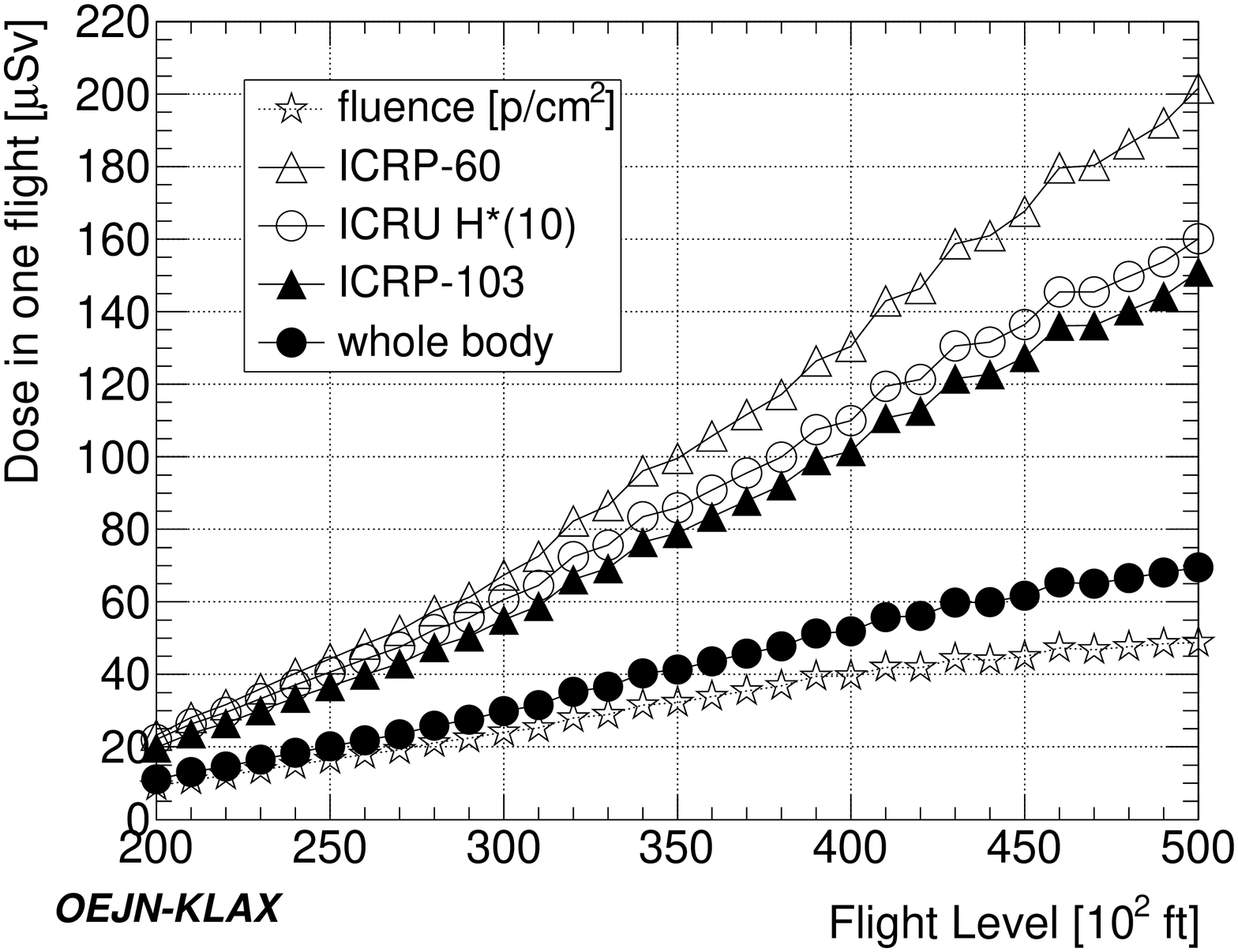}
		\label{fig:OEJN-KLAX_doses}
	}  
	&
	\subfloat[SCEL-YSSY]{
		\includegraphics[width=3.8cm]{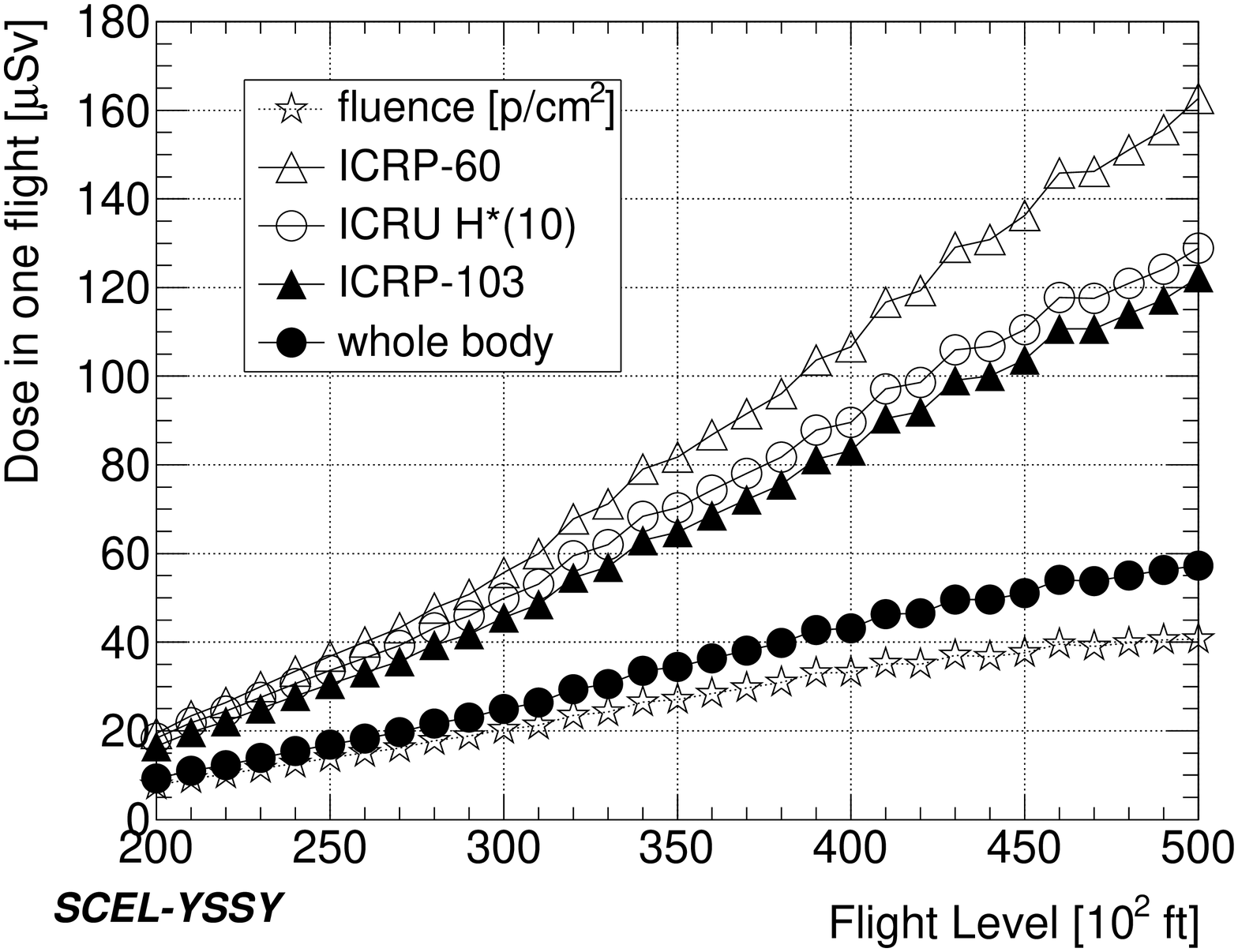}
		\label{fig:SCEL-YSSY_doses}
	} 
	&
  	\subfloat[WSSS-KEWR]{	
		\includegraphics[width=3.8cm]{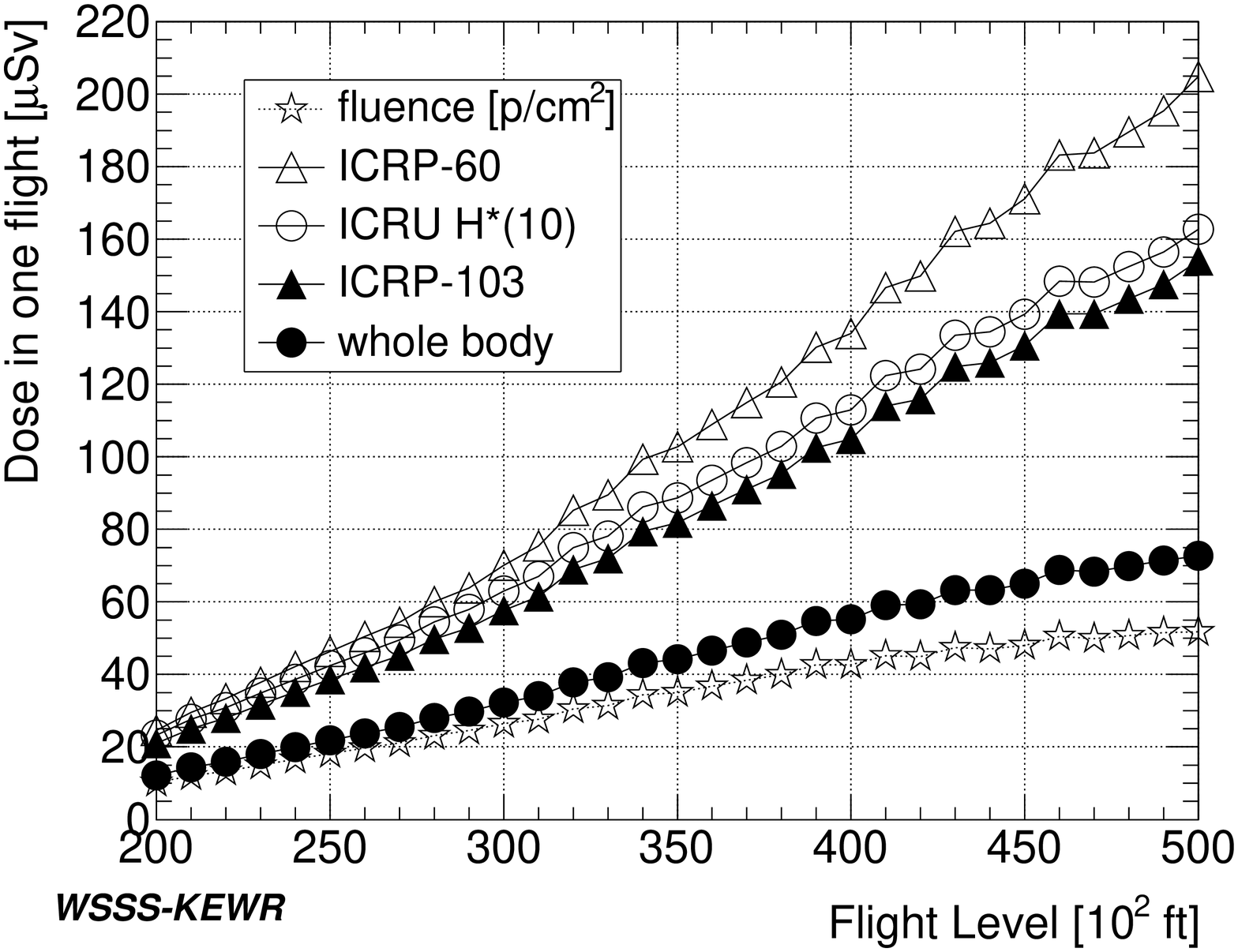} 
		\label{fig:WSSS-KEWR_doses}
	} 
	\\
  	\subfloat[KEWR-WSSS]{	
		\includegraphics[width=3.8cm]{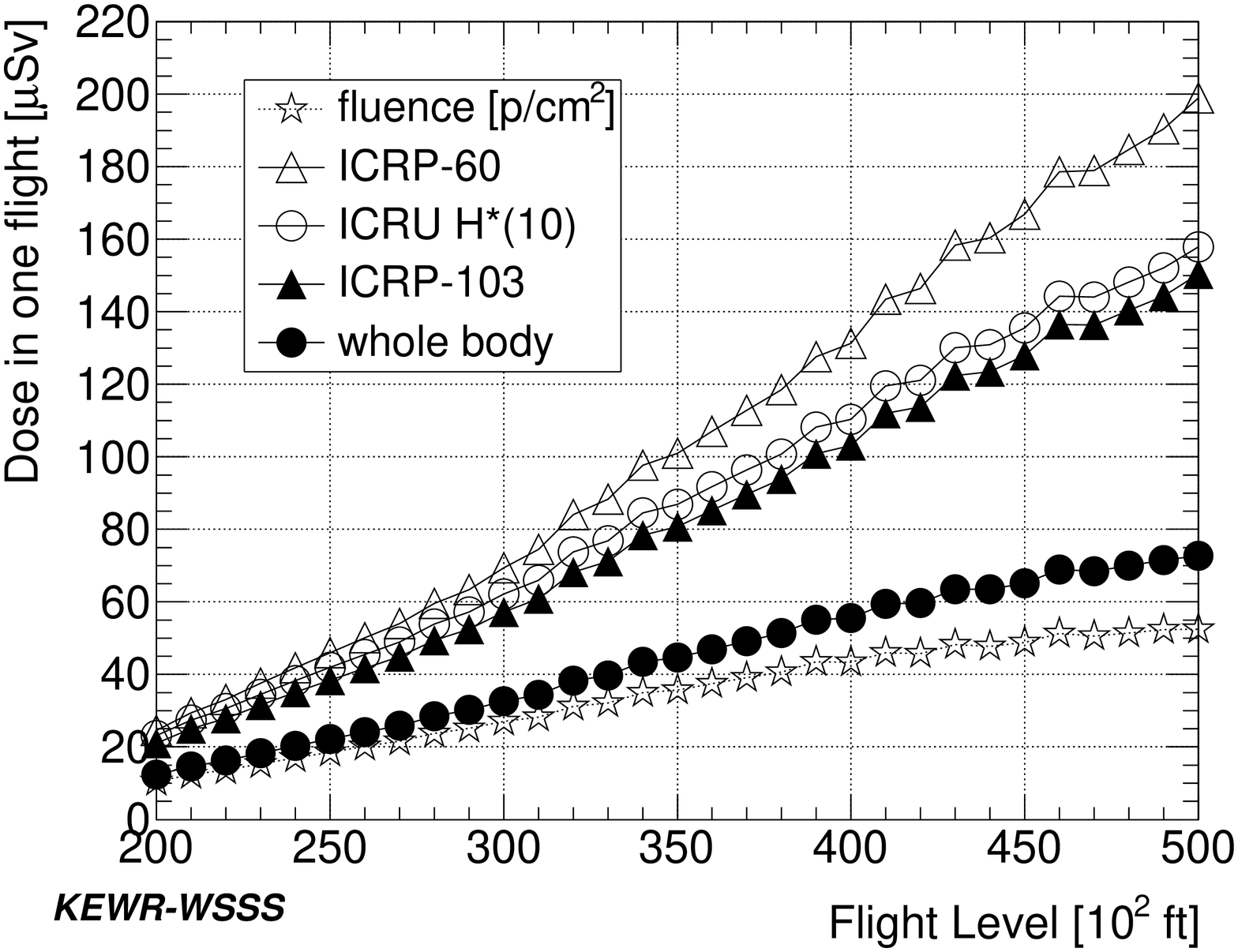} 
		\label{fig:KEWR-WSSS_doses}
	}
	&
  	\subfloat[SAWG-SAWB]{	
		\includegraphics[width=3.8cm]{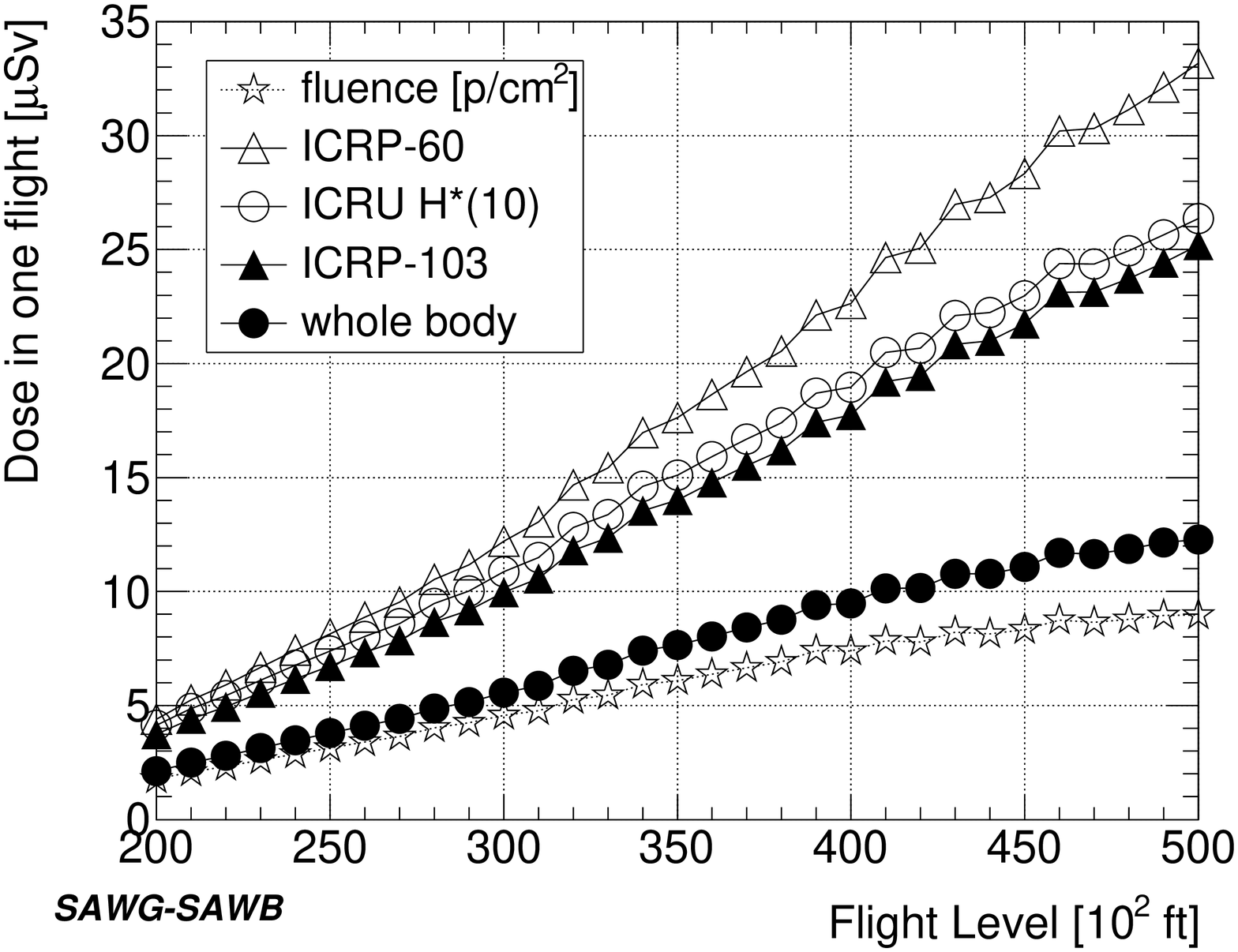} 
		\label{fig:SAWG-SAWB_doses}
	}
	&
  \subfloat[KLAX-WSSS]{	
		\includegraphics[width=3.8cm]{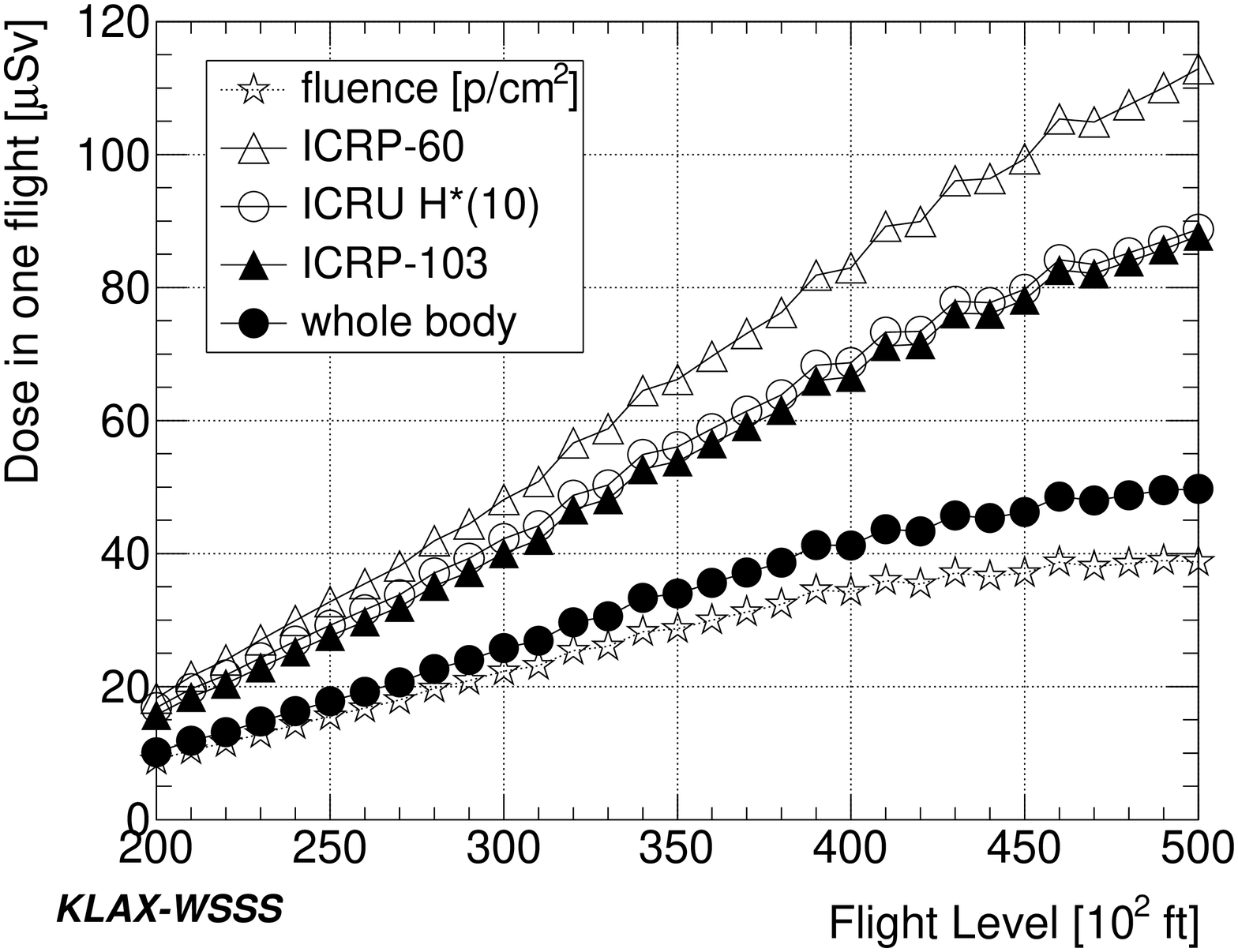} 
		\label{fig:KEWR-WSSS_doses}
	}
	\\
 \subfloat[YPPH-EGLL]{	
		\includegraphics[width=3.8cm]{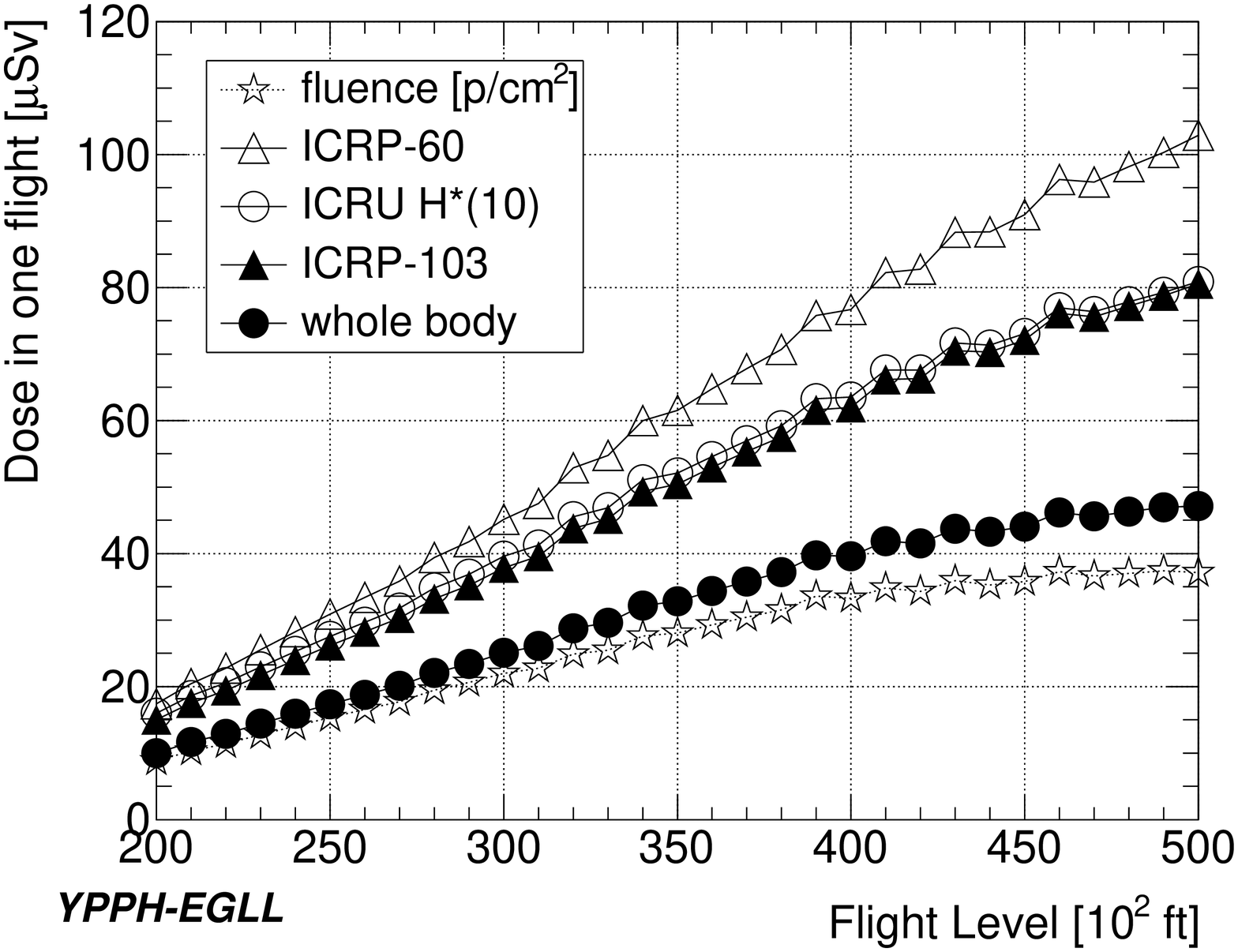} 
		\label{fig:YPPH-EGLL_doses}
	}
	&
 	\subfloat[FAOR-KATL]{	
		\includegraphics[width=3.8cm]{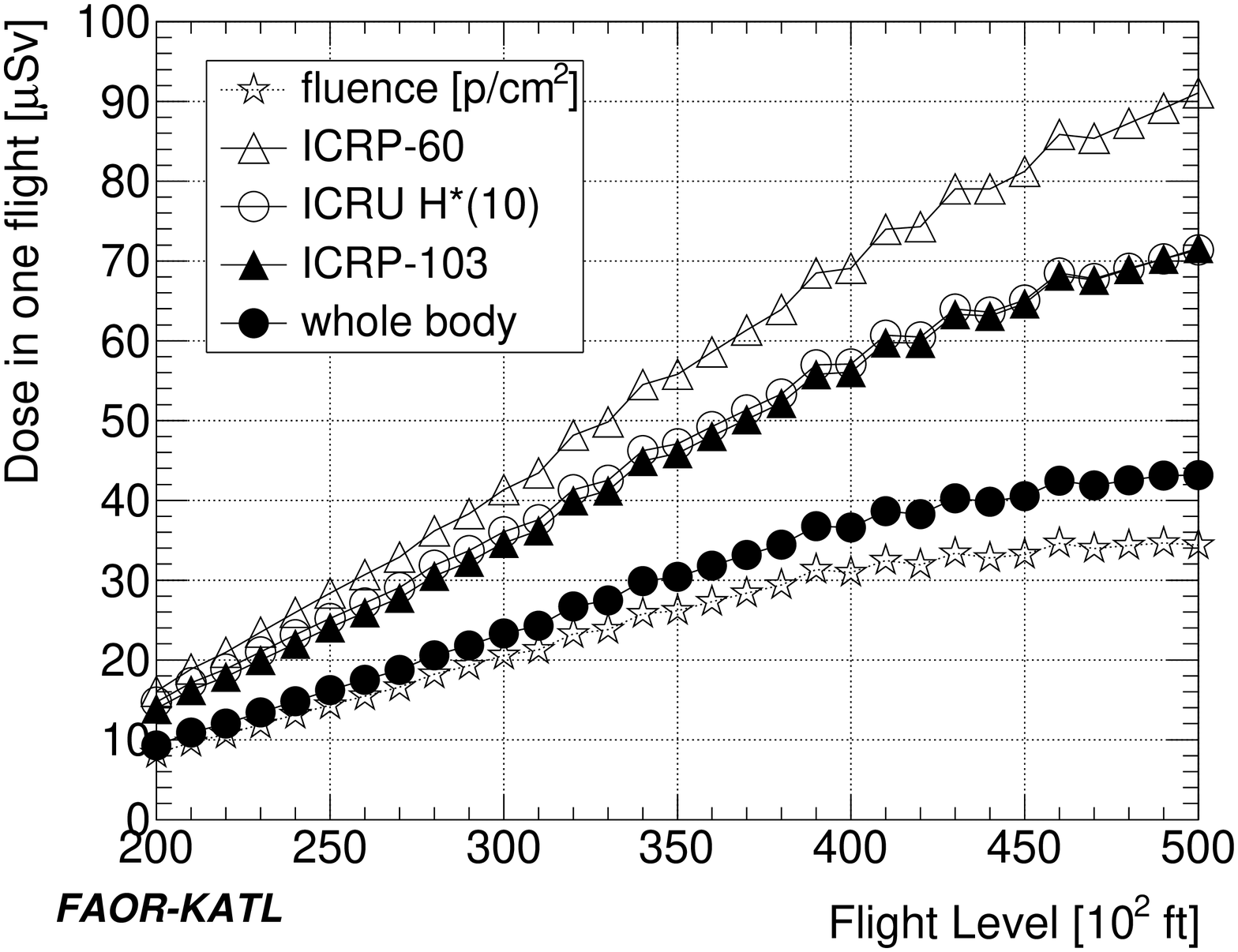}
		\label{fig:FAOR-KATL_doses}
	}
	&
  	\subfloat[KSFO-WSSS]{	
  	\includegraphics[width=3.8cm]{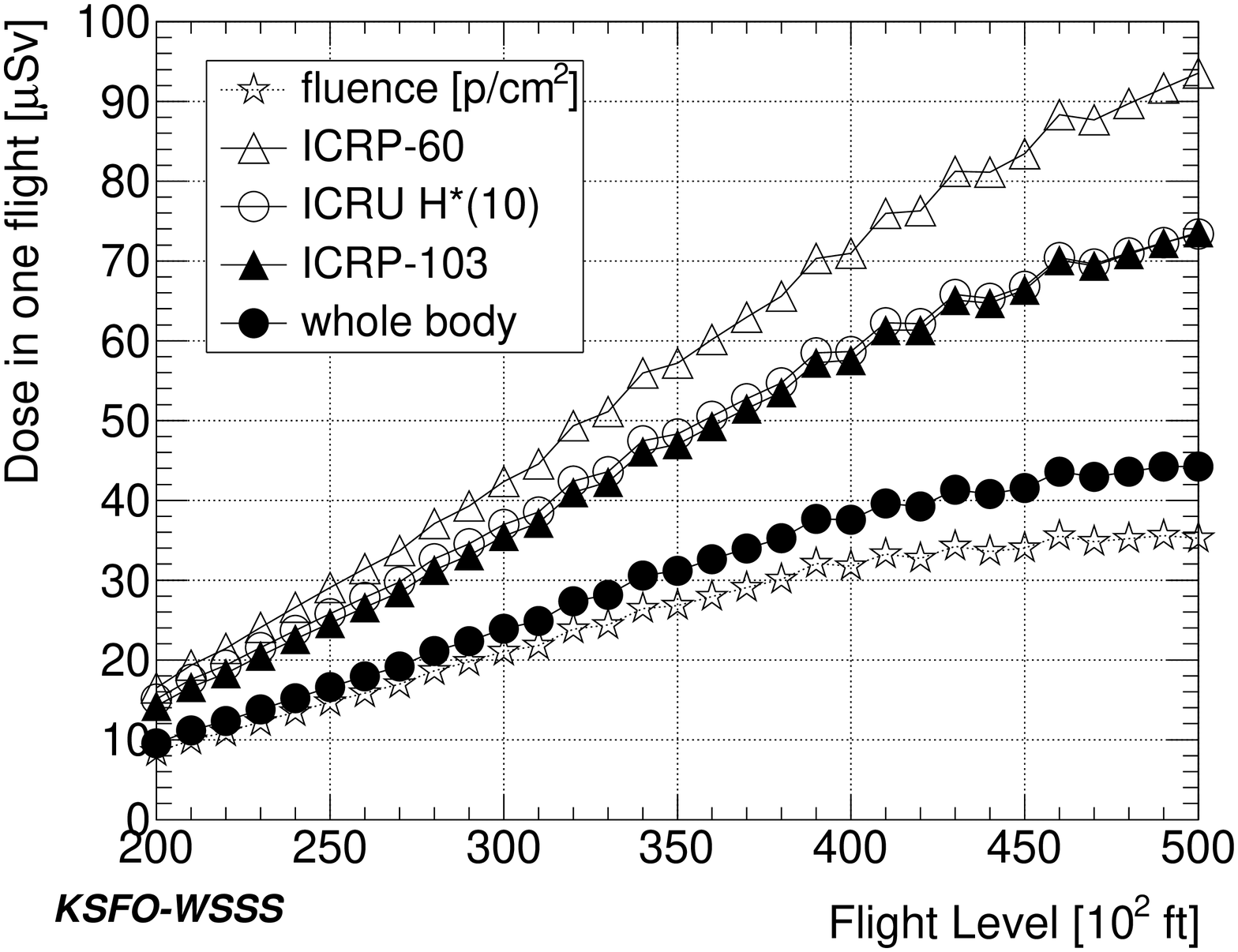} 
		\label{fig:KSFO-WSSS_doses}	
	}
	\\
	\subfloat[NZAA-OTBD]{
		\includegraphics[width=3.8cm]{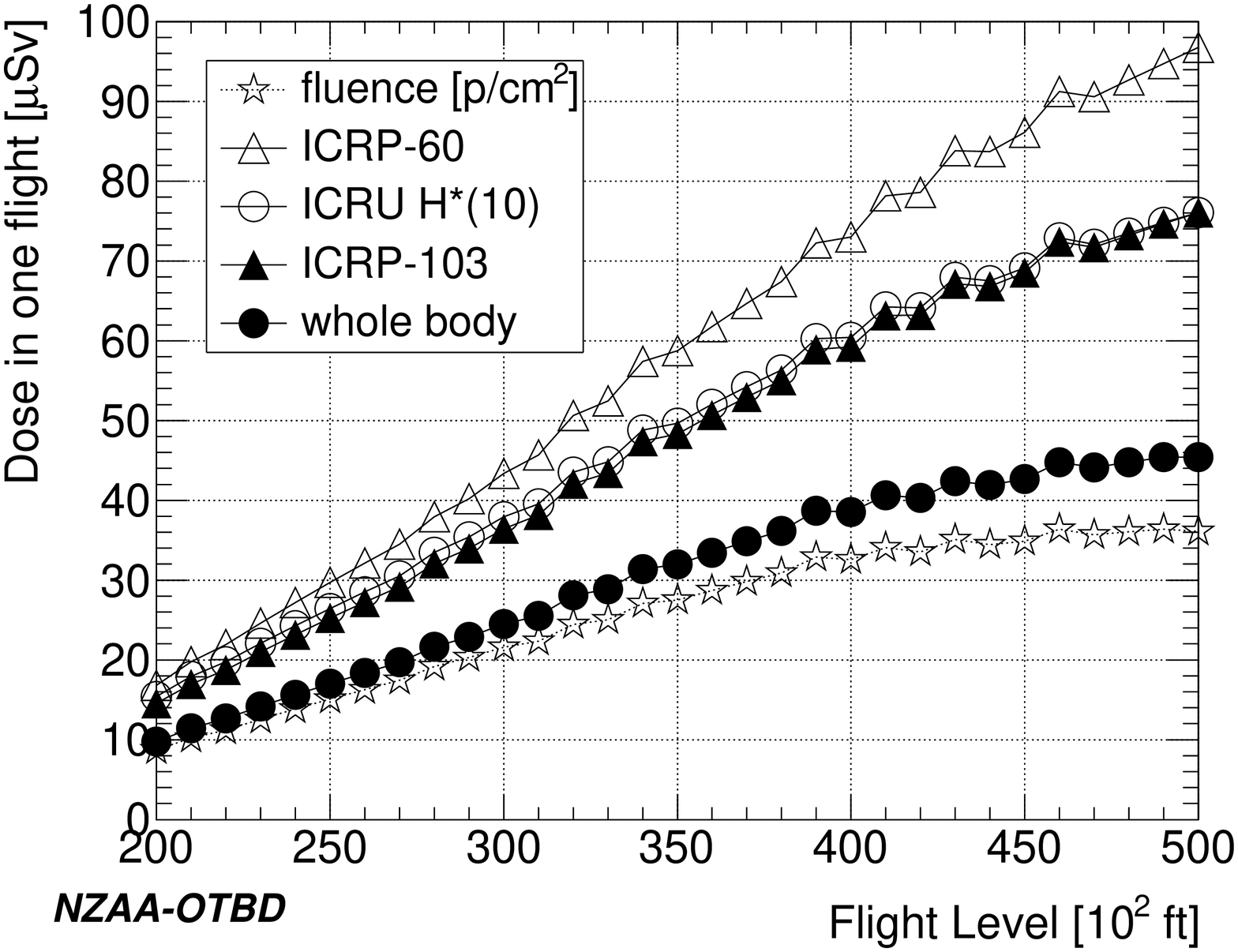}
		\label{fig:NZAA-OTBD_doses}
	}  
	&
  	\subfloat[KIAH-YSSY]{	
		\includegraphics[width=3.8cm]{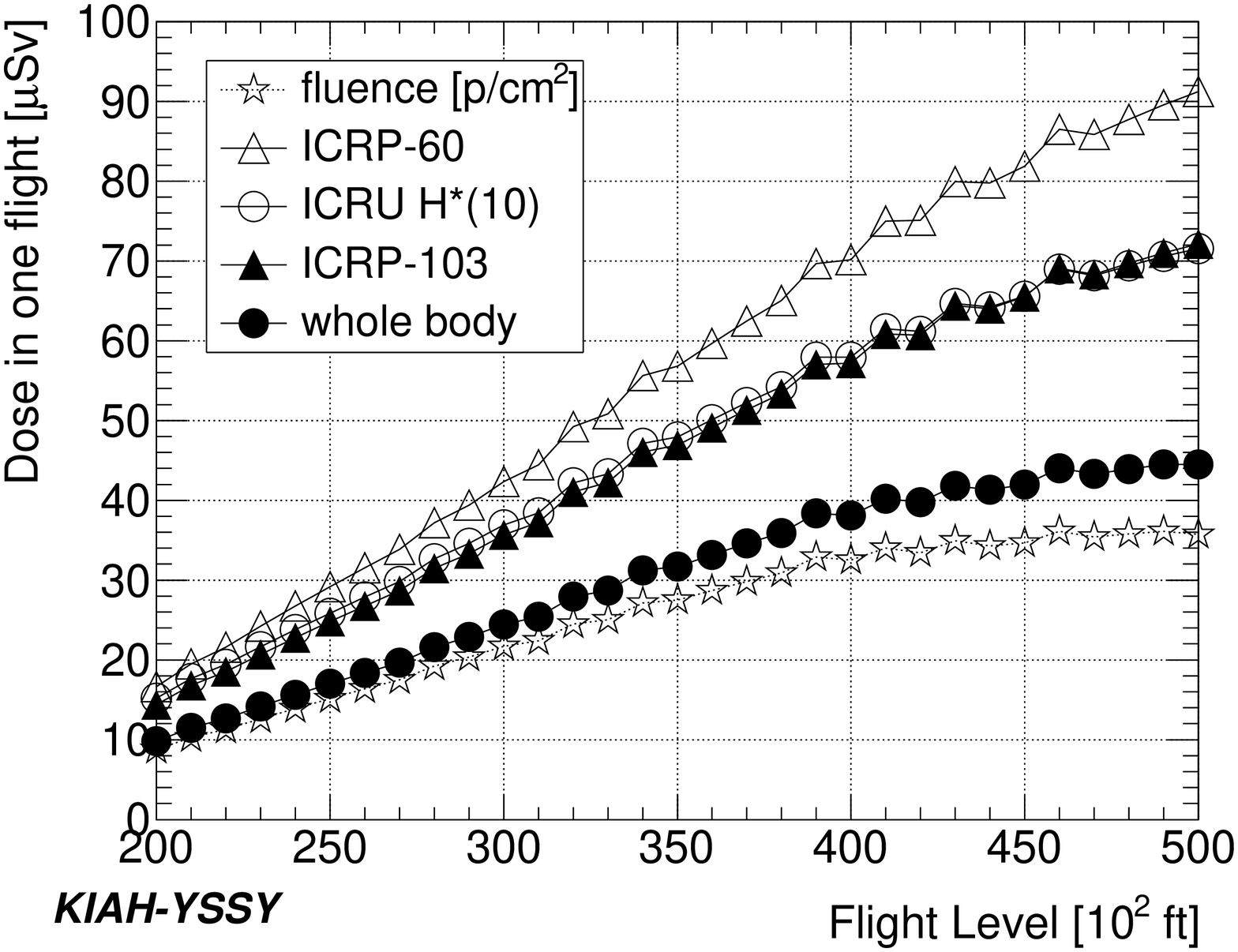}
		\label{fig:KIAH-YSSY_doses}
	}
	&
	\subfloat[KDFW-YSSY]{
		\includegraphics[width=3.8cm]{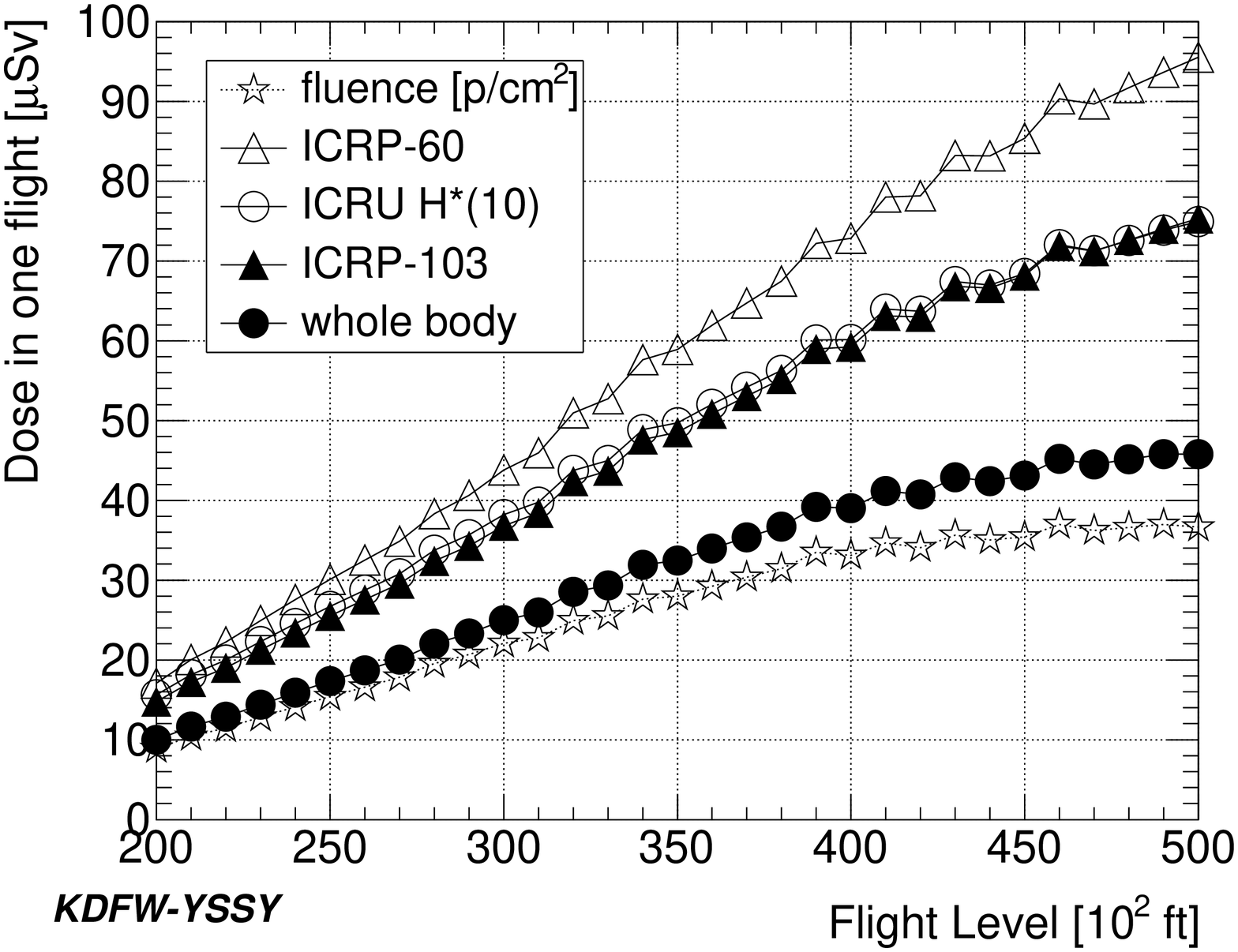}
		\label{fig:KDFW-YSSY_doses}
	} 
	\end{tabular}
  \caption{Simulation results for one flight in the aerial route specified. The abscissa presents the altitude in units of flight level $\UFL$,
i. e. in hundreds of feet. The ordinate presents the radiation dose for each one options available in CARI-7A.}
  \label{fig:oneflight}
\end{figure}

\begin{figure}[h!]
	\centering
	\begin{tabular}{ccc}
	\subfloat[OTBD-KLAX]{
		\includegraphics[width=3.0cm]{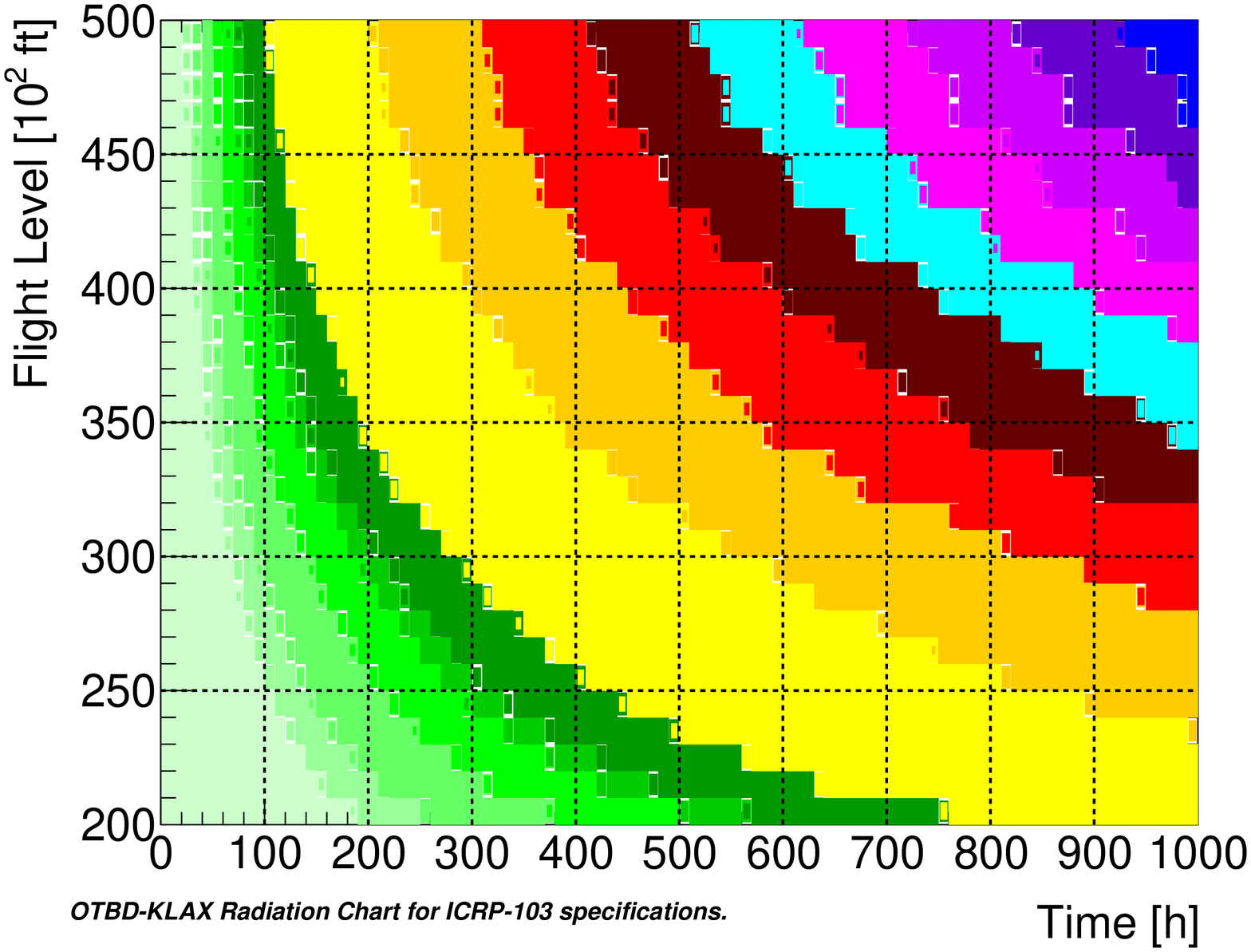}
		\label{fig:OTBD-KLAX_map1000}
	} 
	&
	\subfloat[OMAA-KLAX]{
		\includegraphics[width=3.0cm]{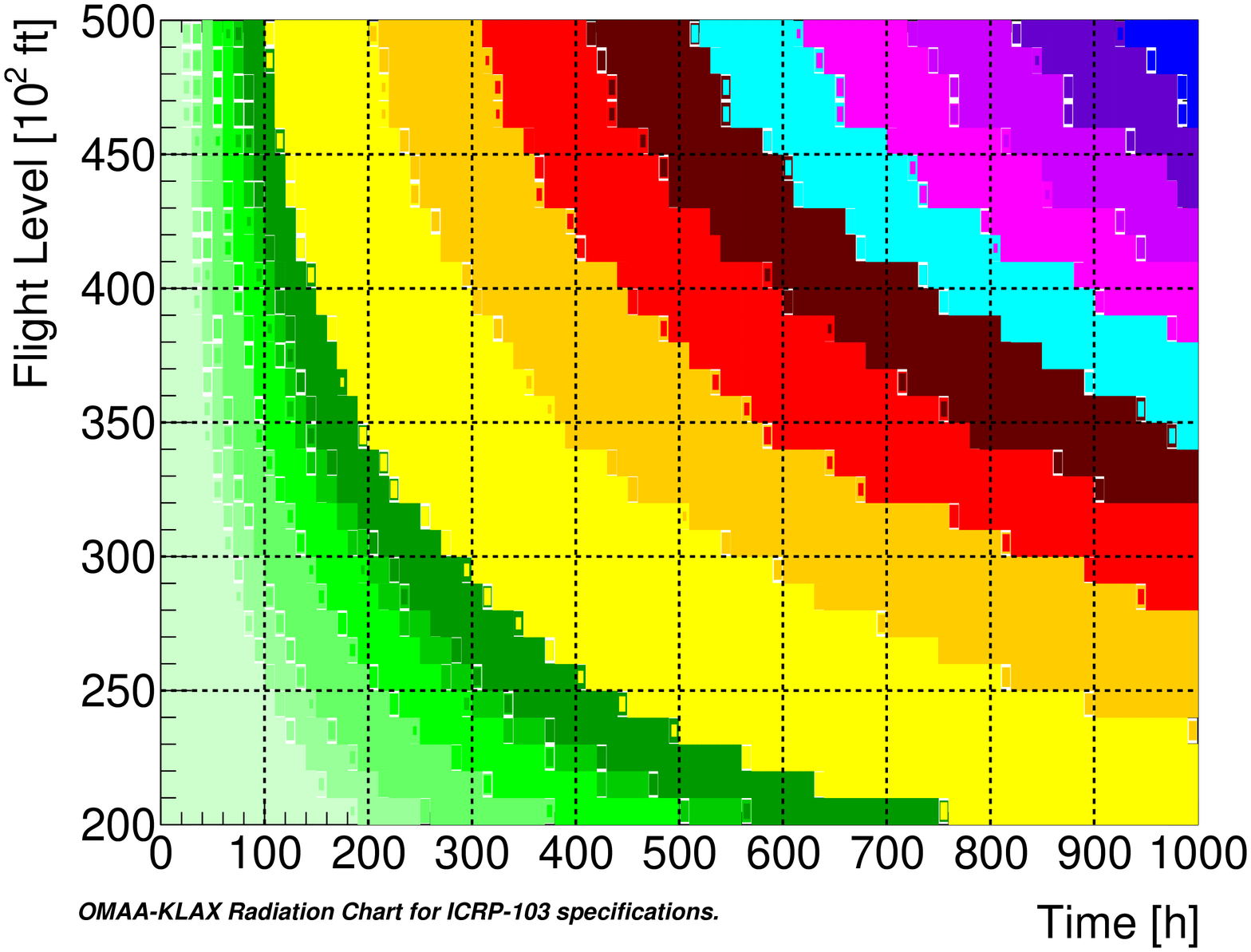}
		\label{fig:OMAA-KLAX_map1000}
	} 
	&
	\subfloat[OMDB-KLAX]{
		\includegraphics[width=3.0cm]{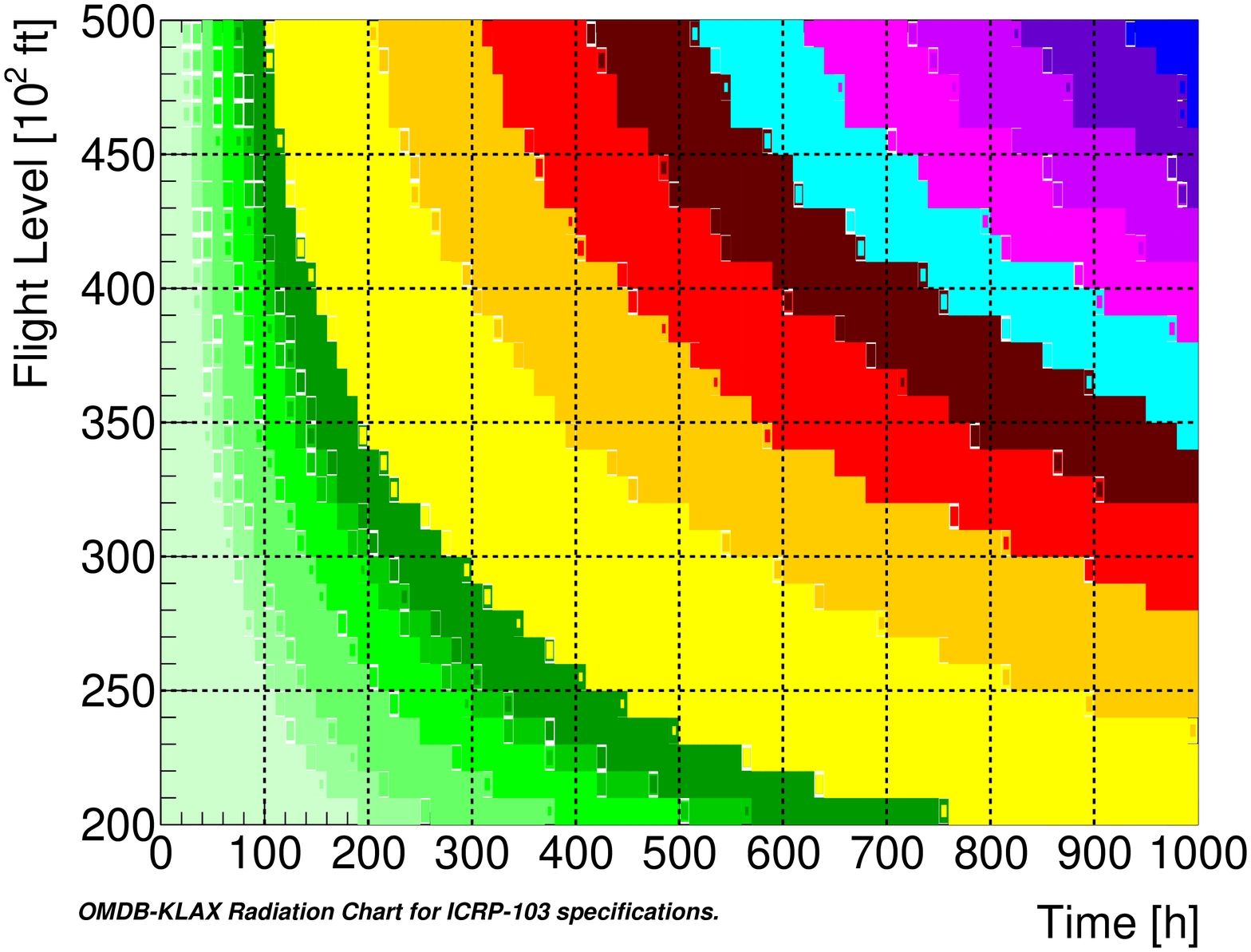}
		\label{fig:OMDB-KLAX_map1000}
	} 
	\\
	\subfloat[OEJN-KLAX]{
		\includegraphics[width=3.0cm]{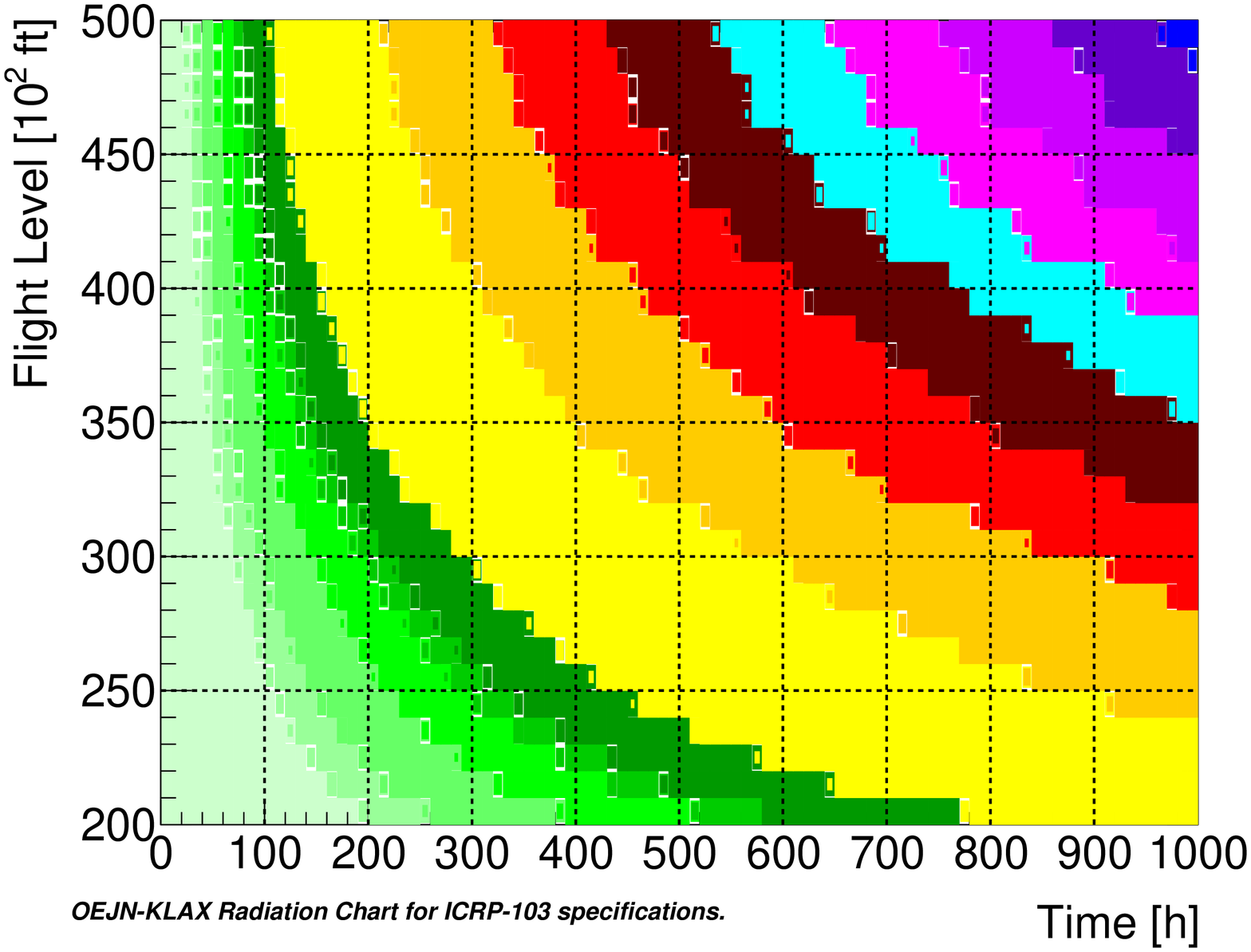}
		\label{fig:OEJN-KLAX_map1000}
	}  
	& 
	\subfloat[SCEL-YSSY]{
		\includegraphics[width=3.0cm]{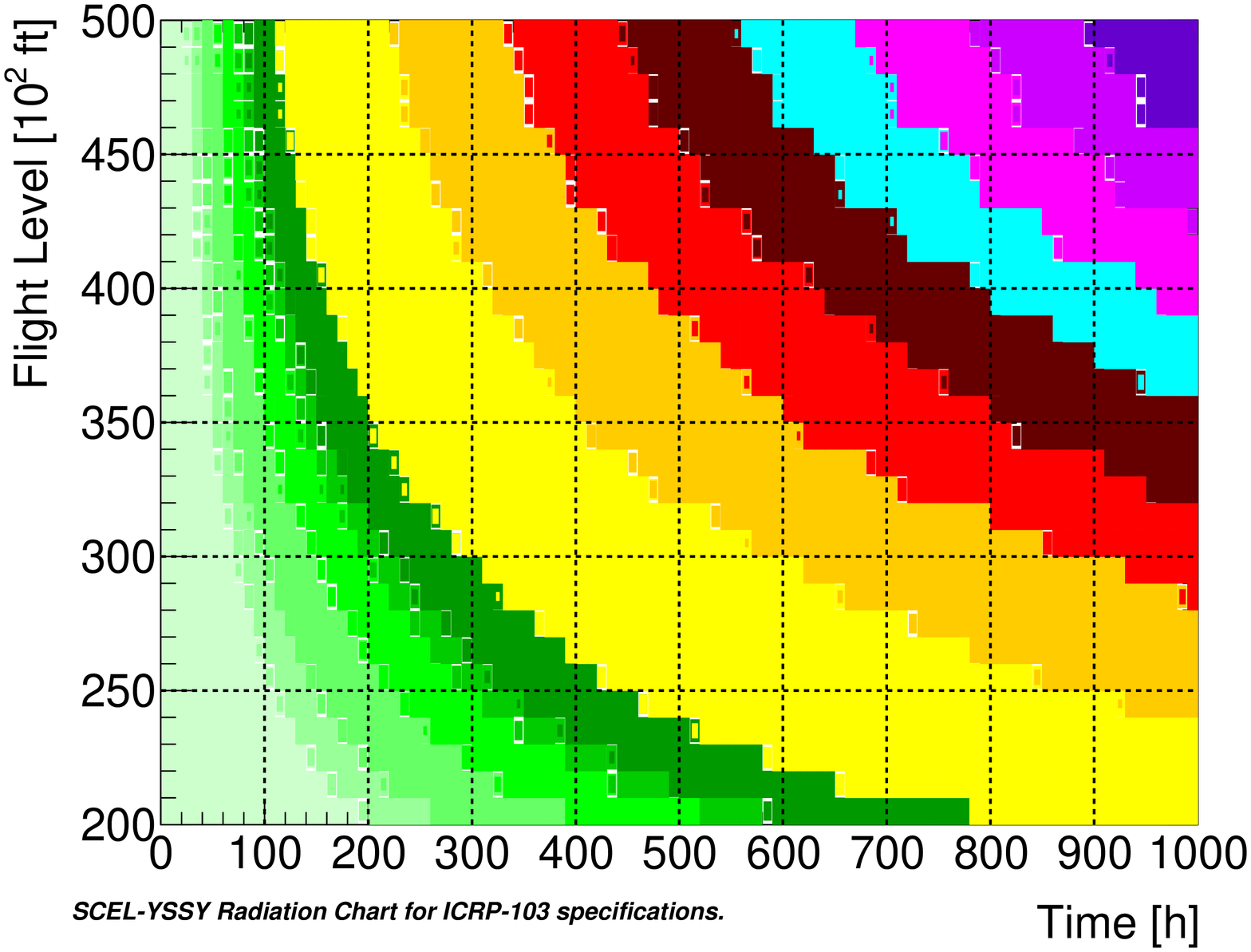}
		\label{fig:SAWG-SAWB_map1000}
	} 
	&
	\subfloat[WSSS-KEWR]{
		\includegraphics[width=3.0cm]{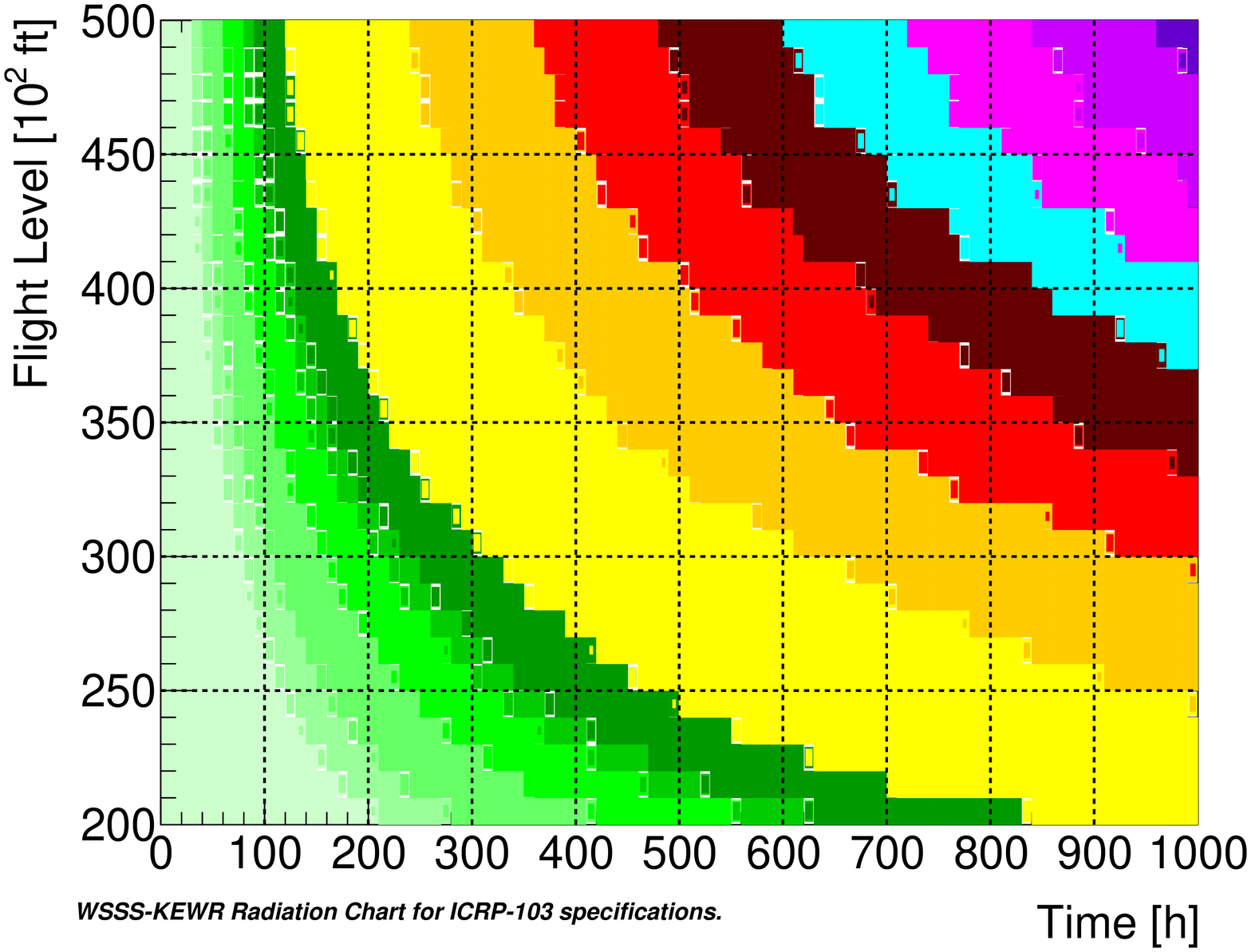}
		\label{fig:WSSS-KEWR_map1000}
	} 
	\\
	\subfloat[KEWR-WSSS]{
		\includegraphics[width=3.0cm]{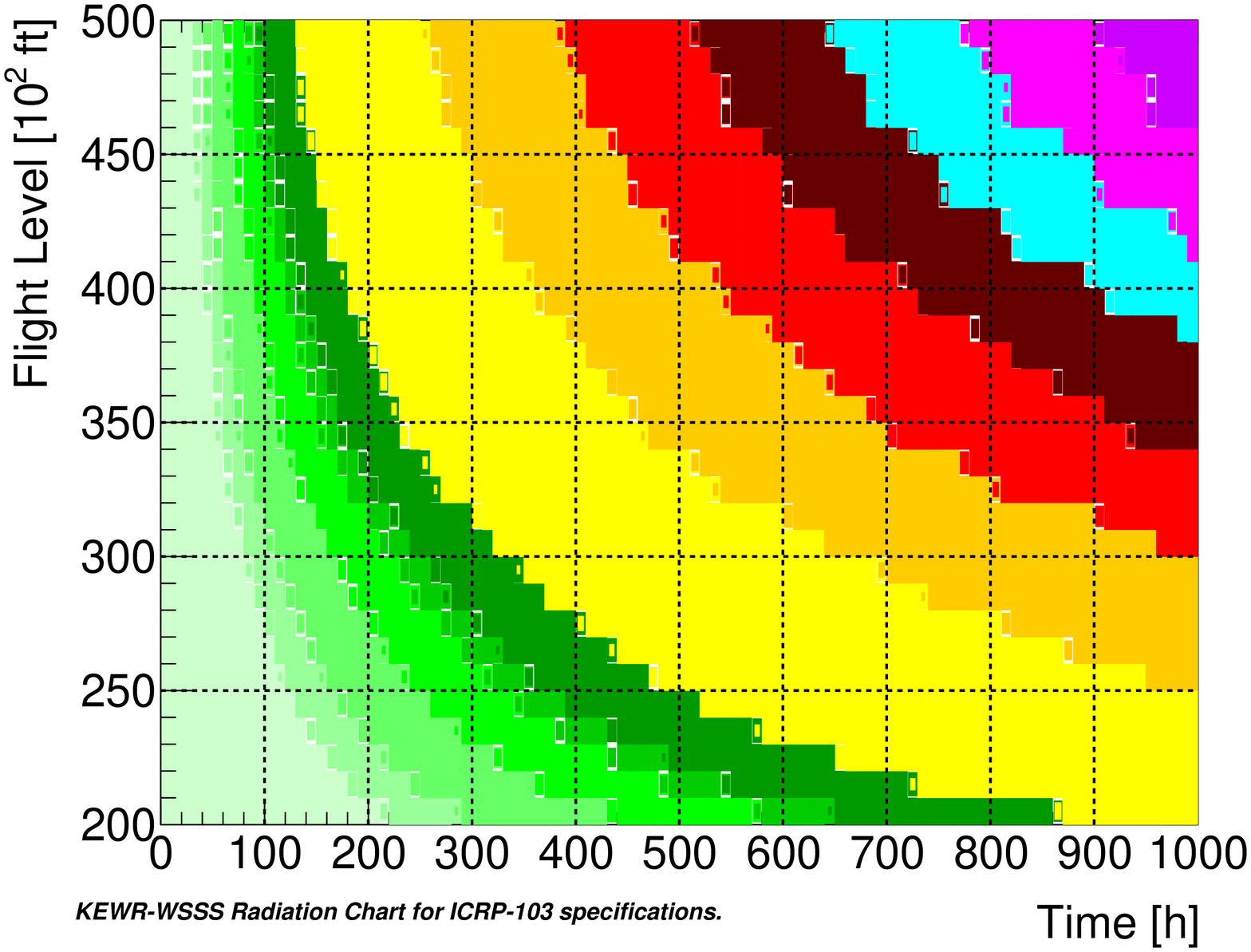}
		\label{fig:KEWR-WSSS_map1000}
	} 
	&
	\subfloat[SAWG-SAWB$^{*}$]{
		\includegraphics[width=3.0cm]{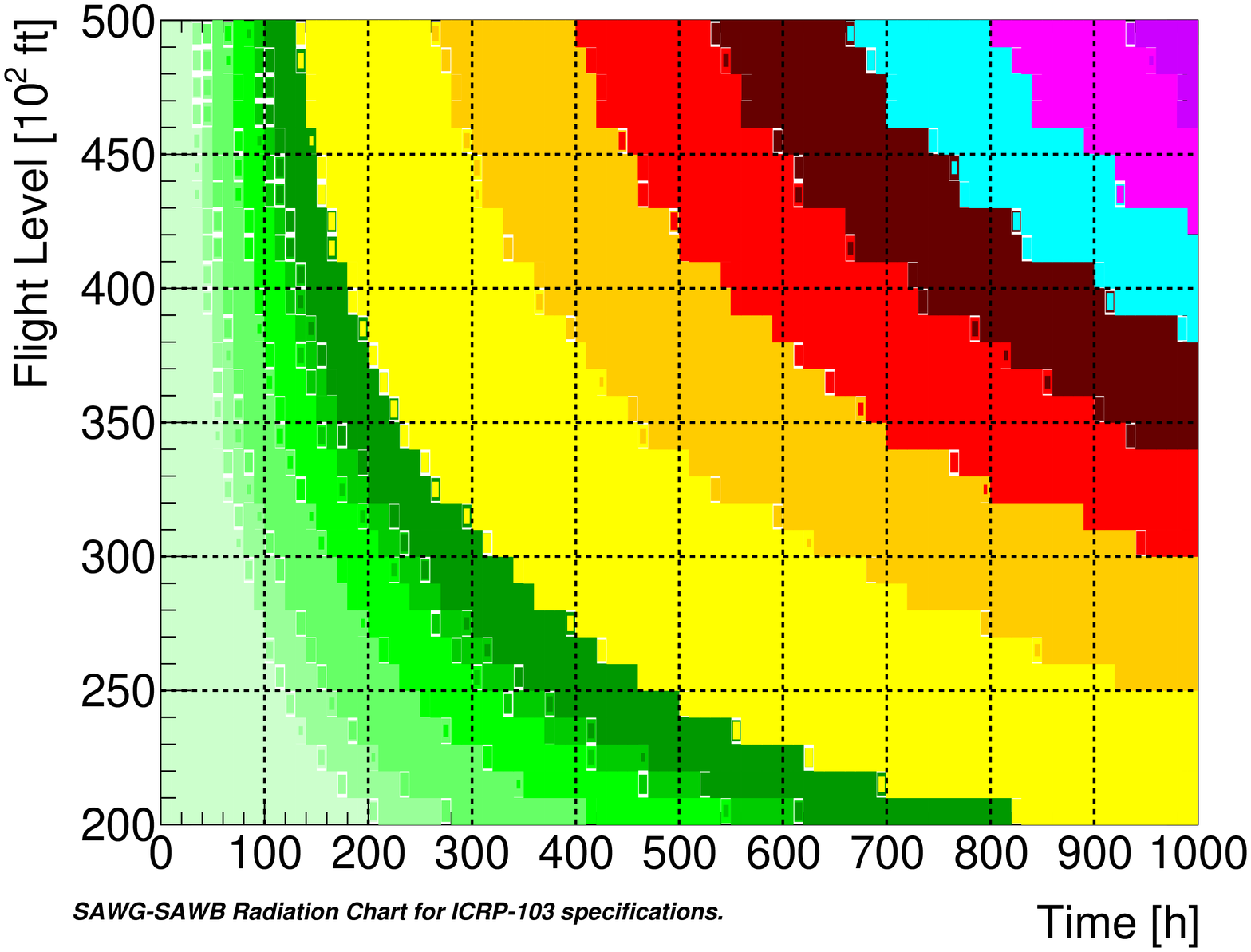}
		\label{fig:SAWG-SAWB_map1000}
	} 
	&
	\subfloat[KLAX-WSSS]{
		\includegraphics[width=3.0cm]{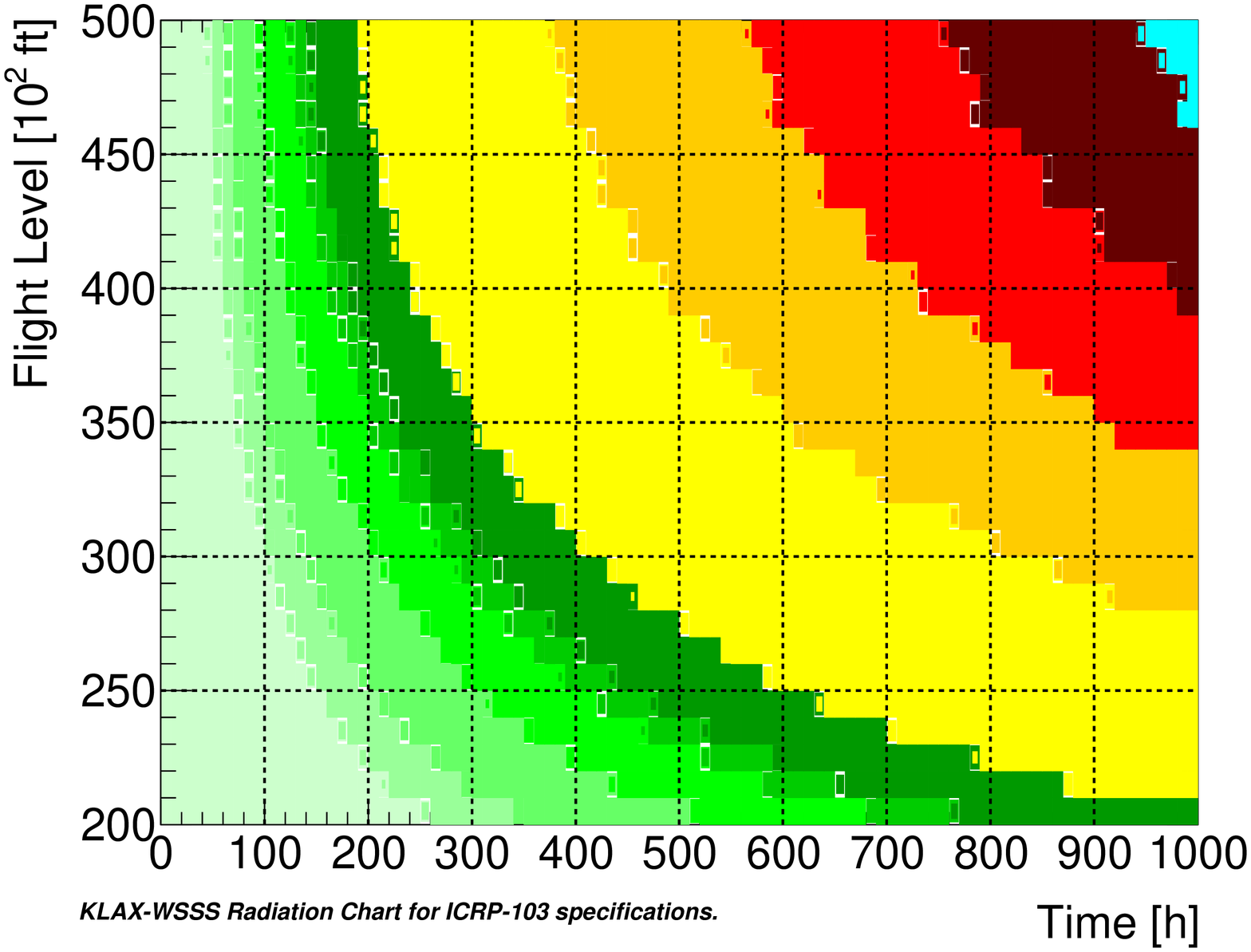}
		\label{fig:KLAX-WSSS_map1000}
	} 
	\\
	\subfloat[YPPH-EGLL]{
		\includegraphics[width=3.0cm]{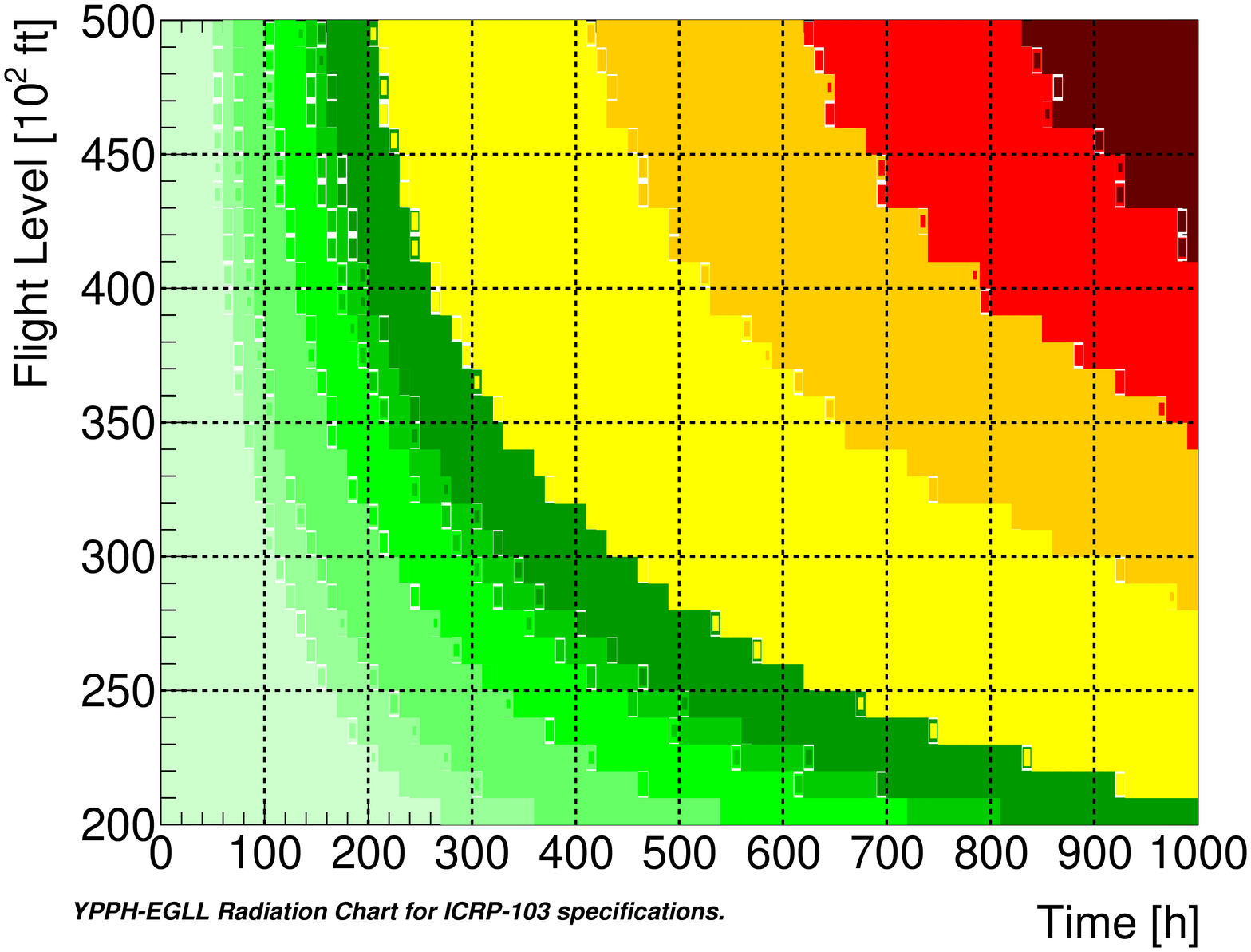}
		\label{fig:YPPH-EGLL_map1000}
	} 
	&
	\subfloat[KSFO-WSSS]{
		\includegraphics[width=3.0cm]{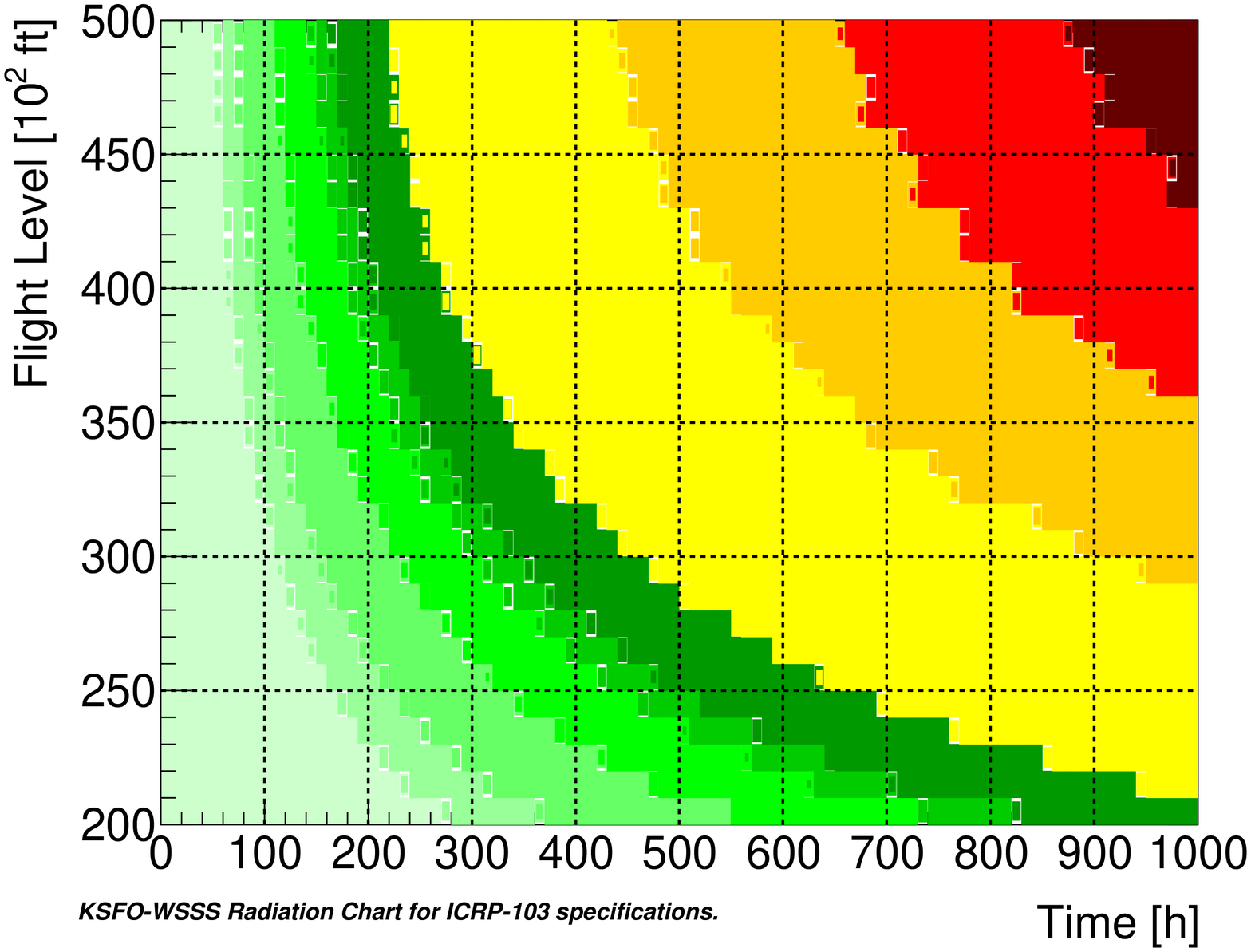}
		\label{fig:KSFO-WSSS_map1000}
	}  
	&
	\subfloat[FAOR-KATL]{
		\includegraphics[width=3.0cm]{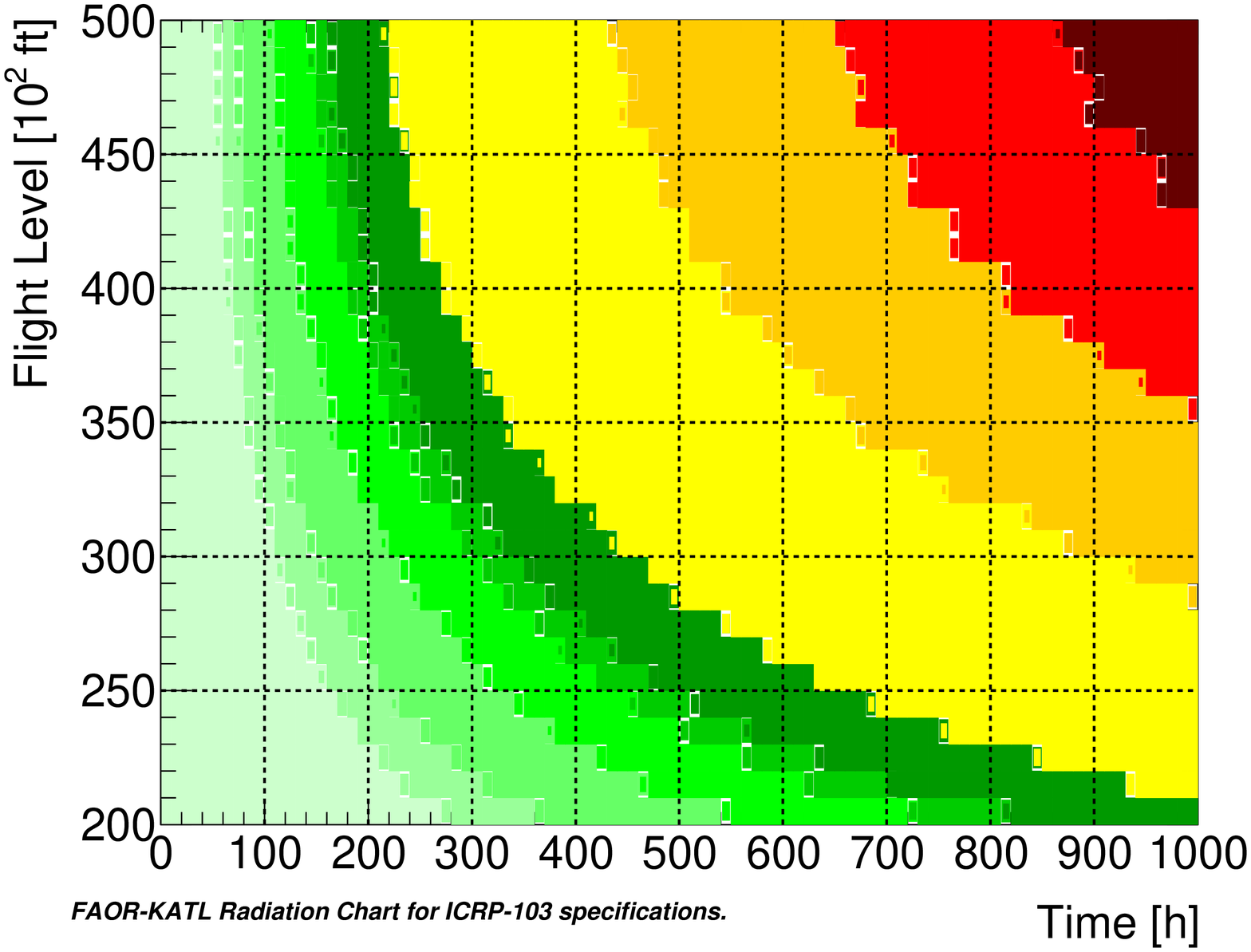}
		\label{fig:FAOR-KATL_map1000}
	} 
	\\
	\subfloat[NZAA-OTBD]{
		\includegraphics[width=3.0cm]{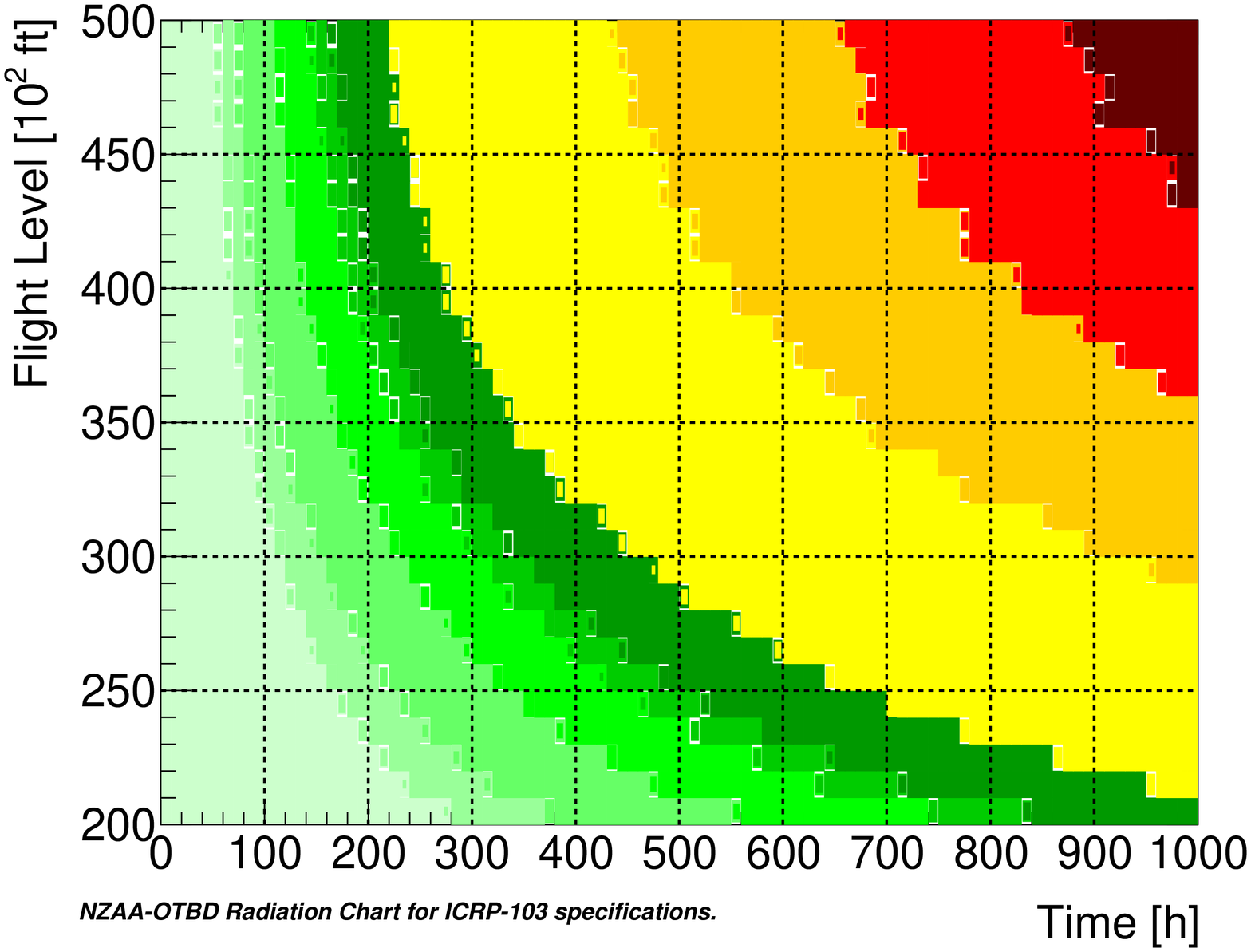}
		\label{fig:NZAA-OTBD_map1000}
	} 
	&
	\subfloat[NZAA-OMDB]{
		\includegraphics[width=3.0cm]{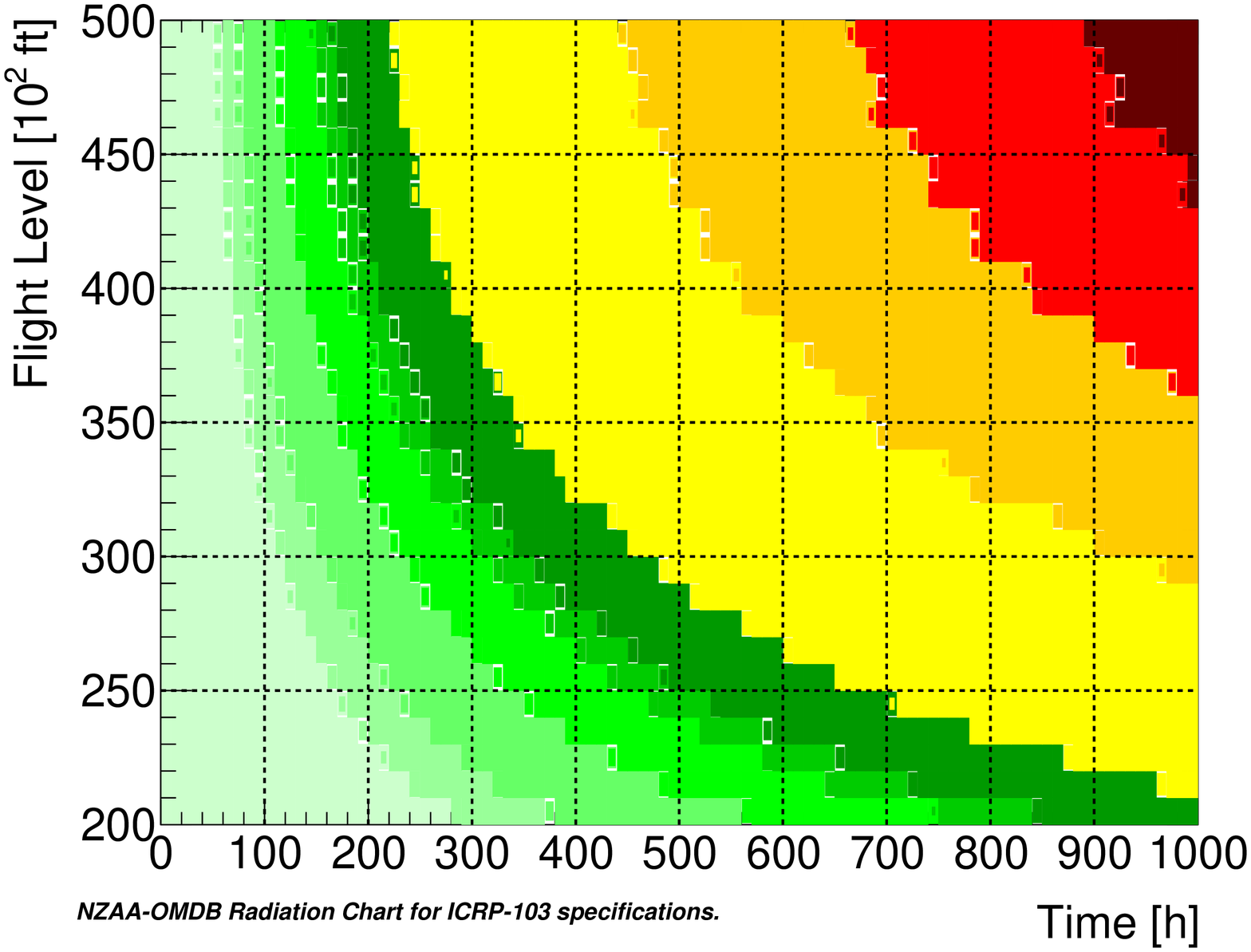}
		\label{fig:NZAA-OMDB_map1000}
	} 
	&
	\subfloat[KDFW-YSSY]{
		\includegraphics[width=3.0cm]{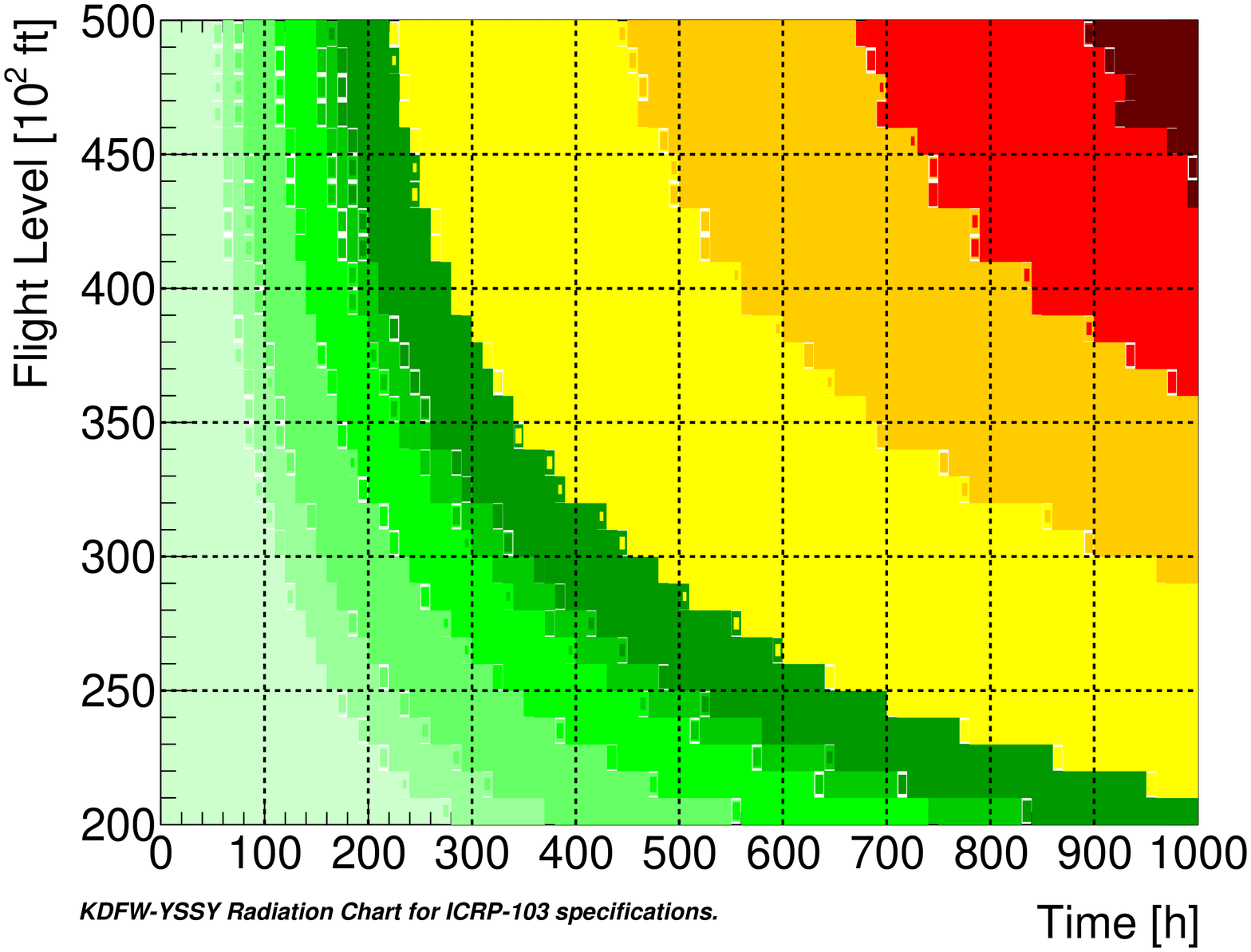}
		\label{fig:KDFW-YSSY_map1000}
	} 
	\\
	\subfloat[KIAH-YSSY]{
		\includegraphics[width=3.0cm]{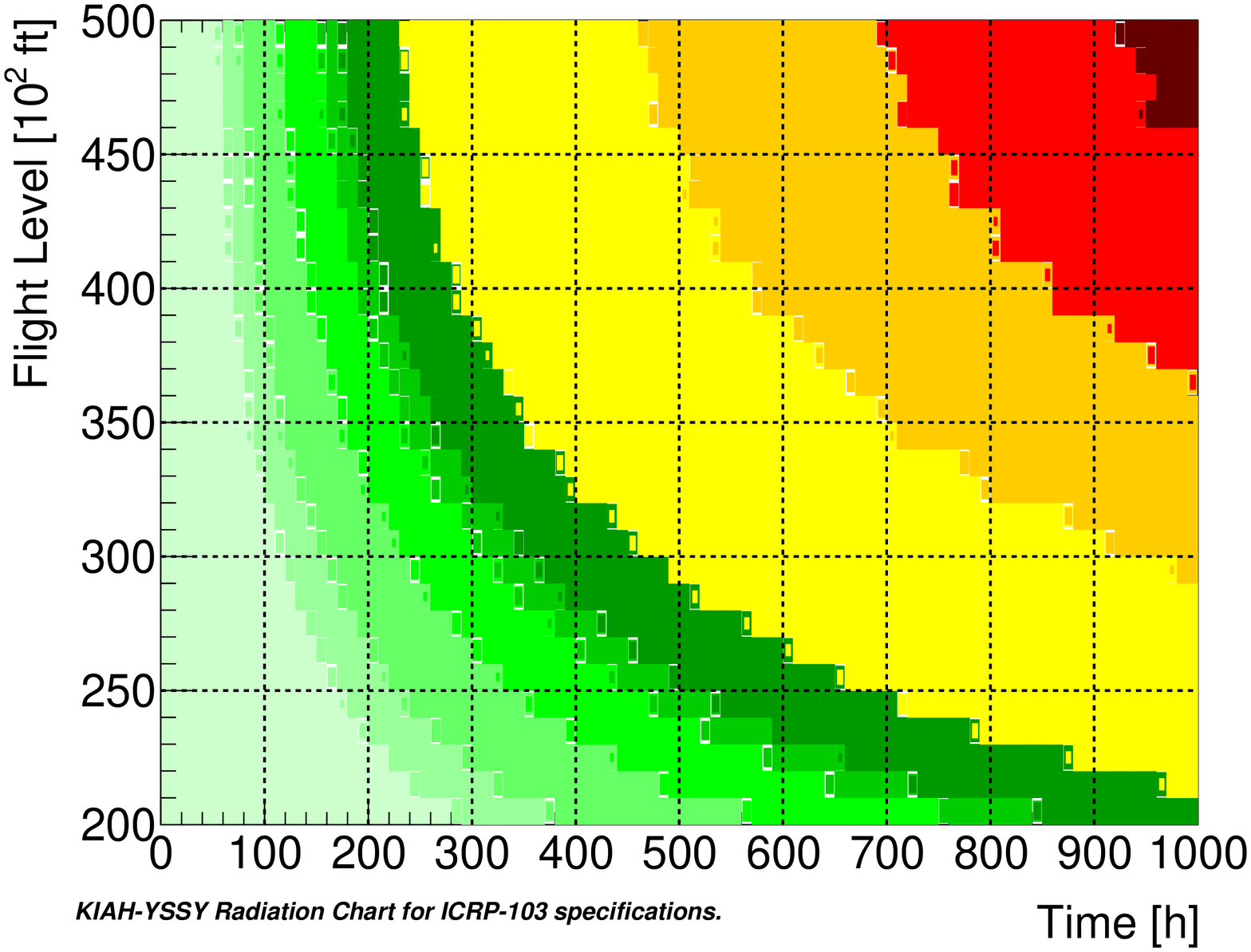}
		\label{fig:KIAH-YSSY_map1000}
	} 
	&

	&
	\subfloat[Radiation colors legend.]{
		\includegraphics[width=3.8cm]{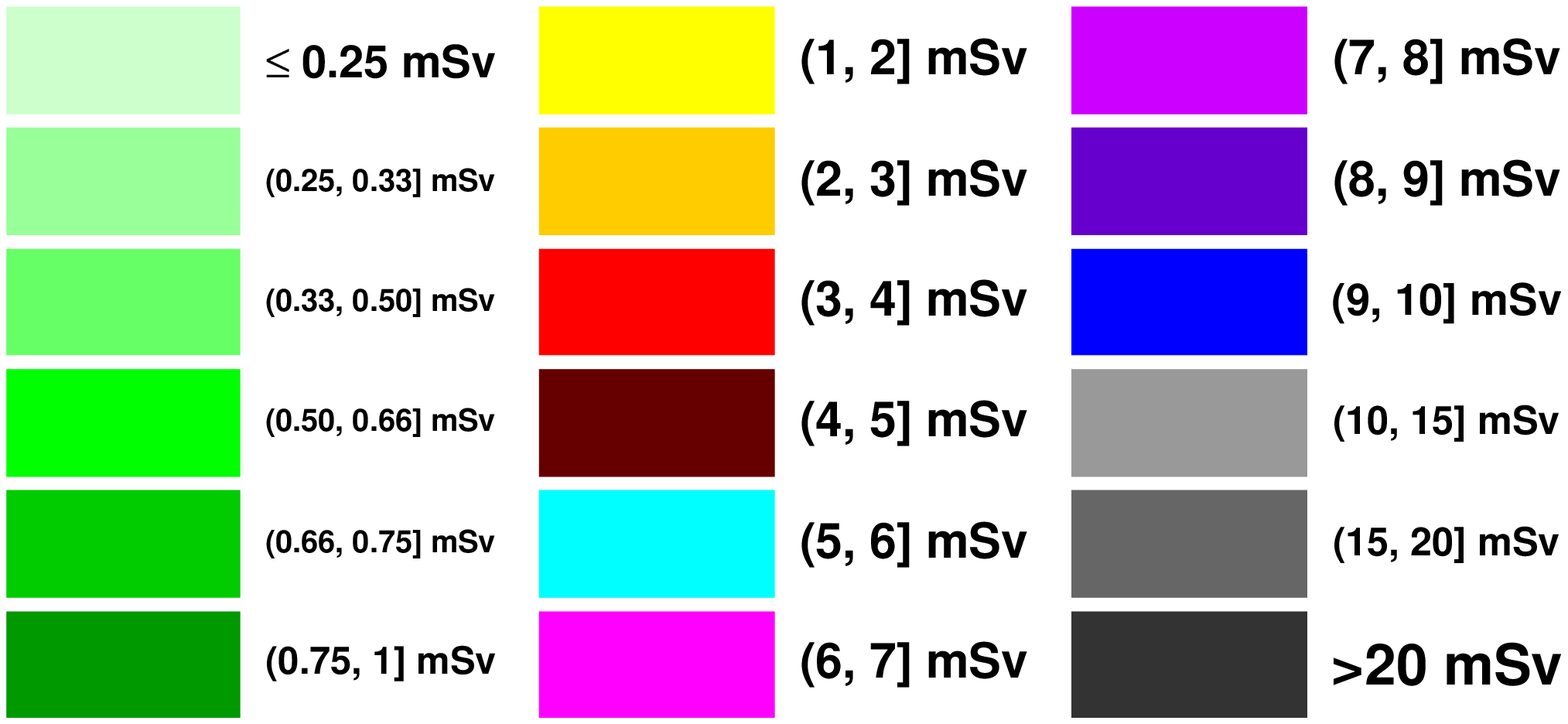}
		\label{subfig:legend}
	} 
	\\
	\end{tabular}
\caption{ICRP-103 Radiation Dose Charts for Aviation as a function of the altitude and flight time for every route in the sample. 
}
\label{fig:maps}
\end{figure}

\begin{figure}[h!]
	\centering
	\begin{tabular}{ccc}
	\subfloat[OTBD-KLAX]{
		\includegraphics[width=3.0cm]{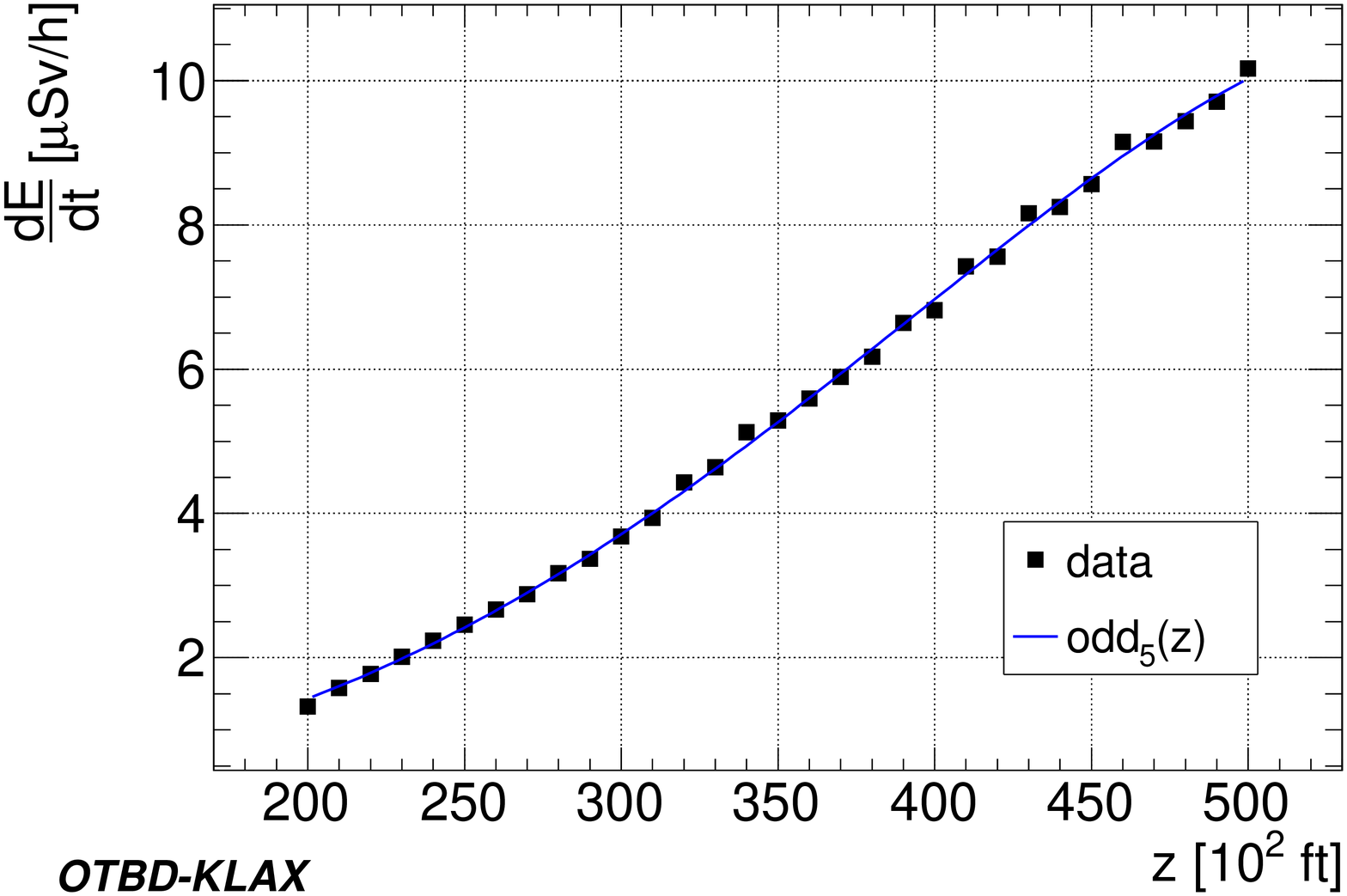}
		\label{fig:OTBD-KLAX_fit}
	} 
	&
	\subfloat[OMAA-KLAX]{
		\includegraphics[width=3.0cm]{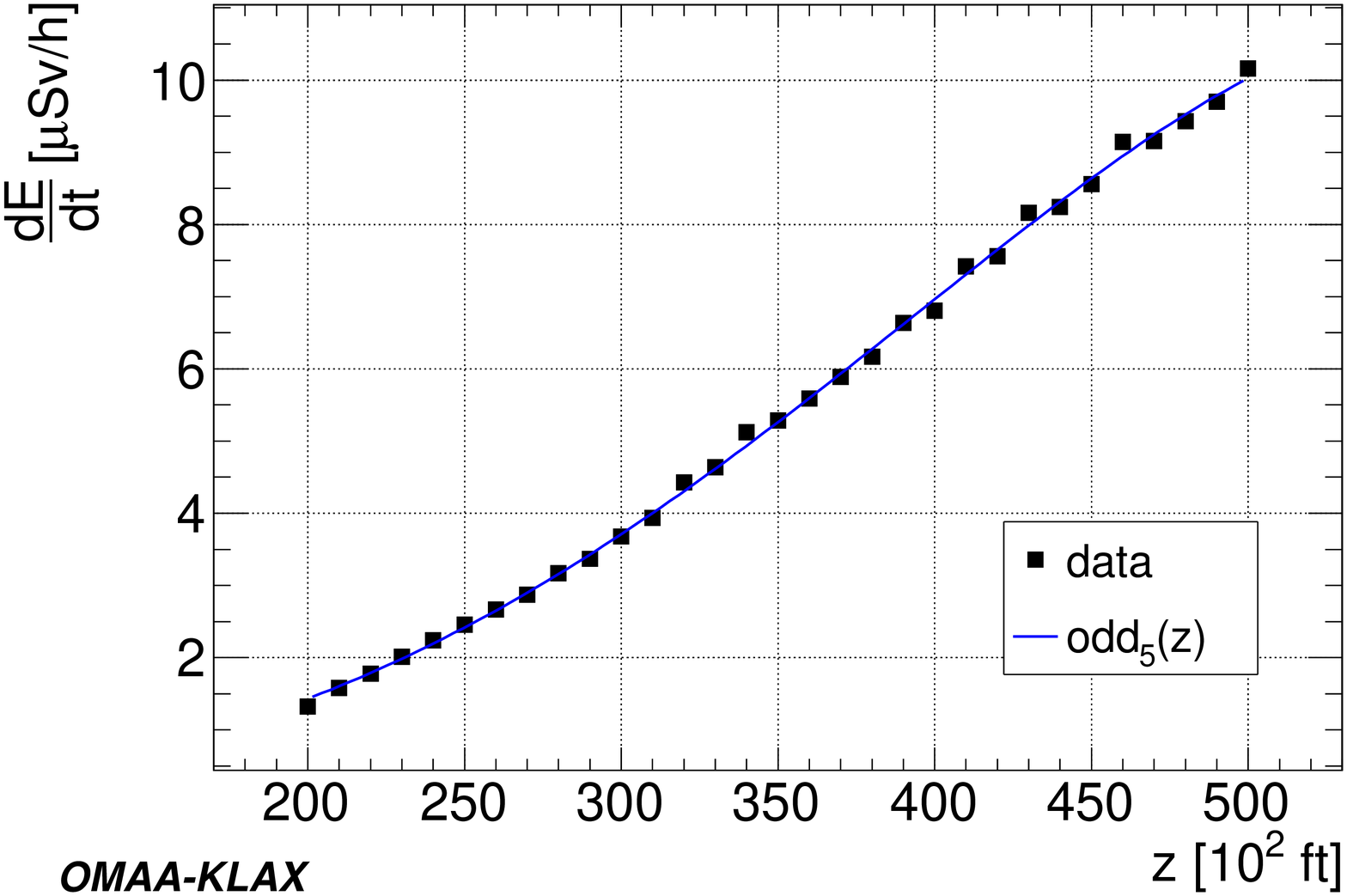}
		\label{fig:OMAA-KLAX_fit}
	} 
	&
	\subfloat[OMDB-KLAX]{
		\includegraphics[width=3.0cm]{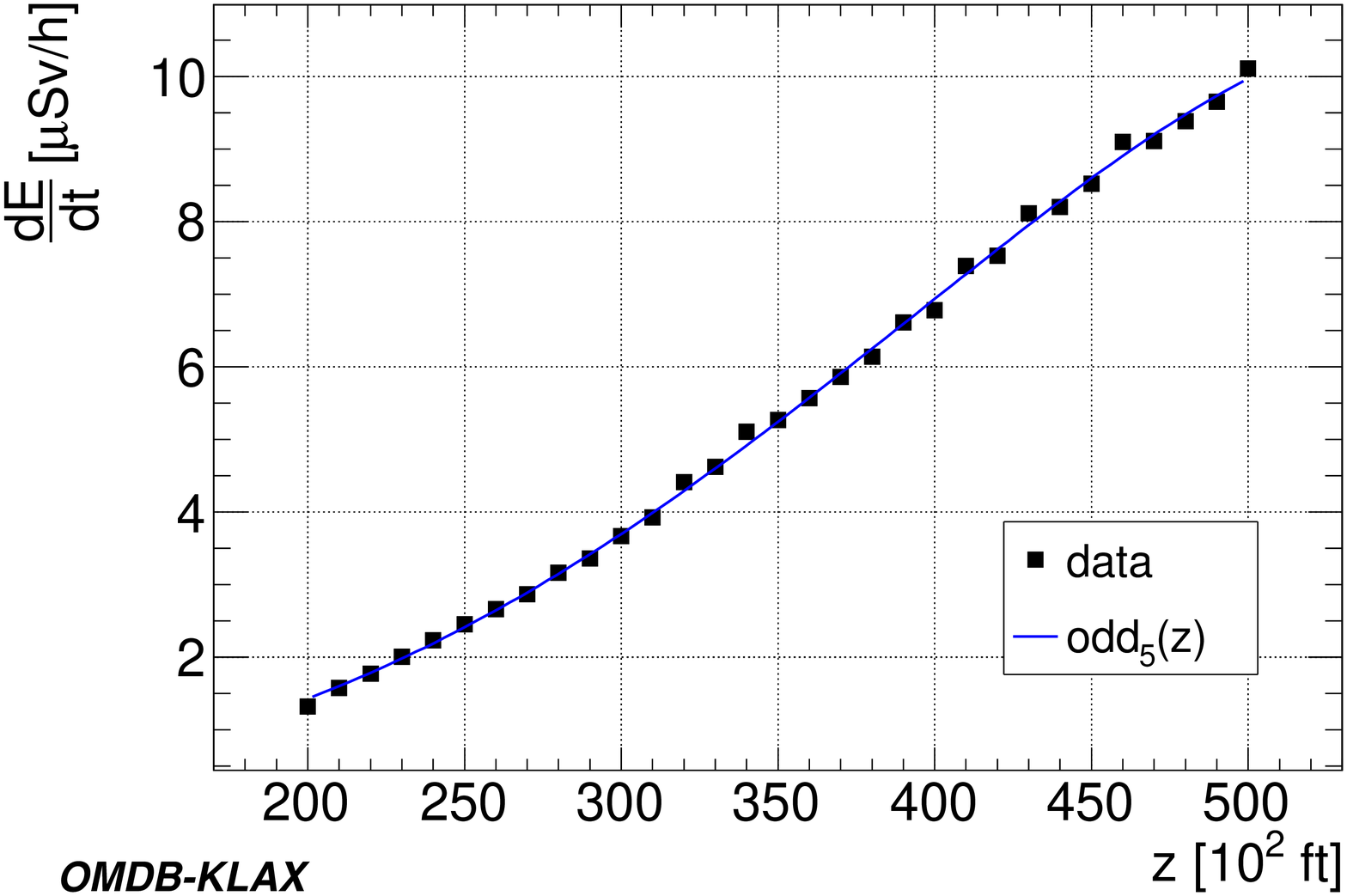}
		\label{fig:OMDB-KLAX_fit}
	} 
	\\
	\subfloat[OEJN-KLAX]{
		\includegraphics[width=3.0cm]{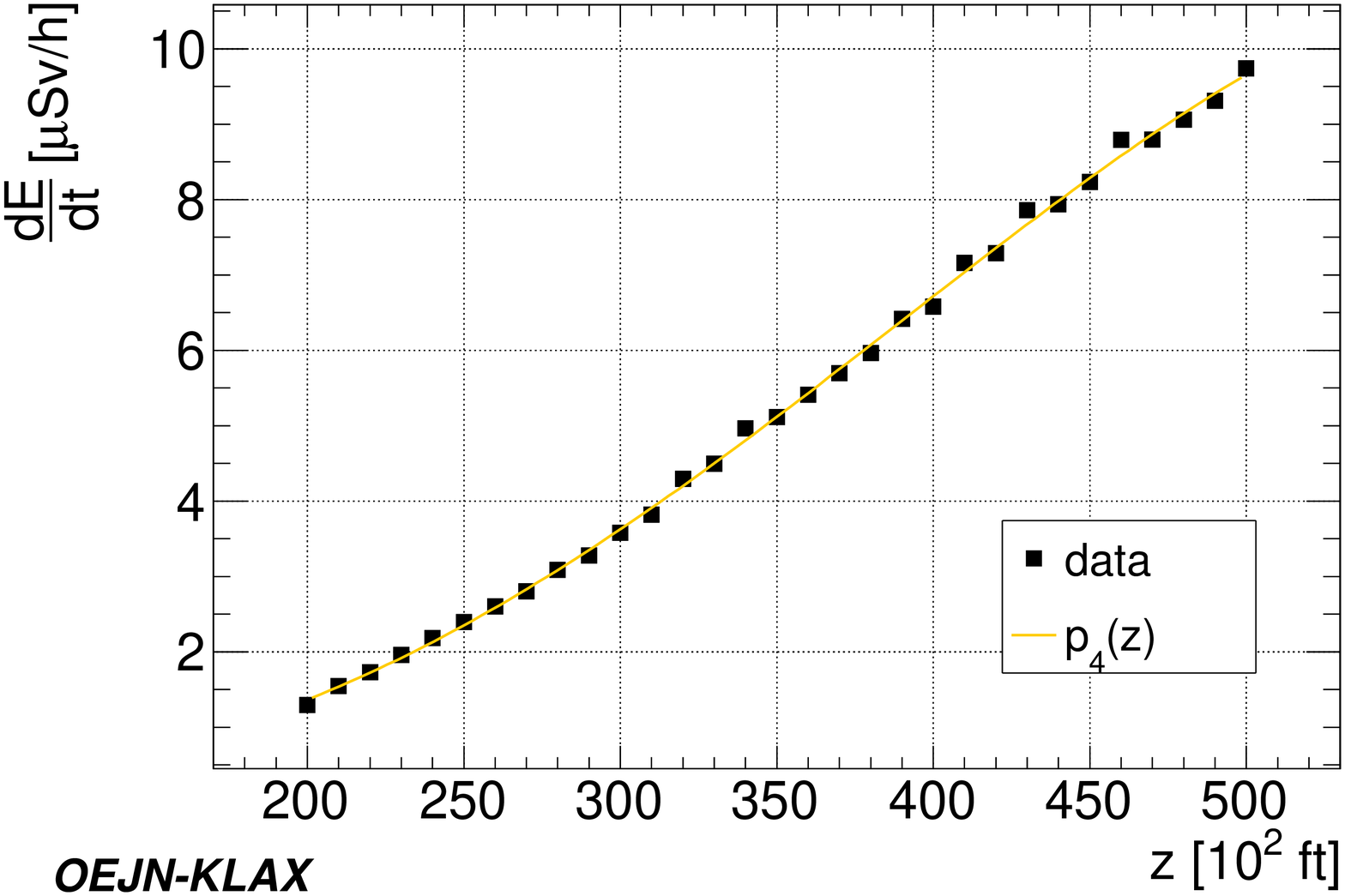}
		\label{fig:OEJN-KLAX_fit}
	}  
	& 
	\subfloat[SCEL-YSSY]{
		\includegraphics[width=3.0cm]{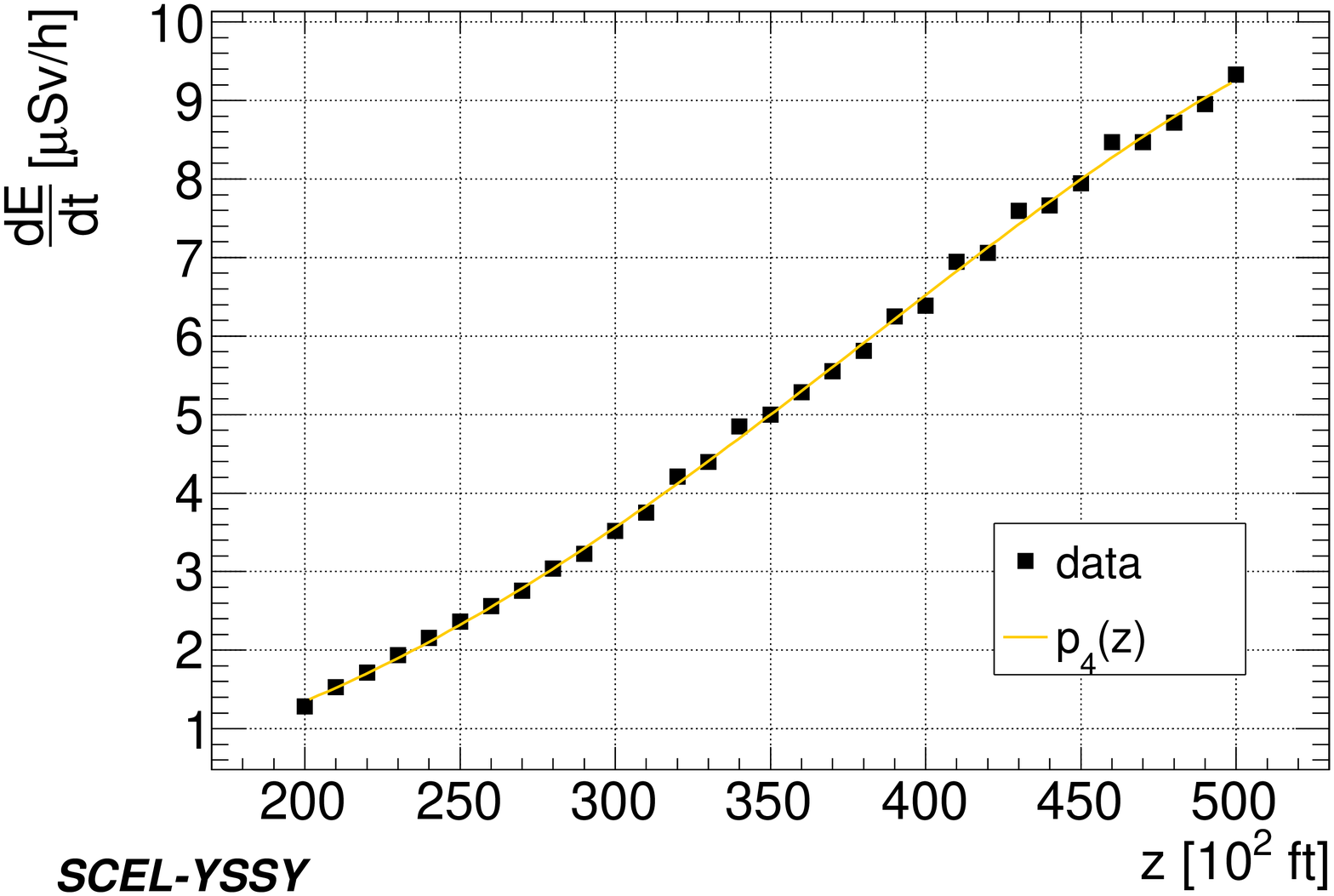}
		\label{fig:SAWG-SAWB_fit}
	} 
	&
	\subfloat[WSSS-KEWR]{
		\includegraphics[width=3.0cm]{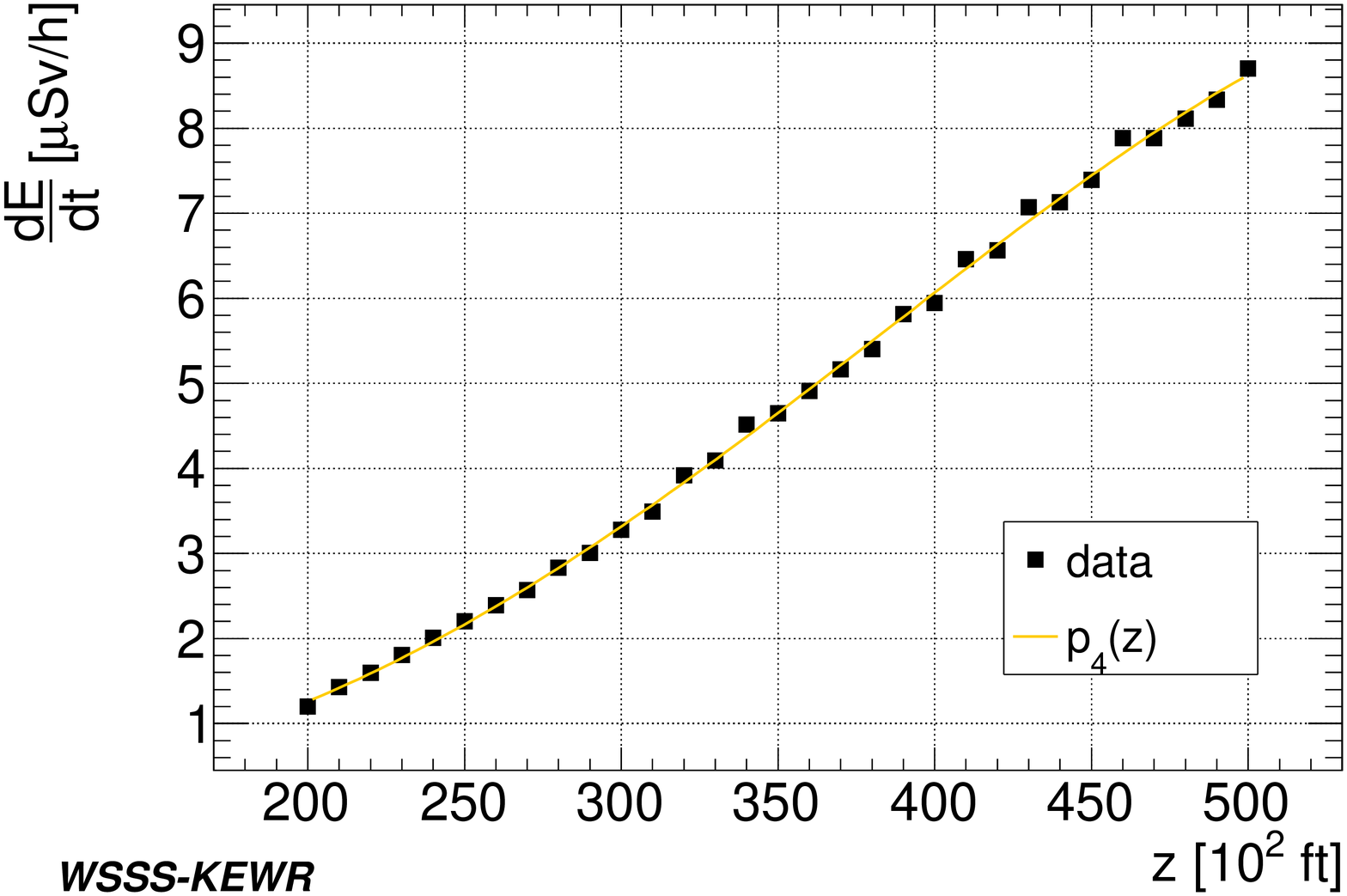}
		\label{fig:WSSS-KEWR_fit}
	} 
	\\
	\subfloat[KEWR-WSSS]{
		\includegraphics[width=3.0cm]{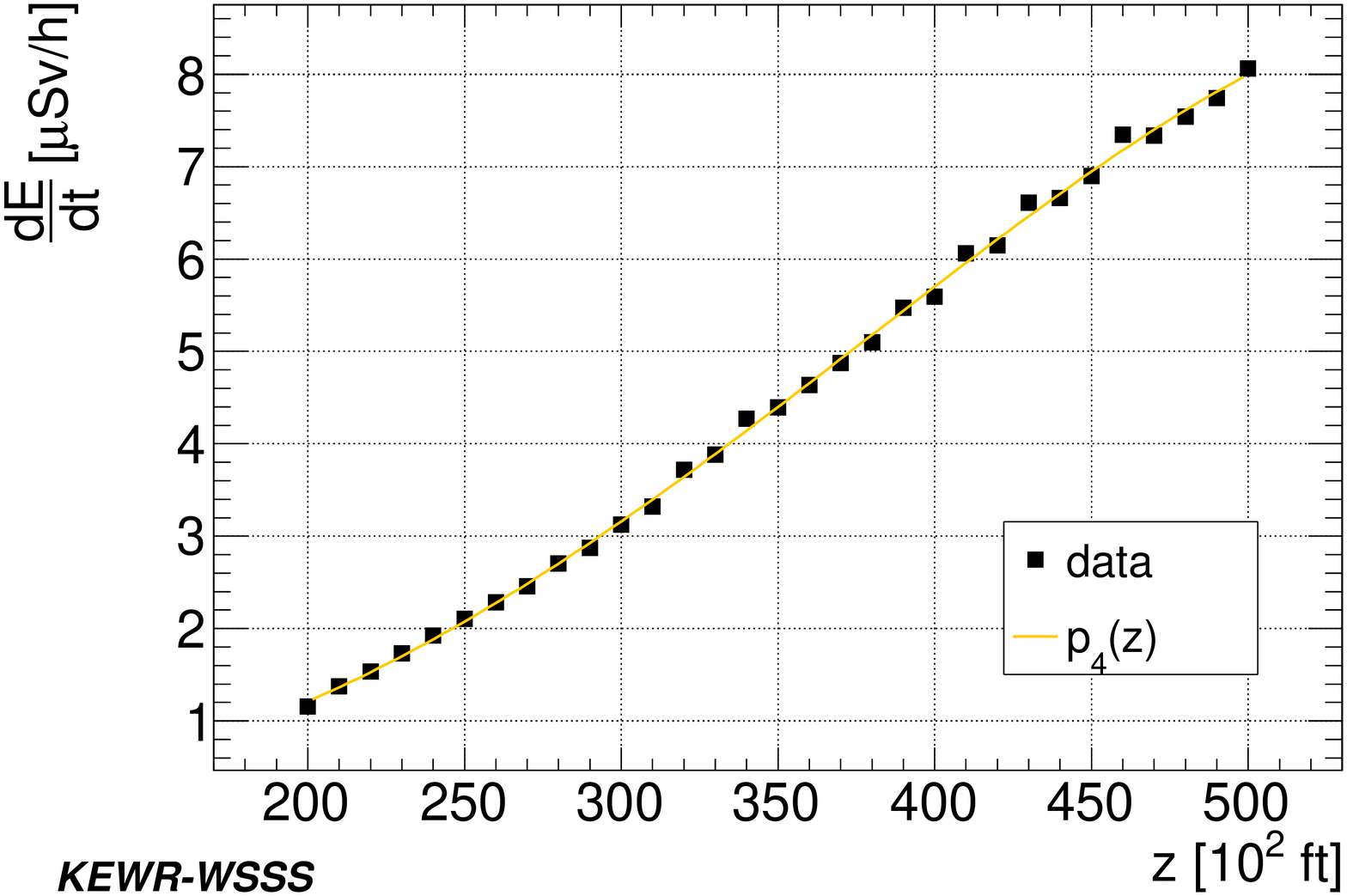}
		\label{fig:KEWR-WSSS_fit}
	} 
	&
	\subfloat[SAWG-SAWB$^{*}$ at Hercules C-130 cruise speed.]{
		\includegraphics[width=3.0cm]{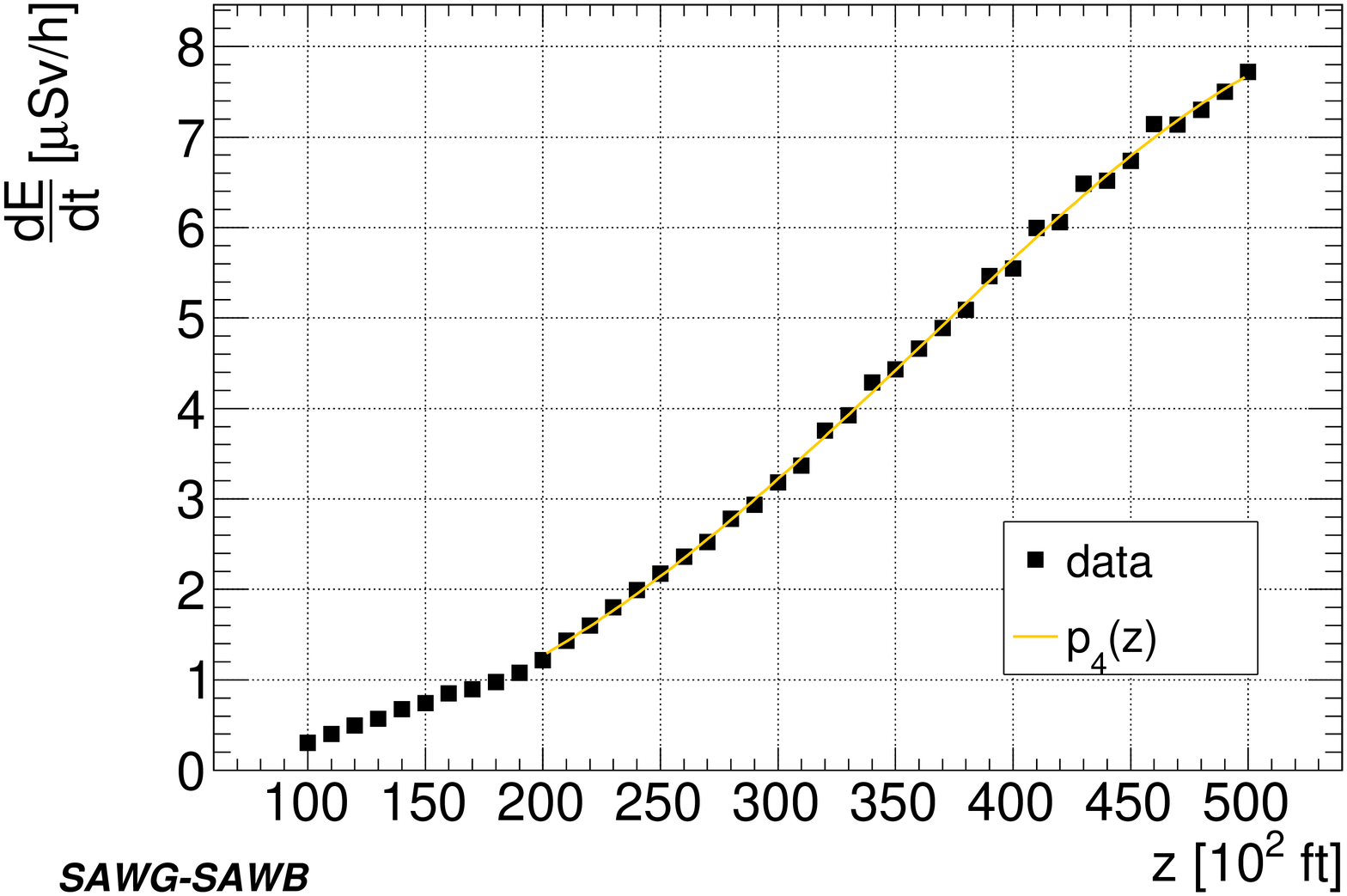}
		\label{fig:SAWG-SAWB_fit}
	} 
	&
	\subfloat[KLAX-WSSS]{
		\includegraphics[width=3.0cm]{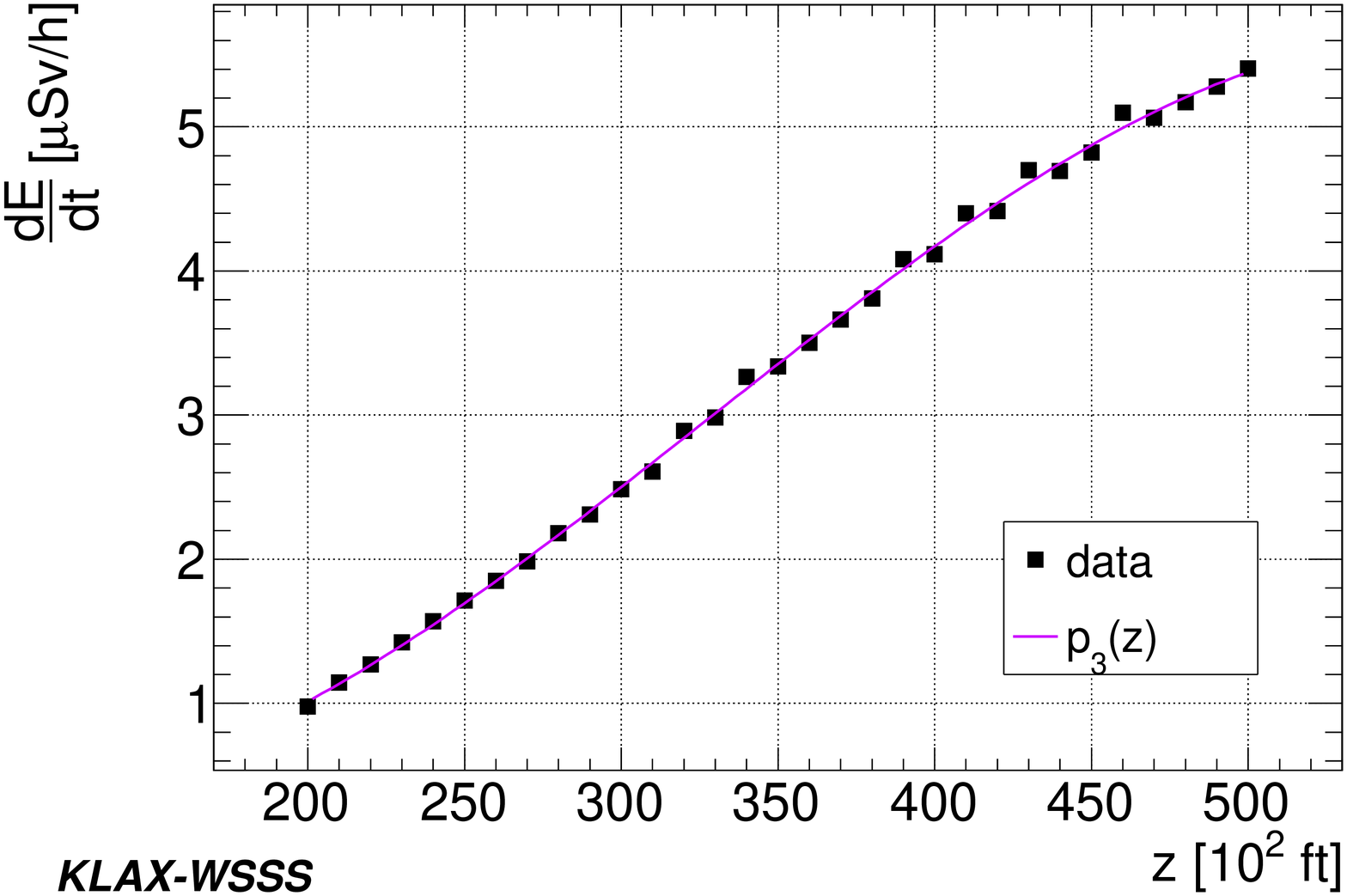}
		\label{fig:KLAX-WSSS_fit}
	}  
	\\
	\subfloat[YPPH-EGLL]{
		\includegraphics[width=3.0cm]{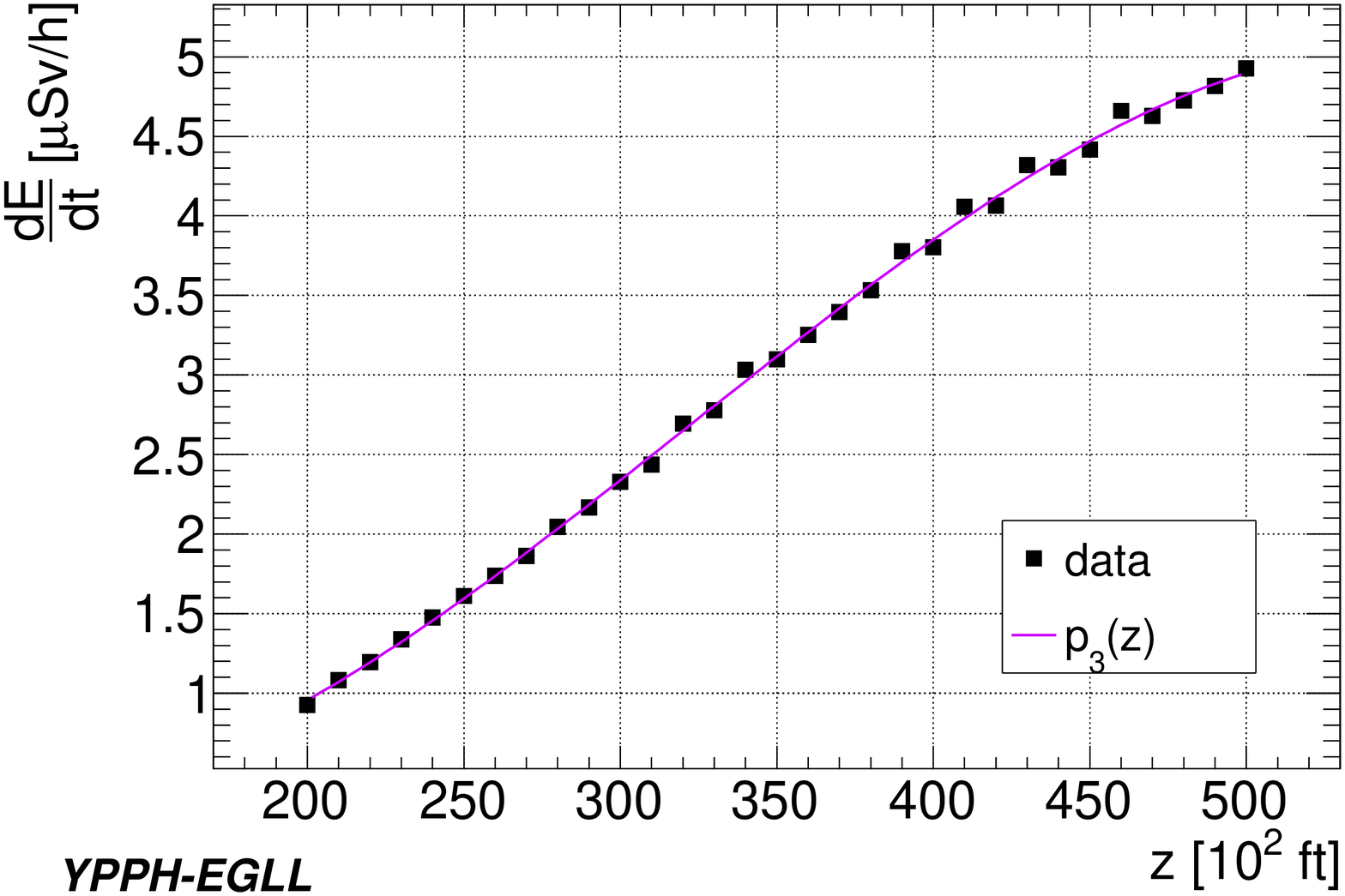}
		\label{fig:YPPH-EGLL_fit}
	} 
	&
	\subfloat[FAOR-KATL]{
		\includegraphics[width=3.0cm]{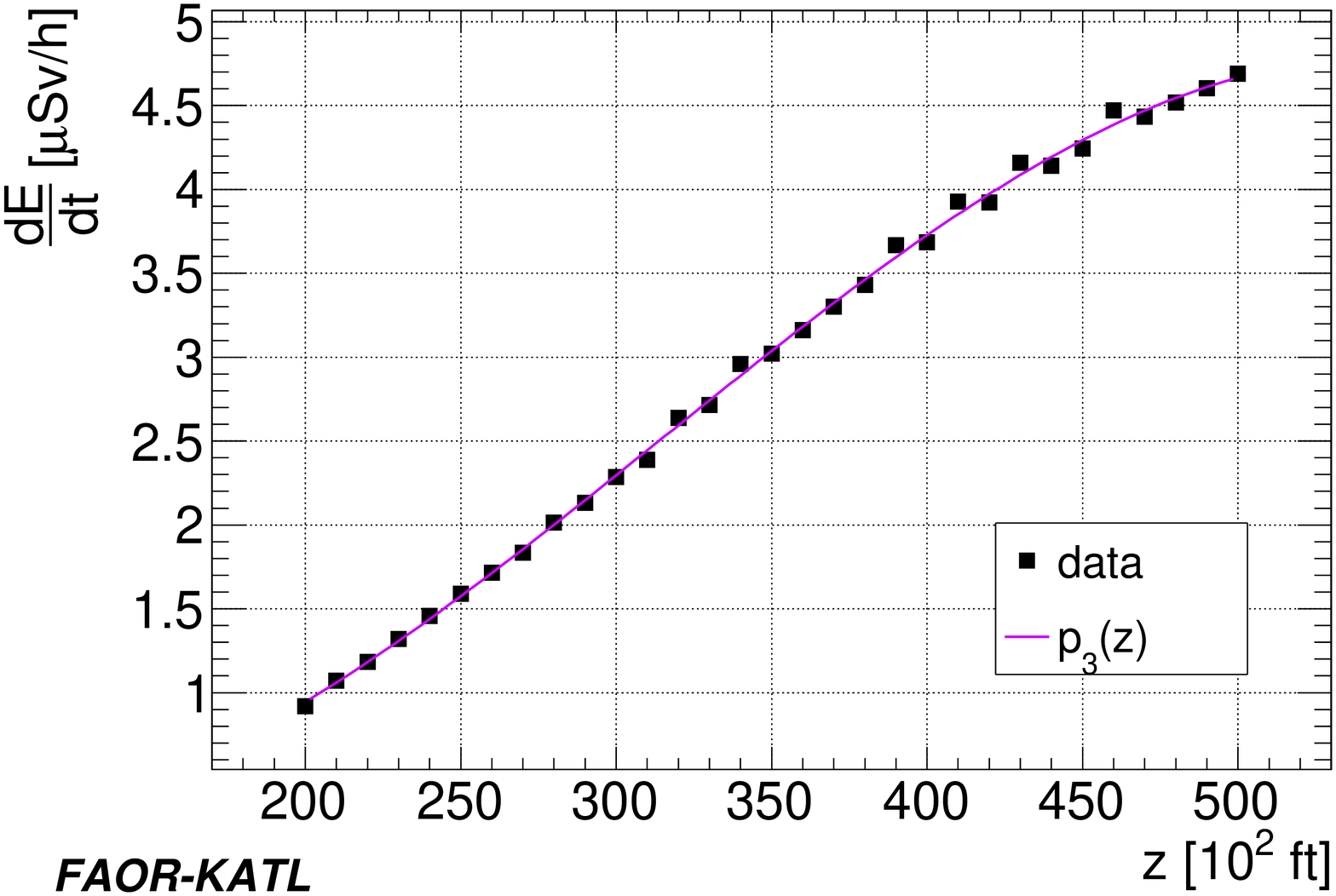}
		\label{fig:FAOR-KATL_fit}
	} 
	&
	\subfloat[KSFO-WSSS]{
		\includegraphics[width=3.0cm]{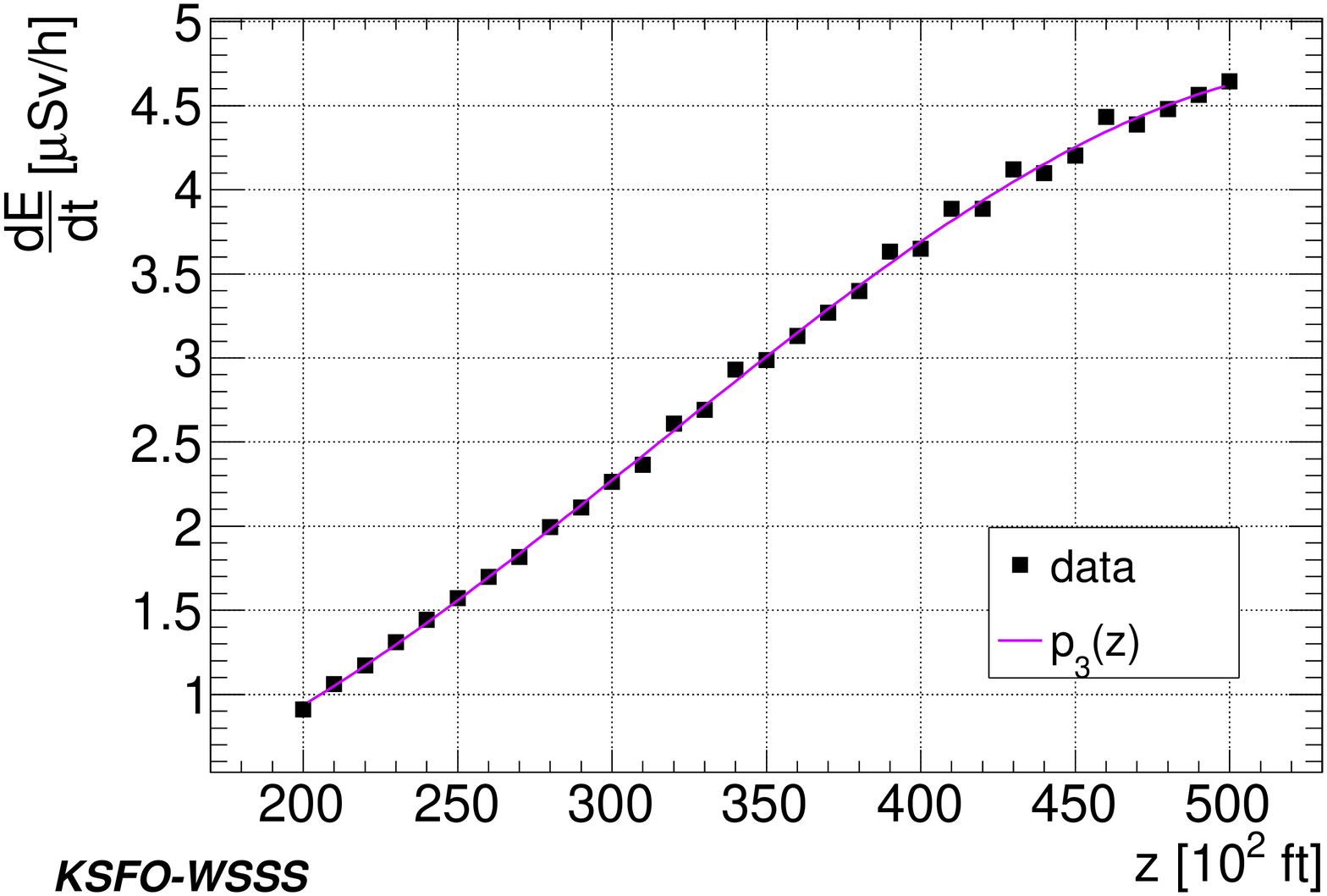}
		\label{fig:KSFO-WSSS_fit}
	} 
	\\
	\subfloat[NZAA-OTBD]{
		\includegraphics[width=3.0cm]{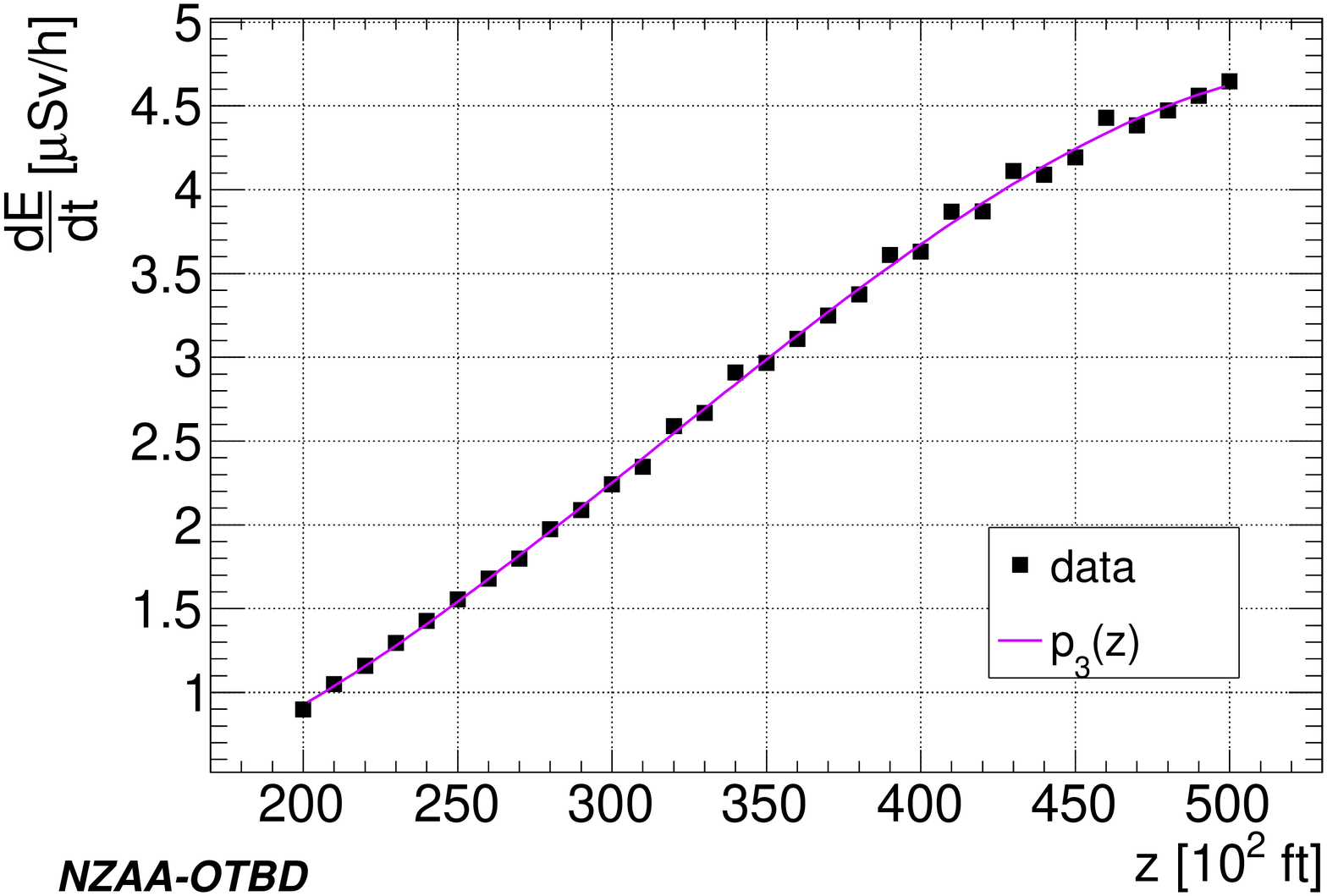}
		\label{fig:NZAA-OTBD_fit}
	} 
	&
	\subfloat[NZAA-OMDB]{
		\includegraphics[width=3.0cm]{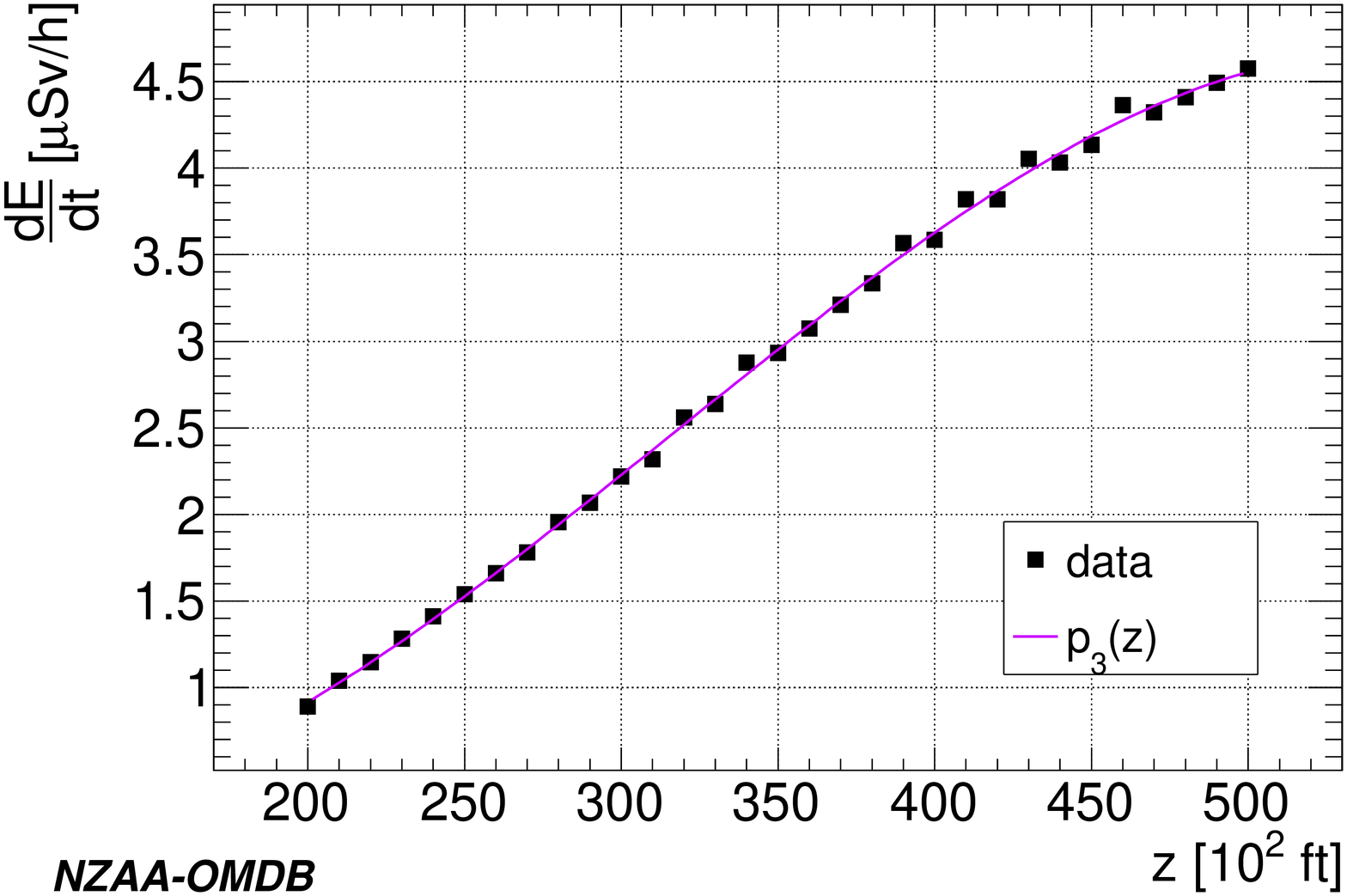}
		\label{fig:NZAA-OMDB_fit}
	} 
	&
	\subfloat[KDFW-YSSY]{
		\includegraphics[width=3.0cm]{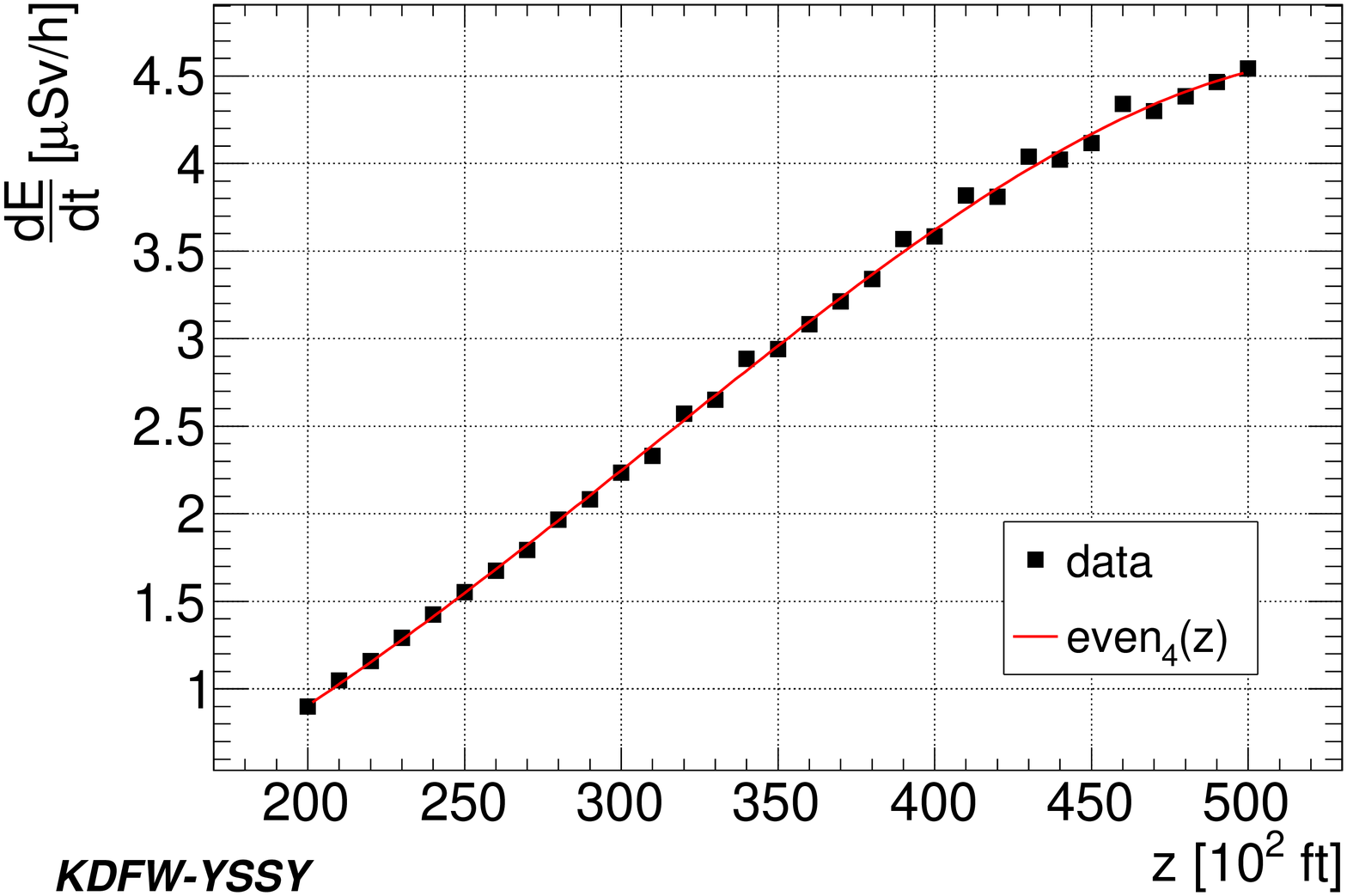}
		\label{fig:KDFW-YSSY_fit}
	} 
	\\
	\subfloat[KIAH-YSSY]{
		\includegraphics[width=3.0cm]{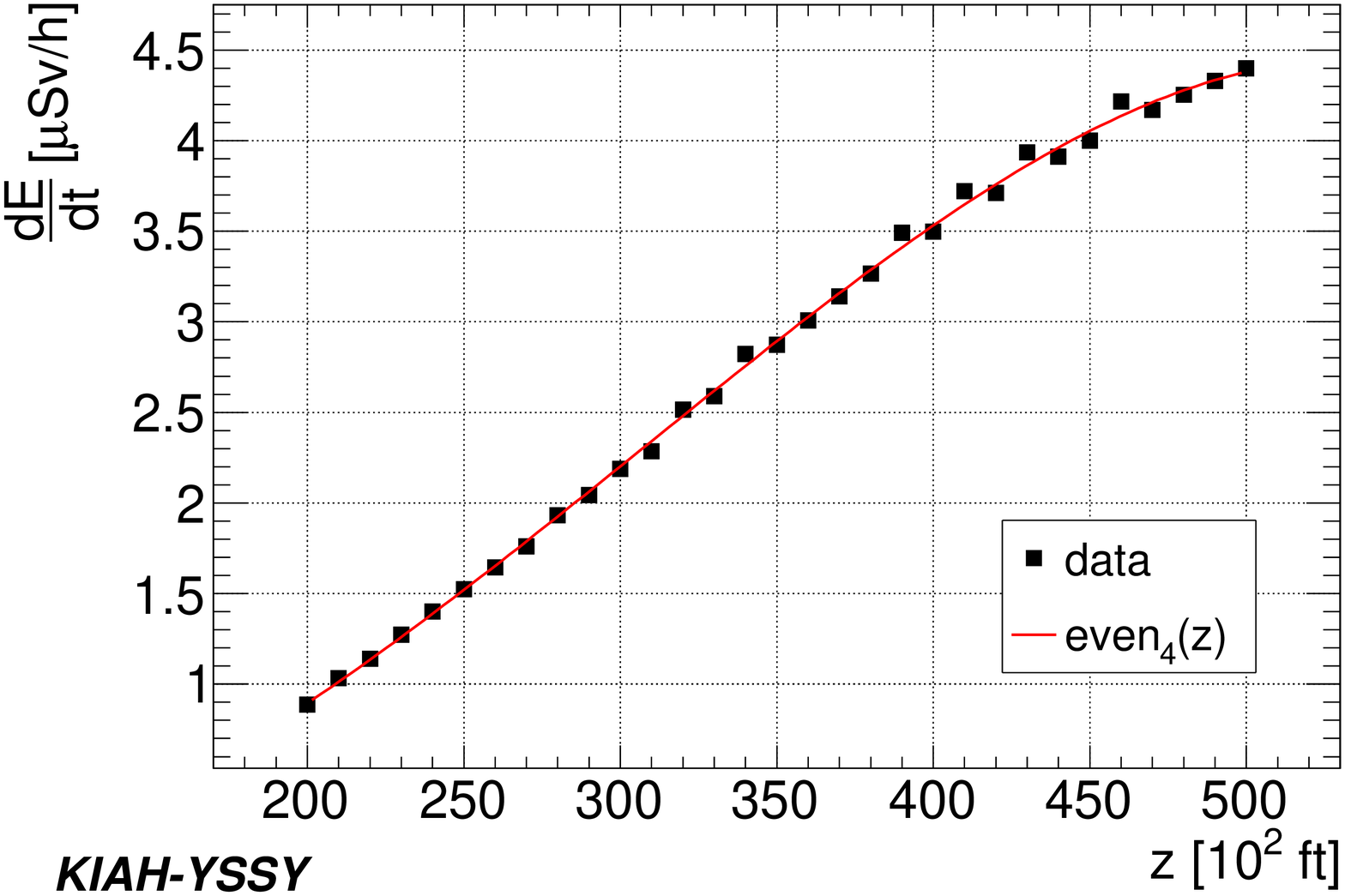}
		\label{fig:KIAH-YSSY_fit}
	} 
	&
	\subfloat[SAWG-SAWB$^{**}$ at Hercules C-130 cruise speed.]{
		\includegraphics[width=3.0cm]{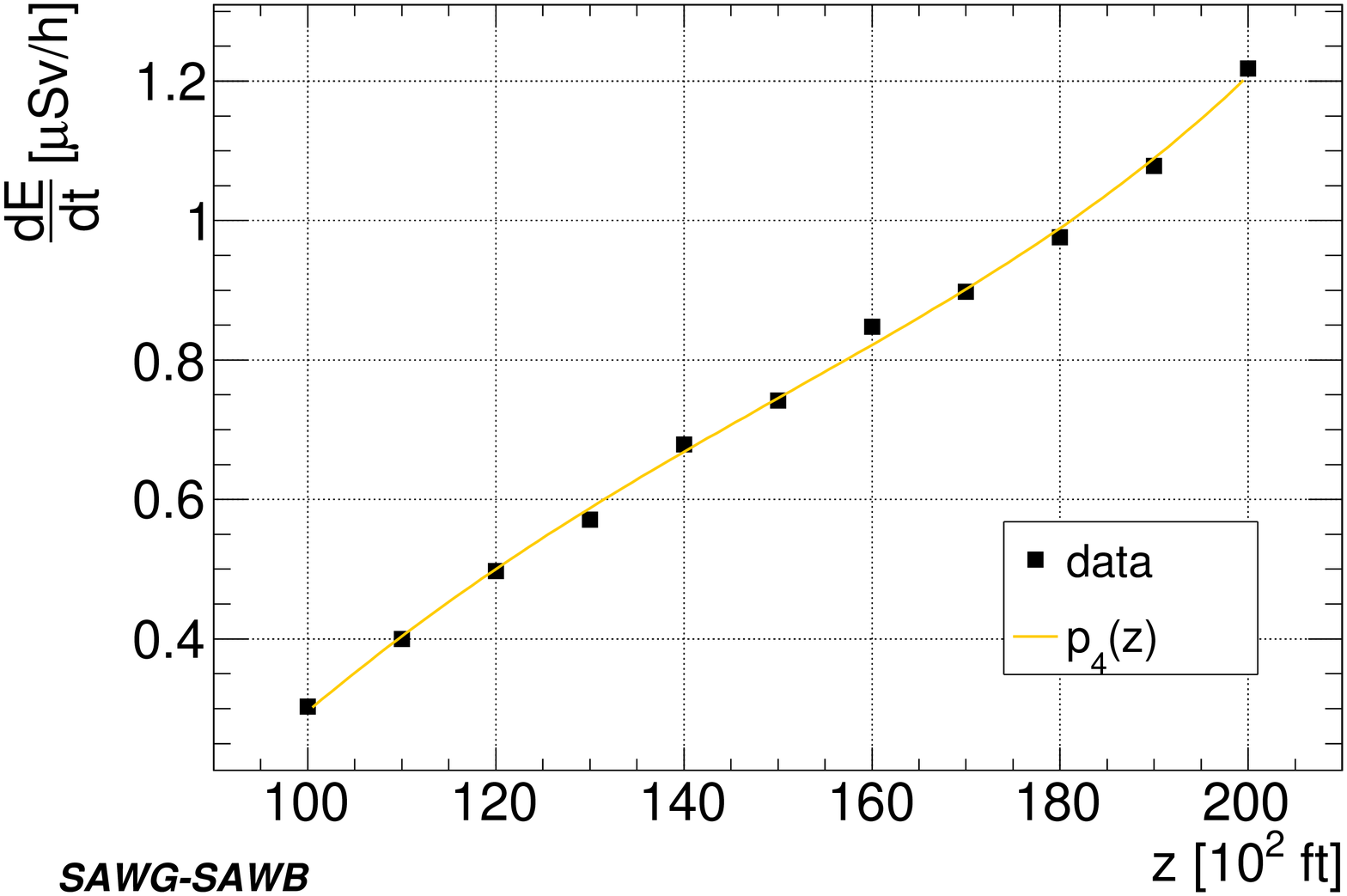}
		\label{fig:SAWG-SAWB_fit}
	} 
	&
	\subfloat[SAWG-SAWB$^{\dagger}$ at Twin Otter DHC-60 cruise speed.]{
		\includegraphics[width=3.0cm]{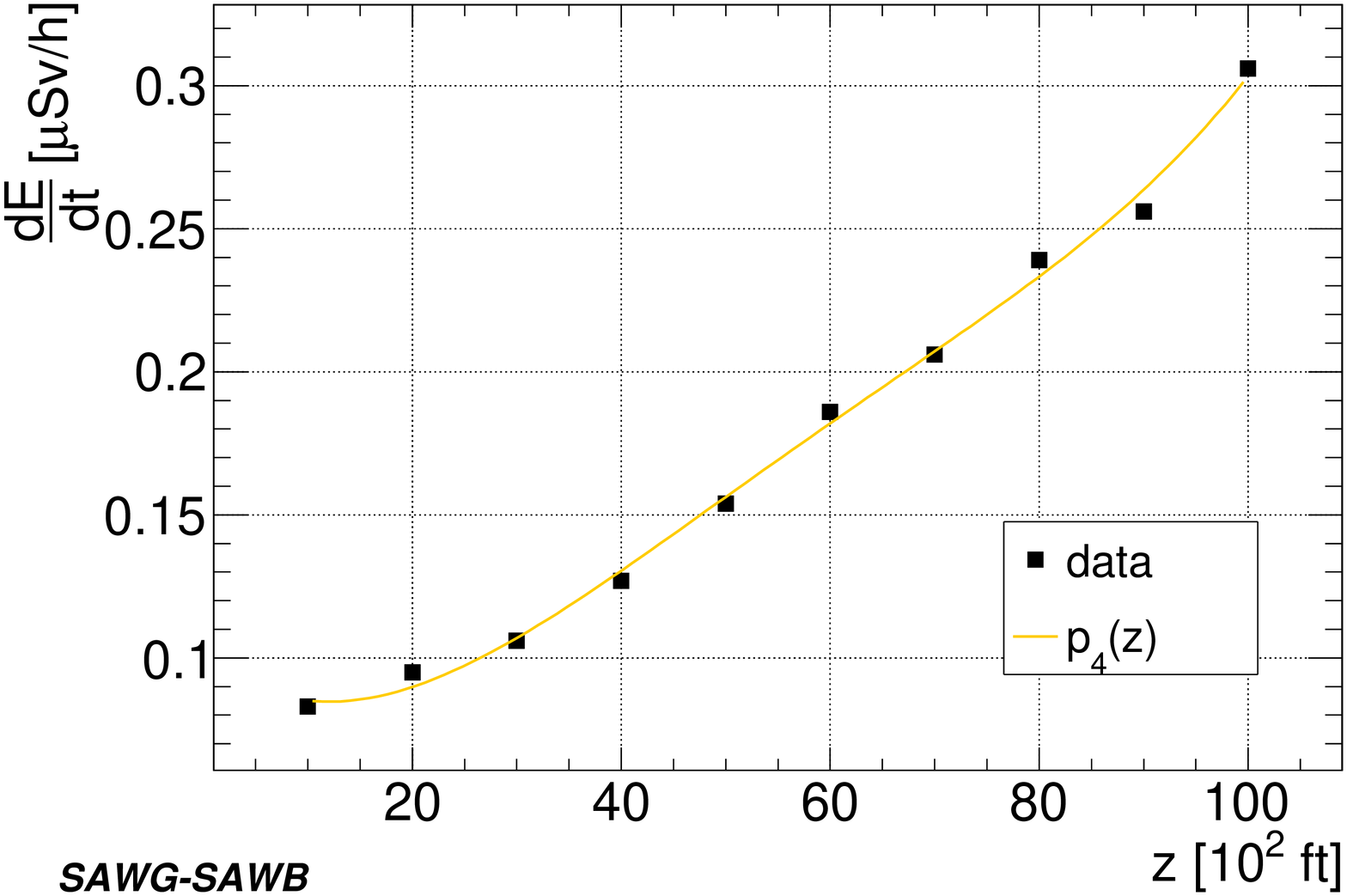}
		\label{fig:SAWG-SAWB_fit}
	} 
	\\
	\end{tabular}
\caption{ICRP-103 EARDR best fit as a function of altitude $z$ plots. The cosmic aeroradiation polynomials. 
}
\label{fig:plotfits}
\end{figure}

Figure~\ref{fig:oneflight} plots the simulation results in one flight $O^{route}_{FL}$ for 
the particle fluence and the absorbed doses calculated by CARI-7A, for ICRP-60, ICRP-103 definitions, 
ICRU $H^{*}(10)$, and whole body.
In these plots we can see the flight level (in hundred of feet) on the abscissa and 
the fluence or absorbed doses (fluence in $particles/\UcmZ^{2}$, the doses in $\UuSvZ$) on the ordinate, 
where white stars are for fluence, black circles are for whole body, 
black triangles represent ICRP-103, white circles represent $H^{*}(10)$,
and white triangles are for ICRP-60.
\\\\We can take the number of hours in a year 8766 as the maximal value for a variable that represents the net duration of flights 
in each aerial flight route, let's say $t$.
As pilots does not fly the entire year, we need to change the maximal value by $1000\Uh$ which is the maximum number of flight hours per year
allowed by the FAA.
Then we divide it by $\tau^{route}_{FL}$ to get the number of flights at a given FL:
$N^{flights}_{FL} = t/\tau^{route}_{FL}$, necessary to be performed in order to complete its theorized and allowed time of flights. 
In this way, we can vary $t$ from zero to one thousand hours in steps of one hour. As $N^{flights}_{FL}$ resulted by definition in a linear function of $t$, i
it gives us another reason to plot the radiation dose charts Figure \ref{fig:maps} for $t \in [0,1000]\Uh$. 
\\\\From Figure~\ref{fig:oneflight} we note that ICRP-103 and ICRU $H^{*}(10)$ have almost the same values.
Since we have already known the value of the absorbed radiation dose according to the ICRP-103 specifications (the black triangles)
for one flight at a given FL, this is $O^{route}_{FL}$ plotted in Figure~\ref{fig:oneflight},
we can finally obtain the absorbed radiation dose as
\begin{equation}
E^{route}_{FL}(t) 
= N^{flights}_{FL} \cdot O^{route}_{FL} = \bigg( \frac{t}{\tau^{route}_{FL}} \bigg) \cdot O^{route}_{FL} = \frac{dE}{dt} \; t,
\label{eq:main}
\end{equation}
where $dE/dt \equiv O^{route}_{FL} / \tau^{route}_{FL}$.
\\\\As an application example at FL=430, we look at  
Table~\ref{tab:podium} and search ICRP-103 for route OMAA-KLAX, we call it 
$O^{{\scriptsize \textrm{OMAA-KLAX}}}_{430} = 128.5 \UuSv$, and also take the time elapsed in that flight 
\\$\tau^{{\scriptsize \textrm{OMAA-KLAX}}}_{430}=15.75\Uh$ and inserting all in (\ref{eq:main}), 
we have $E^{{\scriptsize \textrm{OMAA-KLAX}}}_{430}(t) = \left( 8.1587 \UuSv/\UhZ \right)\cdot t$, 
which is an amazing result! 
We are going to exploit it later, but now we can say it stands for any $t$ and is the replacement of the charts
Figure~\ref{fig:maps} for $t>1000\Uh$ (but only at FL=430). 
We can also regard $8.1587 \UuSv/\UhZ > 5 \UuSv/\UhZ$!
\\\\But the easiest way to search the absorbed radiation dose is looking at the radiation dose maps directly in Figure~\ref{fig:maps}.
In Figures~\ref{fig:maps}(a)-\ref{fig:maps}(p) we could see the radiation dose isocurves in the plane altitude versus time, $z$ vs $t$. 
The colors are explained in Figure~\ref{fig:maps}(q),
there the green color extends for $E_{ICRP-103}$ from $0$ til $1\UmSv$, where there are six zones, from light green to dark green:
$E<0.25\UmSv$,
$0.25<E<0.33\UmSv$,
$0.33\UmSv<E<0.5\UmSv$,
$0.5\UmSv<E<0.66\UmSv$,
$0.66\UmSv<E<0.75\UmSv$,
$0.75\UmSv<E<1\UmSv$.
The whole green zone is the accepted value for humans at ground level during a year, see Table~\ref{tab:doses}.
Yellow color is for $1\UmSv<E<2\UmSv$,
orange color is for $2\UmSv<E<3\UmSv$,
red color is for $3\UmSv<E<4\UmSv$,
dark red color is for $4\UmSv<E<5\UmSv$, until here we get the limit for workers that perform their activities in areas of radiation.
From now on the following colors represent values of EARD that are outside the limits of radiological protection, these are:
Cyan is for $5\UmSv<E<6\UmSv$,
magenta is for $6\UmSv<E<7\UmSv$,
violet is for $7\UmSv<E<8\UmSv$,
violet2 is for $8\UmSv<E<9\UmSv$,
blue is for $9\UmSv<E<10\UmSv$,
gray1 is for $10\UmSv<E<15\UmSv$, 
gray2 is for $15\UmSv<E<20\UmSv$, and
gray3 is for $E>20\UmSv$.
\\\\The (\ref{eq:main}) plotted for dose option ICRP-103 in Figure~\ref{fig:maps} was the main goal of this work,
because it can be very useful for health concerned pilots and passengers.
However we did an additional analysis, returning to the amazing result 
$E^{{\scriptsize \textrm{OMAA-KLAX}}}_{430}(t) = \left( 8.1587 \UuSv/\UhZ \right)\cdot t$, 
we generalize the procedure and take the transformation (\ref{eq:main}) for all the simulated data
and we could be able to obtain $dE/dt = f(z)$ for all routes, 
where $z$ represents the alitude in units of $\UFLZ$.
We use the reduced chi squared criterion 
$\tilde{\chi}^{2} = \chi^{2}/\nu$, where $\nu$ is the fit number of degrees of freedom,
in order to find the best fit for a lot of trial functions, 
thus we can be able to find the fit function with the 
minimum reduced chi squared and take it as the best fit.
Although we have tested trial functions of the form
\begin{equation}
dE/dt = k_{0} z^{k_{1}},
\label{eq:power}
\end{equation}
and 
\begin{equation}
dE/dt = k_{0} + k_{1} z^{k_{2}},
\label{eq:powershift}
\end{equation}
the best results are found to be the following function families: the $n$-th order polynomials with $n=3,4$  
\begin{equation}
dE/dt = p_{n}(z),
\label{eq:pn}
\end{equation}
the 5-th order odd polynomial
\begin{equation}
dE/dt =  p_{5}(z)_{odd} =  k_{1} z + k_{3} z^{3} + k_{5} z^{5},
\label{eq:odd5}
\end{equation}
and finally the 4-th order even polynomial
\begin{equation}
dE/dt =  p_{4}(z)_{even} = k_{0} + k_{2} z^{2} + k_{4} z^{4}.
\label{eq:even4}
\end{equation}
The best fit functions or the aeroradiation polynomials for all routes can be seen in Table~\ref{tab:fits}. 
Plots for the best fit function for each route can be seen in Figure~\ref{fig:plotfits}, see also Figure~\ref{fig:globo}.
Now we choose the specific point at flight level 430 to do some inspection on the physical variables, 
in~Table~\ref{tab:podium} we can see 
the podium for the routes in descendent order of irradiation, column 1 is for the route, column 2 
is $E$ from CARI-7A ICRP-103, column 3 is for CARI-7A $H^{*}(10)$, while the column 4 shows a picture in flight time at $t=1000\Uh$
instead of a fixed FL430 it shows the actual FL at which we can surpase $5\UmSv$ in $E$. 
We also show the particle content at FL430.
Table~\ref{tab:fluenceparticle1} shows the particles fluence computed with CARI-7A 
for the five routes with the leading EARDR, these are the quantities $\int dL \; L\, F_{T}(L)$.
Using Tables~\ref{tab:fluenceparticle1} with the help of \cite{ICRP116} and \cite{ICRP123},
one may obtain the absorbed effective radiation doses
for each kind of radiation particle and for the total contribution,
however, CARI-7A does it internally for us.
Table~\ref{tab:EARDparticle1} shows the percentage contribution of each kind of particle to the ICRP-103 EARD.
\begin{table}[h!]
\centering
\begin{tabular}{clc}
\hline
ROUTE & $dE/dt$ in $[\UuSvZ/\UhZ]$ as a function of $z$ in $[\UFLZ]$ & $\tilde{\chi}^{2}$ \\
\hline
OTBD-KLAX & $(20 \pm 2 )\times 10^{-4}\,z + (139 \pm 3 )\times 10^{-9}\,z^{3} + (-267 \pm 9 )\times 10^{-15}\,z^{5}$ & 0.0088009 \\
OMAA-KLAX & $(20 \pm 2 )\times 10^{-4}\,z + (138 \pm 3 )\times 10^{-9}\,z^{3} + (-267 \pm 9 )\times 10^{-15}\,z^{5}$ & 0.00877444 \\
OMDB-KLAX & $(20 \pm 2 )\times 10^{-4}\,z + (138 \pm 3 )\times 10^{-9}\,z^{3} + (-266 \pm 9 )\times 10^{-15}\,z^{5}$ & 0.00873154 \\
\hline
OEJN-KLAX & $(78 \pm 7 )\times 10^{-2} + (-88 \pm 4 )\times 10^{-4}\,z + (418 \pm 9 )\times 10^{-7}\,z^{2}$ & 0.00799637  \\
          & $+(126 \pm 2 )\times 10^{-9}\,z^{3}+ (-206 \pm 3 )\times 10^{-12}\,z^{4}$ & \\
SCEL-YSSY & $(64 \pm 7 )\times 10^{-2} + (-77 \pm 3 )\times 10^{-4}\,z + (406 \pm 9 )\times 10^{-7}\,z^{2}$ & 0.00712103  \\
          & $+ (119 \pm 2 )\times 10^{-9}\,z^{3}+ (-200 \pm 3 )\times 10^{-12}\,z^{4}$ & \\
WSSS-KEWR & $(56 \pm 6 )\times 10^{-2} + (-68 \pm 3 )\times 10^{-4}\,z + (374 \pm 8 )\times 10^{-7}\,z^{2}$ & 0.00628343  \\
          & $+ (108 \pm 2 )\times 10^{-9}\,z^{3}+ (-182 \pm 3 )\times 10^{-12}\,z^{4}$ & \\
KEWR-WSSS & $(40 \pm 6 )\times 10^{-2} + (-54 \pm 3 )\times 10^{-4}\,z + (349 \pm 7 )\times 10^{-7}\,z^{2}$ & 0.00531334  \\
          & $+ (97 \pm 2 )\times 10^{-9}\,z^{3}+ (-169 \pm 3 )\times 10^{-12}\,z^{4}$ & \\
SAWG-SAWB$^{*}$ & $(31 \pm 5 )\times 10^{-2} + (-46 \pm 3 )\times 10^{-4}\,z + (351 \pm 7 )\times 10^{-7}\,z^{2}$ & 0.00327901  \\
          & $+ (93 \pm 2 )\times 10^{-9}\,z^{3}+ (-173 \pm 3 )\times 10^{-12}\,z^{4}$ & \\
SAWG-SAWB$^{**}$ & $(-147 \pm 4 )\times 10^{-2} + (221 \pm 4 )\times 10^{-4}\,z + (-5 \pm 2 )\times 10^{-6}\,z^{2}$ & 0.000265652  \\
          & $+ (-59 \pm 1 )\times 10^{-8}\,z^{3}+ (198 \pm 5 )\times 10^{-11}\,z^{4}$ & \\
SAWG-SAWB$^{\dagger}$ & $(10 \pm 2 )\times 10^{-2} + (-3 \pm 1 )\times 10^{-3}\,z + (15 \pm 4 )\times 10^{-5}\,z^{2}$ & 0.0000220056  \\
          & $+ (-18 \pm 6 )\times 10^{-7}\,z^{3}+ (8 \pm 3 )\times 10^{-9}\,z^{4}$ & \\
\hline
KLAX-WSSS & $(115 \pm 38 )\times 10^{-2} + (-174 \pm 38 )\times 10^{-4}\,z + (105 \pm 10 )\times 10^{-6}\,z^{2}$ & 0.00224627  \\
          & $+ (-107 \pm 10 )\times 10^{-9}\,z^{3}$ & \\
YPPH-EGLL & $(93 \pm 38 )\times 10^{-2} + (-149 \pm 35 )\times 10^{-4}\,z + (94 \pm 10 )\times 10^{-6}\,z^{2}$ & 0.00184954  \\
          & $+ (-98 \pm 9 )\times 10^{-9}\,z^{3}$ & \\
FAOR-KATL & $(77 \pm 1 )\times 10^{-2} + (-133 \pm 1 )\times 10^{-4}\,z + (214 \pm 4 )\times 10^{-7}\,z^{2}$ & 0.00173408  \\ 
          & $+ (-95 \pm 9 )\times 10^{-9}\,z^{3}$ & \\
KSFO-WSSS & $(77 \pm 37 )\times 10^{-2} + (-132 \pm 33 )\times 10^{-4}\,z + (891 \pm 1 )\times 10^{-7}\,z^{2}$ & 0.00174351  \\ 
          & $+ (-94 \pm 9 )\times 10^{-9}\,z^{3}$ & \\
NZAA-OTBD & $(82 \pm 37 )\times 10^{-2} + (-136 \pm 34 )\times 10^{-4}\,z + (897 \pm 100 )\times 10^{-7}\,z^{2}$ & 0.00175141  \\ 
          & $+ (-94 \pm 9 )\times 10^{-9}\,z^{3}$ & \\
NZAA-OMDB & $(79 \pm 36 )\times 10^{-2} + (-132 \pm 33 )\times 10^{-4}\,z + (881 \pm 1 )\times 10^{-7}\,z^{2}$ & 0.00169213  \\ 
          & $+ (-93 \pm 9 )\times 10^{-9}\,z^{3}$ & \\
\hline
KDFW-YSSY & $(-38 \pm 3 )\times 10^{-2} + (-346 \pm 4 )\times 10^{-7}\,z^{2} + (9 \pm 1 )\times 10^{-12}\,z^{4}$ & 0.0016514  \\ 
KIAH-YSSY & $(-37 \pm 3 )\times 10^{-2} + (-346 \pm 4 )\times 10^{-7}\,z^{2} + (9 \pm 1 )\times 10^{-12}\,z^{4}$ & 0.00161769  \\ 
\hline
\end{tabular}
\caption{Cosmic aeroradiation polynomials defined in (\ref{eq:pn}), (\ref{eq:odd5}), and (\ref{eq:even4}). 
Each plot can be viewed from Figure \ref{fig:plotfits} All these fits were made by using the HEP data analysis platform ROOT version 5.34.38 \cite{ROOT}.}
\label{tab:fits}
\end{table}
\begin{table}[h!]
\centering
\begin{tabular}{ccccc}
\hline
 Route & ICRP-103 $[\UuSvZ]$ & $H^{*}(10)$ $[\UuSvZ]$ & $\tau^{route}_{430}$ $[\UhZ]$ & FL $| E>5\UmSv$ in 1 FAA $\UyrZ$\\
\hline
 OTBD-KLAX & 126.300 & 136.100 & 15.472 & 343 \\ 
 OMAA-KLAX& 128.500 & 138.490 & 15.750 & 344  \\
 OMDB-KLAX& 127.570 & 137.460 & 15.714 & 344 \\
 OEJN-KLAX&  121.530 & 130.560 & 15.465 & 346  \\
 SCEL-YSSY$^{*}$ & 99.074 & 105.900 & 13.043 & 350  \\
 WSSS-KEWR& 124.860 & 133.470 & 17.664  & 363  \\
 KEWR-WSSS& 122.390 & 130.060 & 18.511 & 373  \\
 SAWG-SAWB$^{*}$ & 20.851 & 22.116 & 3.2159 & 375 \\ 
 KLAX-WSSS& 76.093 & 77.940 & 16.194 & 465 \\
 YPPH-EGLL& 70.481 & 71.672 & 16.317 & $-$ \\
 FAOR-KATL& 63.327 & 63.974 & 15.231 & $-$ \\
 KDFW-YSSY& 66.840 & 67.386 & 16.545 & $-$ \\ 
 NZAA-OTBD& 65.569 & 66.293 & 16.179 & $-$ \\
 NZAA-OMDB& 67.147 & 67.965 & 16.333 & $-$ \\
 KSFO-WSSS& 65.079 & 65.771 & 15.788 & $-$ \\
 KIAH-YSSY& 64.382 & 64.691 & 16.366 & $-$ \\
\hline
\end{tabular}
\caption{Radiation dose values at $z=430\UFL$ after one flight on Airbus 380, except SAWG-SAWB and SCEL-YSSY.
SAWG-SAWB$^{*}$ on Hercules C130, SAWG-SAWB$^{\dagger}$ at $z=100\UFL$ on Twin Otter,
SCEL-YSSY$^{*}$ is for aircraft Boeing~787 the remaining data is for Airbus~380 or Boeing~777 with cruise speed of $907 \Ukm/\UhZ$. 
Flight level for EARD starts to be greater than the standard recommendations, i. e. $E>5\UmSv$, for one FAA allowed year of $1000\Uh$.
Values with dash are greater than $500$ and therefore
out of the range considered in this work.
}
\label{tab:podium}
\end{table}
\begin{table}[h!]
  \centering
\begin{tabular}{cccccc}
\hline
\small{Particle}   & \small{OMAA-KLAX}   & \small{OMDB-KLAX}    & \small{OTBD-KLAX}  & \small{OEJN-KLAX} & \small{SCEL-YSSY}  \\\hline
\Pn              & 93.357     & 92.558     & 91.711     &  86.934     & 69.976       \\
\Pp              & 5.3987     & 5.3532     & 5.3051     &  5.0423     & 4.0883       \\
\PGg             & 709.22     & 705.84     & 697.35     &  869.34     & 578.58       \\
\Pepm            & 18.444     & 18.360     & 18.136     &  17.922     & 15.112       \\
alpha            & 0.025731   & 0.025486   & 0.025301   &  0.023967   & 0.019635   \\
\PGmpm           & 1.5369     & 1.5297     & 1.5119     &  1.4872     & 1.2613       \\
\PGppm           & 0.028985   & 0.028862   & 0.028502   &  0.028250   & 0.023864   \\
\small{Deuteron} & 0.11810    & 0.11711    & 0.11609    &  0.11049    & 0.090169   \\
Triton           & 0.015885   & 0.015736   & 0.015600   &  0.014768   & 0.011919  \\
Total           &  45.591    &  45.362    &  44.827    &  44.080     & 37.016     \\
\hline
\end{tabular}
\caption{Particle fluence in units of $[$particles$/\UcmZ^{2}]$ for flights OMAA-KLAX, OMDB-KLAX, OTBD-KLAX, OEJN-KLAX, SCEL-YSSY at FL 430.}
\label{tab:fluenceparticle1}
\end{table}
%
\begin{table}[h!]
  \centering
\begin{tabular}{cccccc}
\hline
  \small{Particle}   & \small{OMAA-KLAX}   & \small{OMDB-KLAX}    & \small{OTBD-KLAX}  & \small{OEJN-KLAX} & \small{SCEL-YSSY} \\\hline
\Pn              	&	43.85\%	&	43.79\%	&	43.84\%	&	43.22\%	&	42.37\%	\\
\Pp              	&	19.62\%	&	19.60\%	&	19.62\%	&	19.41\%	&	19.17\%	\\
\PGg             	&	13.77\%	&	13.80\%	&	13.77\%	&	14.14\%	&	14.57\%	\\
\Pem             	&	7.08\%	&	7.10\%	&	7.08\%	&	7.29\%	&	7.53\%	\\
\Pep             	&	7.00\%	&	7.02\%	&	7.01\%	&	7.21\%	&	7.45\%	\\
alpha            	&	2.99\%	&	2.98\%	&	2.99\%	&	2.96\%	&	2.98\%	\\
\PGmm            	&	1.48\%	&	1.48\%	&	1.48\%	&	1.51\%	&	1.57\%	\\
\PGmp            	&	1.48\%	&	1.48\%	&	1.48\%	&	1.51\%	&	1.57\%	\\
\PGpm            	&	0.11\%	&	0.11\%	&	0.11\%	&	0.11\%	&	0.11\%	\\
\PGpp            	&	0.11\%	&	0.11\%	&	0.11\%	&	0.11\%	&	0.12\%	\\
\small{Deuteron} 	&	0.37\%	&	0.37\%	&	0.37\%	&	0.37\%	&	0.36\%	\\
Triton           	&	0.09\%	&	0.09\%	&	0.09\%	&	0.09\%	&	0.09\%	\\
Total percentages       &	97.95\%	&	97.95\%	&	97.94\%	&	97.94\%	&	97.89\%	\\
Total EARD $[\UuSvZ]$   & 128.500  & 127.570  & 126.300  & 121.530  & 99.074    \\
\hline
\end{tabular}
\caption{Type of particle contribution to the ICRP-103 EARD $E^{route}_{FL}$ for flights OMAA-KLAX, OMDB-KLAX, OTBD-KLAX, OEJN-KLAX, SCEL-YSSY at FL 430.}
\label{tab:EARDparticle1}
\end{table}
%
%

%
%

\section{Discussion}%
From Figure~\ref{fig:oneflight} we could see
the tendence along the time of the ICRP definitions has gone in the way to be more loose, 
being the definitions of ICRP-60 more restrictive compared with ICRP-103.
\\\\Also we can note definitions ICRP-103 and ICRU $H^{*}(10)$ take both almost the same values. 
In particular, on OMAA-KLAX aerial route, for $z \in [200,350]\UFL$ the discrepancy between ICRU $H^{*}(10)$ and ICRP-103 varies between 0 and $9\UuSv$, while
for $z \in [350,500]\UFL$ the discrepancy varies between 8 and $10\UuSv$. 
For all the other routes the discrepancies are smaller than showed for OMAA-KLAX.
\\\\We have found finer results in $z$ for the EARDR $\dot{E}(z)$ for our sample of routes,
than those presented in the big works ICRU-84 \cite{ICRU84}, 
EURADOS \cite{EURADOS} or ICRP-132 \cite{ICRP132},
where they use their own sample of flight aerial routes. So we could show the dependence of the absorbed radiation dose
for CARI-7A ICRP-103 specifications as a function of the flight level 
and the flight time in hours, where we have taken the maximum number of flight hours for pilots 
to be one thousand as recommended by the FAA.
At first sight the algorithm to calculate the absorbed effective radiation doses from the ICRP-103
\cite{ICRP103} is fairly acceptable. But when we pay attention to the geometry for the reference human phantom, 
the first thing that we have to face is that the radius of the spherical phantom has value $15\Ucm$ 
and density $1\Ug/\UcmZ^{3}$, therefore we obtain a phantom mass of $14.137 \Ukg$. 
Thus, they are definitely losing the sense of a good approximation for the masses, cross-section, volume, etc. of the reference human phantom,
but this is balanced by the algorithm with another variables that help to compute the statistical radiation and tissue weights shown in Tables~
\ref{tab:radiationweights}, \ref{tab:tissueweights}
in the ALARA\footnote{ALARA: As Low As Reasonably Achievable.} scheme.
\\\\Nevertheless, we adhered to the ICRP-103 recommendations and after comparing the reference thre\-shold values for radiation ex\-po\-su\-re 
from \cite{ICRP103}, this is $E<5\UmSv$ for workers in or near to radiation areas, 
with the observed in the radiation charts for aviation Figure~\ref{fig:maps}, after flying $1000\Uh$ in a year,
we could see in Table~\ref{tab:podium} the flight level at which the value for the EARD
violates the recommendations, i. e. $E>5\UmSv$
for routes OTBD-KLAX, OMAA-KLAX, OMDB-KLAX, OEJN-KLAX, SCEL-YSSY, WSSS-KEWR, KEWR-WSSS, SAWG-SAWB and KLAX-WSSS. 
However we are in the safe zone $E<10\UmSv$ indicated by ICRP-132 \cite{ICRP132}. 
\\\\Requiring $E>5\UmSv$ for one FAA year of $1000\Uh$ to violate the requirements is equivalent to require
$dE/dt > \frac{5\UmSv}{1000\Uh} = 5\UuSv/\UhZ$.
From Figure \ref{fig:plotfits} the last condition (thought just after completing $1000\Uh$ of flight at an averaged flight level) 
can be checked easily along the $y$-axis and we put attention at the data values yielding above $5\UuSv/\UhZ$. We note
for the routes OTBD-KLAX, OMAA-KLAX, OMDB-KLAX, we reach $10\UuSv/\UhZ$ just at $z=50000\Uft$. 
\\\\This study is not conclusive because we are not taking into account in this analysis neither the aircraft shielding nor  
the internal atmosphere with more than one human phantom inside, but it had shown new results and perspectives for this classic problem
to determine the effective absorbed radiation dose by a spherical phantom composed of human-like material that is inside an aircraft without shield
that fly in the Earth atmosphere.
In this scenario, another technique to compute the Cherenkov contribution applied on another set of routes can be seen in \cite{Quinonez2017}.

\section{Conclusions}
Counterintuitively the flights with major duration are not the more irradiated ones.
Instead, long geodetic transpolar flights are more irradiated than long geodetic flights that does not fly over the poles.
\\\\After one flight we found for all the routes at flight level values considered here obey:
\begin{equation}
E_{ICRP103} \leq H^{*}(10).
\label{eq:EH10}
\end{equation}
Therefore, if we allow it, everything we have concluded here for $E_{ICRP103}$ will stand for $H^{*}(10)$ approximately.
\\\\The radiation dose after one flight Figure \ref{fig:oneflight}, and the radiation dose charts for ICRP-103 specifications in a recommended FAA year Figure~\ref{fig:maps}, 
and consequently the cosmic aeroradiation polynomials Figure~\ref{fig:plotfits},
when no considering the wind or the Earth rotation, are invariants under the interchange of aircrafts. 
This means when we plot the same figures for two different aircrafts with similar velocity profiles, the plots shows the same.
The interchangeable aircrafts considered in this study are: Airbus~380 and Boeing~777 ($v_c = 907\Ukm/\UhZ$), Boeing~787 ($v_c = 913\Ukm/\UhZ$).
\\\\With base on Figure~\ref{fig:oneflight} data, we were able to construct the radiation dose charts for aviation and plot them in Figure~\ref{fig:maps}, 
which was the main goal of this work. With these charts users can easily check for the EARD having the information of the average flight level and analyzing in depth the simulation data for one flight in each route at every altitude  $z$ Figure~\ref{fig:oneflight},
we could derive (\ref{eq:main}). Applying this for all the simulated data we get numerical tables for the EARDR $dE/dt$ as function of altitude $z$ in the 
range $20000\Uft$ to $50000\Uft$. By requiring reduced chi squared minimum and
doing the fits according to functions (\ref{eq:power}), (\ref{eq:powershift}), (\ref{eq:pn}), (\ref{eq:odd5}) and (\ref{eq:even4}),
we could find the cosmic aeroradiation polynomials displayed in Table \ref{tab:fits} and plotted in Figure \ref{fig:plotfits}.
\\\\We did a ranking with the aerial flight routes according to their irradiation rates, 
the first four of this ranking have in common that all start from
middle east and arrive at Los Angeles passing over the true north pole of the Earth.
These flights are: OTBD-KLAX, OMAA-KLAX, OMDB-KLAX, OEJN-KLAX. 
Routes OTBD-KLAX, OMAA-KLAX, OMDB-KLAX are best adjusted each one with a 5-th order odd polynomial, 
OEJN-KLAX is best adjusted by a 4-th order polynomial but the 5-th order odd polynomial
is the second best option for it according to the reduced chi squared criterion.
The fifth flight route in the ranking is SCEL-YSSY that starts from Santiago arrives at Sydney
passing over the Antarctica.
Sixth and seventh corresponds to WSSS-KEWR (pacific side) and KEWR-WSSS (atlantic side) are the longest non stop routes in the world.
The eighth is SAWG-SAWB which is the shortest duration flight in the sample with flight time equals to $3.2159\Uh$ (see Table~\ref{tab:podium}),
but presented a high EARDR. 
Thus opening the possibility to find another short duration flights with high EARDR.
OEJN-KLAX, SCEL-YSSY, WSSS-KEWR, KEWR-WSSS and SAWG-SAWB are best adjusted with a 4-th order polynomial.
Ninth position is for KLAX-WSSS and the EARDR is best adjusted with a 3-th order polynomial,
but the second best option for KLAX-WSSS is a 4-th order polynomial. 
The following positions in our ranking are routes very close in duration time or in the end-points to KLAX-WSSS resulting that have a similar behavior 
in their cosmic aeroradiation polynomials, i. e. all are best described with a 3-th order polynomial, these are
tenth YPPH-EGLL, eleventh FAOR-KATL, twelfth KSFO-WSSS, thirteenth NZAA-OTBD and fourteenth NZAA-OMDB. Fifteenth and sixteenth
are best described with a 4-order even polynomial and they are KDFW-YSSY and KIAH-YSSY respectively; but the second best option for the last routes are 
3-th order polynomials. Therefore we had just characterized geographically and kinematically the cosmic radiation absorption rates at Earth atmosphere in terms of 
the structure and order of polynomials in the altitude, see Table \ref{tab:fits}, Figures \ref{fig:plotfits} and \ref{fig:globo}. 
\\\\In the most to the right column of Table~\ref{tab:podium}, we can see the flight level whichin nine aerial routes actually 
begin to violate the ICRP103-recommendations \cite{ICRP103} after flying $1000\Uh$ in a year as recommended by the FAA. 
This conclusion did not take into account the airplane shielding.
\\\\With the leading EARDR flight route OTBD-KLAX, we can show for the samples considered here a thumb rule to calculate the EARD 
and estimate $H^{*}(10)$ (\ref{eq:EH10})
as a function of the altitude $z$ and the flight time $t$ of any route in the world according to the
fitter power function (\ref{eq:power}), resulting the fit parameters with their errors are $k_{1}=(3.85 \pm 0.89)\times 10^{-5}\frac{\UuSvZ}{\UhZ\cdot\UFLZ^{2}}$ and $k_{2}=2.01\pm 0.03$,
i. e.
\begin{equation}
E \leq \bigg(12 \frac{\UpSvZ}{\UsZ\cdot\UkmZ^{2}}\bigg) \; t \; z^{2} = \bigg(3.8\times10^{-5} \frac{\UuSvZ}{\UhZ\cdot\UFLZ^{2}}\bigg) \; t \; z^{2}. 
\label{eq:thumb}
\end{equation}
Finally we note the beauty of (\ref{eq:thumb}) because its dependence is linear in flight time and quadratic in altitude.

\section*{Acknowledgements}
Author FQ gladly acknowledge the support provided by the Fuerza A\'erea Colombiana via CODALTEC contract-00214 of 2017.
FQ also wants to acknowledge to the Universidad Industrial de Santander for the teaching contract June-July 2018 and
Universidad Manuela Beltr\'an for the teaching contract August-December 2018. 
LAN gratefully acknowledge the permanent support of Vicerrector\'ia de Investigaci\'on y Extensi\'on de la Universidad Industrial de Santander.
EAC acknowledge the constant support of Fuerza A\'erea Colombiana.
PAOH acknowledge the support of Vicerrector\'ia Acad\'emica y de Investigaci\'on de la Universidad Santo Tom\'as with
the project No. GIFIMECACBASICP12017.
VSBG and JLGA acknowledge the support of Vicerrector\'ia de Investigaci\'ones de la Universidad de Pamplona with
acta No. 003 of 22/11/2017.\\
%
\bibliographystyle{unsrt}

\end{document}